\documentclass[twocolumn,twocolappendix]{aastex631}
\usepackage{color}
\usepackage{amsmath,amssymb}
\usepackage{graphicx}

\received{}
\revised{}
\accepted{}
\submitjournal{ApJS}

\shorttitle{Systematic study of preSN- and early SN neutrino emission}
\shortauthors{Kato et al.}

\begin{document}

\title{Comprehensive Neutrino Light Curves and Spectra: From Presupernova Evolution to Early Supernova Phase}

\correspondingauthor{Chinami Kato}
\email{chinami.kato@phys.s.u-tokyo.ac.jp}

\author{Chinami Kato}
\affiliation{Department of Physics, Graduate School of Science, The University of Tokyo, 7-3-1 Hongo, Bunkyo-ku, Tokyo 113-0033, Japan}

\author{Hiroki Nagakura}
\affiliation{Division of Science, National Astronomical Observatory of Japan, 2-21-1 Osawa, Mitaka, Tokyo 181-8588, Japan}

\author{Akira Ito}
\affiliation{Department of Physics and Applied Physics, School of Advanced Science \& Engineering, Waseda University, Tokyo 169-8555, Japan}

\author{Ryosuke Hirai}
\affiliation{Astrophysical Big Bang Laboratory (ABBL), Pioneering Research Institute, RIKEN, Wako, Saitama 351-0198, Japan}
\affiliation{School of Physics and Astronomy, Monash University, Clayton, Victoria 3800, Australia}
\affiliation{OzGrav, Australian Research Council Centre of Excellence for Gravitational Wave Discovery, Australia}

\author{Shun Furusawa}
\affiliation{College of Science and Engineering, Kanto Gakuin University, Kanagawa 236-8501, Japan}

\author{Takashi Yoshida}
\affiliation{Yukawa Institute for Theoretical Physics, Kyoto University, Kitashirakawa-Oiwakecho, Sakyo-Ku, Kyoto 606-8502, Japan}

\author{Ryuichiro Akaho}
\affiliation{Department of Physics and Applied Physics, School of Advanced Science \& Engineering, Waseda University, Tokyo 169-8555, Japan}

\begin{abstract}
We present the first systematic study of neutrino emissions from massive stars, continuously tracking the late evolutionary stages through the early core-collapse supernova phase. Using progenitor and supernova models, we analyze the neutrino luminosities and spectra for progenitors with initial masses of 10--40~$M_\odot$. Our systematic analysis reveals that the compactness parameter ($\xi_{2.5}$) and carbon-oxygen core mass ($M_{\text{CO}}$) exhibit strong correlations with neutrino emission. In the presupernova phase, the time-integrated number of neutrinos correlates with $\xi_{2.5}$ when integrated over the final day and with $M_{\text{CO}}$ for longer durations. For the early supernova phase ($<200$ ms postbounce), the neutrino properties are relatively insensitive to the specific stellar evolution code used, allowing for a reliable extraction of physical correlations. We confirm that the neutrino emission features, including the electron neutrino burst properties and accretion-powered luminosity of other species, reflect the progenitor's compactness. An evaluation of the observational feasibility for a nearby progenitor using a false alarm rate approach suggests that these correlations can persist even under practical detection conditions. Such a joint analysis of both phases provides complementary constraints on the internal structure. All calculated time-series data will be made publicly available.
\end{abstract}

\keywords{Massive stars; Core-collapse supernova; Supernova neutrinos}

\section{Introduction}

Massive stars ($M\gtrsim 8~M_\odot$) undergo complex evolution following hydrogen exhaustion, progressing through the successive burning stages of heavy elements. This process leads to the formation of a stratified structure with multiple burning shells surrounding a degenerate iron core, whose gravitational collapse triggers a core-collapse supernova (CCSN). Neutrinos play a crucial role in these final evolutionary stages. During the late presupernova (preSN) phase, thermal neutrino cooling and weak interactions, particularly electron capture, govern the entropy profile and electron fraction ($Y_e$) of the core, driving it toward gravitational instability. However, theoretical predictions for the final core structure suffer from large uncertainties due to poorly constrained input physics, such as mass-loss rates, binary interactions, and mixing processes \citep[e.g.][]{Smith2014,Sana2012,Renzo2017a}. Because conventional electromagnetic observations can only probe the stellar surface, deep interior evolution remains observationally inaccessible. Consequently, these uncertainties result in diverse predictions for the final core compactness and thermodynamic profile, which fundamentally determine the subsequent explosion dynamics \citep[e.g.][]{OConnor2011,Ertl2016,Sukhbold2016}.

Following the core collapse, the central density reaches nuclear saturation, triggering a bounce that launches an outward shock wave. However, energy losses from the photodissociation of infalling heavy nuclei and intense neutrino emission cause this bounce shock to stall within the iron core. At the center, the newly formed proto-neutron star (PNS) generates an enormous neutrino flux.  The absorption of a fraction of these neutrinos behind the stalled shock deposits the energy necessary to revive it, which is the crux of the widely accepted neutrino-driven explosion mechanism. During this early SN stage, the emitted neutrinos are primarily shaped by the internal structure and thermodynamic characteristics of the progenitor’s inner core.

Once the shock wave stalls, however, the dynamics become highly complex. Hydrodynamic instabilities such as Standing Accretion Shock Instability (SASI) and convection emerge \citep[e.g.][]{Takiwaki2014,Mueller2020,Burrows2020,Vartanyan2025}, significantly influencing the explosion's outcome and the resulting neutrino signals. Furthermore, the explodability is highly sensitive to the structure of the progenitor's outer core and Si/O layers, as well as to neutrino reaction rates. While many recent state-of-the-art multidimensional simulations can successfully reproduce explosions \citep[see recent reviews, e.g.][]{Mueller2020,Janka2025,Burrows2021,YAMADA2024}, incorporating the effects of collective neutrino oscillations near the neutrinosphere into self-consistent explosion models remains an active area of research \citep[e.g.][]{nagakura2023,Ehring2023b,Xiong2024,Mori2025,Akaho2026}. Ultimately, because neutrinos dictate the explosion dynamics, future observations are highly anticipated to provide a wealth of invaluable insights.

On the observational front, neutrino detection capabilities have improved dramatically since the landmark detection of SN 1987A \citep{Hirata1987,Bionta1987,Alexeyev1988}. The current and future landscape is defined by massive detector volumes, such as Super-Kamiokande \citep[SK;][]{Abe2014} and the upcoming Hyper-Kamiokande \citep[HK;][]{HKPC2018}, alongside low-energy threshold liquid scintillator detectors like KamLAND \citep{Berger2009} and JUNO \citep{JUNOCollaboration2022}. These facilities are further supported by a diverse global network of next-generation detectors utilizing various target materials and detection channels \citep[e.g.][]{Abe2022,DUNE2022,XenonCollaboration2024}. For a CCSN occurring at the Galactic center, SK is expected to detect several thousand neutrinos, while HK could record tens of thousands of events \citep{AlKharusi2021}. Such guaranteed high-statistics observations will provide an unprecedented opportunity to decipher SN dynamics and rigorously test theoretical models of massive star evolution.

Leveraging these enhanced capabilities, the detection of neutrinos emitted prior to core bounce—so-called presupernova neutrinos—has become a tangible goal \citep{Odrzywolek2004a}. While their low energies and fluxes pose a challenge, theoretical studies have confirmed that nearby progenitors ($\lesssim 1$~kpc) are detectable with current and next-generation detectors \citep[e.g.][]{Kato2020b}. The observation of preSN neutrinos serves two pivotal roles. First, it functions as an early warning for electromagnetic and gravitational wave observatories. For a nearby source like Betelgeuse (200~pc), various detectors could issue alerts days in advance (JUNO: \citealt{Li2020,Abusleme2024}; SK-Gd: \citealt{Simpson2019}; KamLAND: \citealt{Asakura2016}). This lead time ensures preparedness for the subsequent explosion, enabling a comprehensive understanding through multimessenger astronomy. Currently, several neutrino-triggered alert systems have already been implemented \citep{Machado2022, Abe2024a, AlKharusi2021}. Second, preSN neutrinos provide a direct window into the progenitor's deep interior. Previous studies have discussed their potential to identify the progenitor type \citep{Kato2015,Kato2017}, reveal shell burning and convection details \citep{Yoshida2016}, and probe the thermal state near the core \citep{Patton2017}. Unlike indirect constraints derived from SN remnants, which rely on complex explosion models, these neutrinos carry direct signatures of the core's evolutionary stage. To fully exploit this potential, it is essential to construct theoretical models that consistently predict neutrino luminosities and spectra across both the preSN and SN phases.

Advances in neutrino detection technology now enable the continuous monitoring of neutrinos across multiple evolutionary stages of massive stars. Consequently, a systematic and consistent treatment of neutrino emission from the preSN phase through the explosion is of paramount importance. On the theoretical front, numerous studies have investigated the correlations between progenitor properties (e.g. compactness) and SN observables, including explodability, neutrino emission, and gravitational waves \citep[e.g.][]{OConnor2011,Ertl2016,Sukhbold2016,Warren2020,Nagakura2021,Nagakura2023c}. However, these works have primarily focused on the postbounce phase, leaving a gap in consistently linking the preSN phase to the early SN phase. To bridge this gap, we present comprehensive calculations of neutrino luminosities and spectra covering the period from several hundred years prior to collapse up to 200~ms postbounce, utilizing one-dimensional stellar evolution and SN explosion models. Using these simulations, we investigate the correlations between key stellar parameters and neutrino observables, evaluating the potential for both preSN and SN neutrino detection. We demonstrate that the time-integrated neutrino luminosity exhibits strong correlations with the CO core mass ($M_{\text{CO}}$) and the compactness parameter ($\xi_{2.5}$), depending on the integration window. These findings demonstrate that preSN and early SN neutrino signals serve as independent probes of the stellar interior, providing a robust framework for tightly constraining the progenitor's internal structure through future multiphase observations.

The paper is organized as follows. In Section~\ref{ch2}, we describe our methodology for modeling the quasi-static stellar evolution up to the onset of core collapse, the subsequent dynamical evolution, and the estimation of neutrino luminosities and spectra. In Section~\ref{ch3}, we present our main results regarding the time evolution of neutrino emissions from the preSN to the early SN phase. This is followed by a correlation analysis in Section~\ref{ch4}, where we identify the key progenitor parameters strongly related to neutrino emission. In Section~\ref{ch5}, we discuss observational prospects and the potential to infer the internal structure of progenitors using these correlations. Synthesizing the arguments presented so far, Section~\ref{ch6} provides a combined discussion of both preSN and early SN neutrino observations. Finally, Section~\ref{ch7} discusses several uncertainties inherent in our study, and Section~\ref{ch8} summarizes our findings and conclusions.


\begin{figure}[htpb]
    \centering
    \includegraphics[width=\columnwidth,clip]{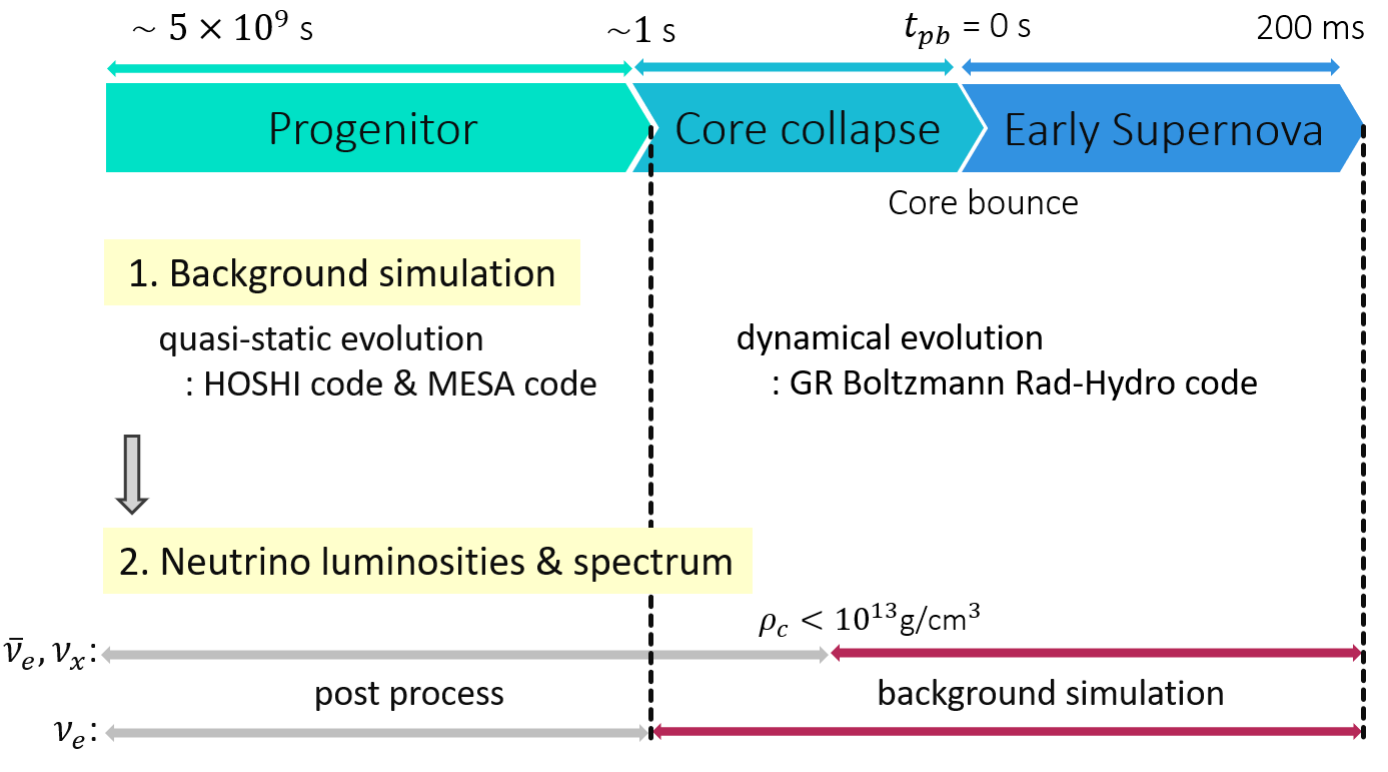}
    \caption{Method outline in this paper.}
    \label{fig:outline}
\end{figure}

\section{Methods} \label{ch2}

In this study, we systematically calculate neutrino luminosities and spectra for 30 models with zero-age main-sequence (ZAMS) masses ranging from 10 to 40 $M_{\odot}$ at solar metallicity. Our calculation methods are summarized in Figure~\ref{fig:outline}. The procedure consists of two main steps: (1) background calculations from the preSN to the early SN phase (Section~\ref{bgsim}), and (2) the derivation of neutrino luminosities and spectra (Section~\ref{lumcalc}).

\subsection{Numerical Setup for Stellar Models and Supernova Models} \label{bgsim}

\subsubsection{Progenitor Phase} \label{subsec:setup}

We compute the quasi-static stellar evolution up to the onset of core collapse using both the HOngo Stellar Hydrodynamics Investigator (HOSHI) code \citep{Takahashi2016, Takahashi2018} and the public stellar evolution code MESA (v24.08.1; \citealt{Paxton2011, Paxton2013, Paxton2015, Paxton2018, Paxton2019, Jermyn2023}). Specifically, we adopt 23 models generated by HOSHI and seven models by MESA.

We first introduce the numerical setup for the HOSHI models. While we refer the reader to \citet{Yoshida2019} for full computational details, we highlight a few key features here. First, the Ledoux criterion is used to identify convectively unstable regions, and mixing within these regions is treated using mixing-length theory. Convective overshooting is modeled with an exponentially decaying coefficient (see eq.~(1) in their paper). We adopt their model M, which applies a lower overshoot parameter ($f_{\mathrm{ov}}=0.01$) prior to core carbon burning. After this phase, convective overshooting is disabled. A 300-species nuclear reaction network is incorporated, and wind mass-loss is also taken into account.

We define the onset of core collapse as the moment when the central temperature exceeds $T_c\sim10^{9.9}$~K. Although this threshold is somewhat arbitrary, we demonstrate in Appendix~\ref{appendix_hydro} that altering this criterion only affects the evolution immediately prior to core bounce, thereby minimally influencing the overall time evolution of the neutrino emission. Most importantly, as shown in Section~\ref{ch4}, the results regarding the total number of neutrinos integrated over a fixed duration remain robust and independent of this definition.

The second set is the MESA models. We mostly adopt settings from the \texttt{20M\_pre\_ms\_to\_core\_collapse} test suite. No mass loss is applied for simplicity. For the nuclear network, we utilize the adaptive nuclear network where isotopes are adaptively added and subtracted from the network based on their abundance and reactions. We again apply a relatively low overshooting parameter to loosely match the $M_\mathrm{ZAMS}$--$M_\mathrm{core}$ relation in the HOSHI models, where $M_\mathrm{ZAMS}$ is the ZAMS mass and $M_\mathrm{core}$ is the final He core mass. Specifically, we use $f_\mathrm{ov}=0.1$ with step overshoot for the overshoot above the H-burning core, and $f_\mathrm{ov}=0.01$ with exponential overshoot for the He-burning core. Only a tiny amount of overshoot is applied for all other convective regions to maintain numerical stability.

The time of core collapse is defined when the Fe core's material has velocities exceeding 100~km/s, which is the default setup in the MESA code\footnote{For 30 and 35~$M_\odot$ progenitors, when the snapshot at the collapse time determined under this condition are used as the initial state for subsequent dynamical calculations with the CCSN code, the Fe core, once unstable, restabilized and the core collapse stops. This is thought to be due to differences in the equations of state used by the MESA and CCSN codes. Therefore, for these models, we performed calculations in the MESA code until the falling matter velocity becomes even faster, using this as the initial condition for the subsequent dynamical calculation. The difference in collapse time compared to the time determined under the default MESA conditions is several hundred milliseconds for both 30 and 35~$M_\odot$ progenitors. As stated in Appendix~\ref{appendix_collapse}, we confirm its impact on the subsequent evolution to be negligible.}. Although the definition used here differs between the HOSHI and MESA models, we confirmed that the time difference in the onset of collapse between the two definitions is extremely short ($\sim$hundreds of milliseconds) and has no effect on subsequent dynamics or neutrino quantities (see Appendix~\ref{appendix_collapse}).

\subsubsection{Dynamical Evolution after Core Collapse}

We simulate the dynamical evolution following core collapse using our general-relativistic Boltzmann radiation-hydrodynamics code \citep{Nagakura2014, Nagakura2017, Nagakura2019b, Akaho2021, Akaho2023}. While the reader is referred to these papers for complete details, we summarize the essential features here. We employ the radial-gauge polar-slicing condition for the spatial metric as in \citet{Akaho2025a} and \citet{Akaho2026}, which is fully general relativistic in 1D. We use the equation of state (EOS) provided by \citet[][hereafter FT EOS]{Furusawa2017}. In this code, the Boltzmann equations are solved fully for neutrino transport. We consider three neutrino species: electron neutrinos ($\nu_e$), electron antineutrinos ($\bar{\nu}_e$), and heavy-lepton neutrinos ($\nu_x$), where the latter collectively represents mu and tau neutrinos and their antineutrinos. We deploy 20 energy grid points logarithmically spaced between 1 and 300~MeV, alongside 10 angular bins in momentum space. The following neutrino reactions are included:

\begin{enumerate}
\item{Emission and absorption: electron captures on nuclei and free nucleons, electron-positron pair annihilations, nucleon-nucleon bremsstrahlung, and their inverse reactions.}
\item{Neutrino scattering: isoenergetic scattering on free nucleons, coherent scattering on nuclei, and nonisoenergetic scattering on electrons and positrons.}
\end{enumerate}
For the reaction rates of charged-current processes and scattering, we primarily adopt the standard set of \citet{Bruenn1985} with extensions from \citet{Sumiyoshi2005}. We note that the chemical composition in nuclear statistical equilibrium (NSE), which is required to calculate neutrino emission from electron captures on nuclei, is consistently evaluated using the FT EOS.

The simulations are terminated at 200~ms postbounce. Up to this point, although multidimensional motions such as prompt convection occur, their impact on the neutrino signals is limited. Therefore, these features can be discussed within the framework of spherically symmetric models, at least qualitatively \citep{Nagakura2021a}. Conversely, once the shock wave stalls, this 1D approximation is no longer sufficient. Multidimensional fluid instabilities, such as SASI and neutrino-driven convection, not only dictate the explodability but also significantly reshape the neutrino signals. Furthermore, collective neutrino oscillations, such as fast flavor and collisional instabilities, alter both the dynamics and the neutrino observables \citep[e.g.][]{nagakura2023,Ehring2023b,Xiong2024, Wang2025a, Akaho2026}. Because performing systematic multidimensional dynamical calculations with full Boltzmann transport incurs prohibitive computational costs, we restrict our analysis of neutrino luminosities to the first 200~ms; investigations of later phases are deferred to future work.

\begin{figure*}[htpb]
    \centering
\includegraphics[width=17cm,clip]{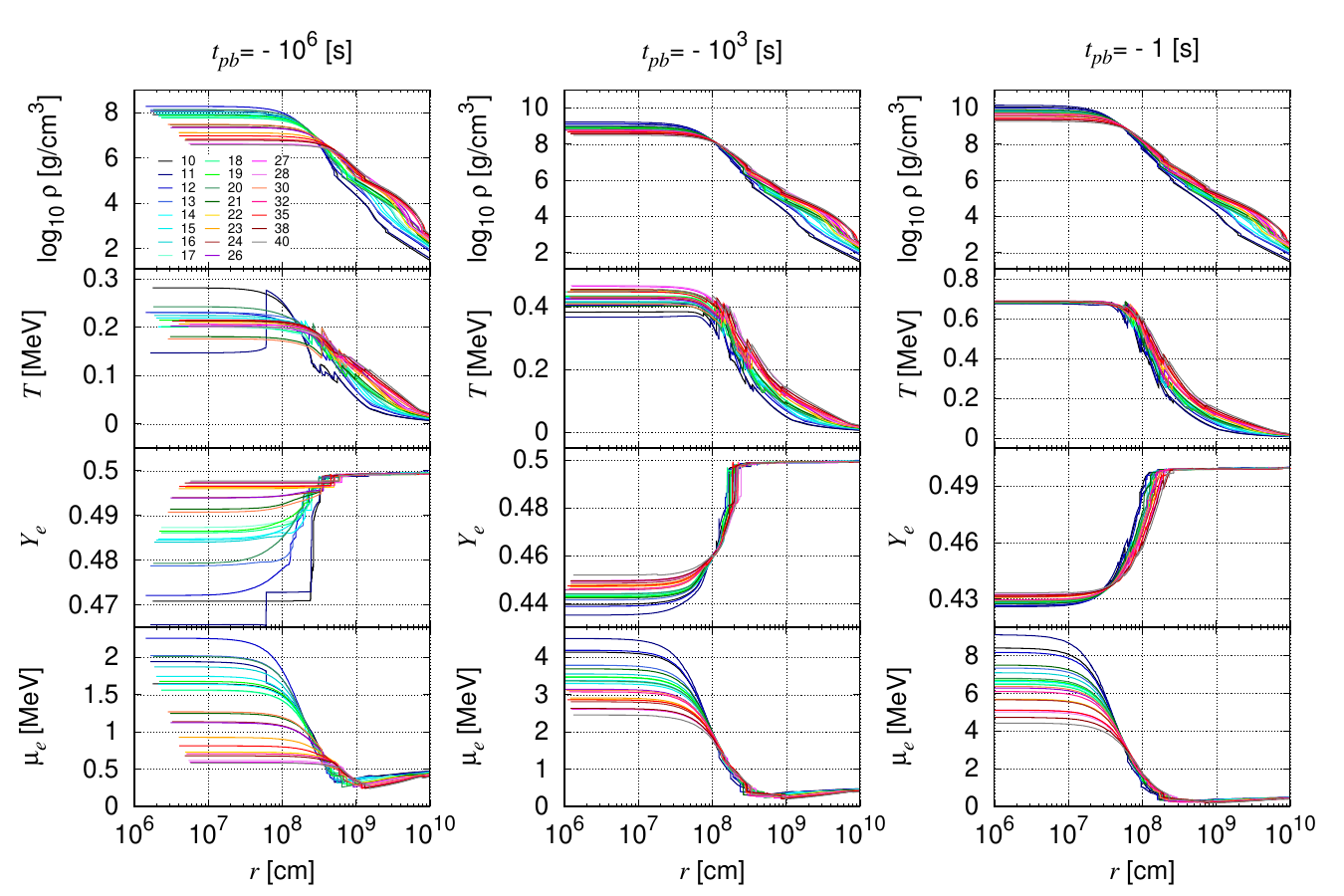}
\includegraphics[width=17cm,clip]{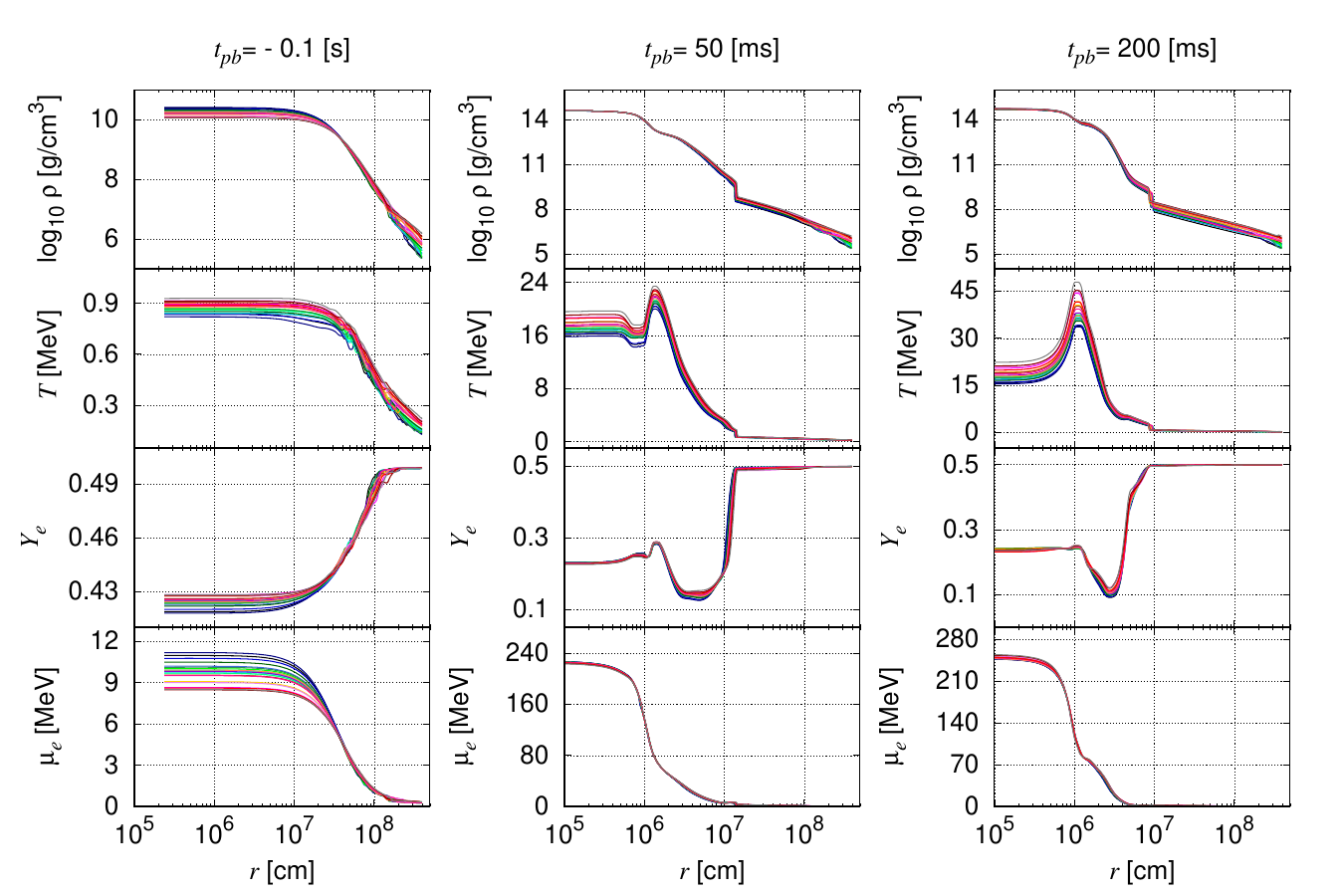}
    \caption{Radial profiles of density, temperature, electron fraction, and electron chemical potential from top to bottom. We show six profiles at $t_{pb}\sim -10^6$~s, $-10^3$~s, $-1$~s, $-0.1$~s, 50~ms and 200~ms. Colors distinguish the progenitor models.}
    \label{fig:progenitor_radial_prof}
\end{figure*}

\subsection{Derivation Method of Neutrino Luminosities and Spectra} \label{lumcalc}

The overall method for deriving neutrino luminosities and spectra largely follows our previous work \citep{Kato2017}. Therefore, only the primary points are reviewed here.

During the progenitor phase, the stellar evolution calculations only track the total neutrino energy loss. Consequently, we calculate the detailed neutrino spectra in a postprocessing step. Six neutrino processes are included: electron-positron pair annihilation (pair), $\beta^{\pm}$ decay by nuclei ($\beta^{\pm}$ decay), $e^{\pm}$ capture by nuclei (EC, PC), and $e^-$ capture by free protons (free p). $\nu_e$ are produced by pair, $\beta^+$ decay, EC, and free p, while $\bar{\nu}_e$ are produced by pair, $\beta^-$ decays, and PC. For $\nu_x$ and $\bar{\nu}_x$, we include only the emission by pair.

We use the same reaction rate of pair as Eqs.1-9 in \cite{Kato2017}, which follows the formulation by \cite{Mezzacappa1993}. Their reaction rates are flavor dependent because the weak coupling constants in the matrix element differ among flavors. The emission rate of the heavy-lepton pair is a few times smaller than the electron one under the same thermal condition \citep{Kato2015}.

For nuclear reactions, we employ the literature tables: FFN \citep{Fuller1982}, ODA \citep{Oda1994}, LMP \citep{Langanke2000}, LMSH \citep{Langanke2003}, and the Tachibana table \citep[TB,][]{Tachibana1995, YOSHIDA2000, Koura2004, Koura2005}. We adopt the data from the table in the following preference: LMSH $>$ LMP $>$ ODA $>$ FFN for $\nu_e$ emission and LMP $>$ ODA $>$ FFN $>$ TB for $\bar{\nu}_e$ emission, respectively. We also employ an approximate formula provided by \citep{Langanke2003} for the EC rate by heavy nuclei where reaction tables are unavailable. For $\bar{\nu}_e$ emission via $\beta^-$ decay, we note that all reaction rates in the TB table are evaluated with Fermi blocking of electrons taken into account by applying a suppression factor $1-f_{\rm eq}(\langle E_e \rangle)$ based on the Fermi-Dirac distribution with the average electron energy $\langle  E_e \rangle$, which is given in the table. The mass fractions $X_i$ in the NSE region ($T>10^{9.7}$~K and $\rho>10^{7.5}$~${\rm g/cm^3}$) are recalculated using the FT EOS \citep{Furusawa2017}. Unlike our previous study, however, we now include neutrino emissions from nonNSE regions. We use the mass fractions in nonNSE regions obtained by nuclear network computations in stellar evolution calculations. As discussed later, this inclusion becomes particularly important during the very early phases of interest. We note that nuclear weak rates still involve uncertainties, and active discussions on their improvements are ongoing (see Section~\ref{ch7}). Continued updates of the weak-rate inputs will therefore be important, and we leave a systematic investigation of their impact for future work.

Because the characteristic timescale of neutrinos during this phase is significantly shorter than the dynamical timescale, neutrino emission and propagation can be treated as stationary. Under this condition, spatially integrating the local neutrino number density ($n_\nu(r,t)$) and spectrum ($dn_\nu/dE_\nu$) yields the neutrino number luminosity ($L^N_\nu(t)$), energy luminosity ($L^E_\nu(t)$), and number spectrum ($dL^N_\nu/dE_\nu$) at any given time:
\begin{eqnarray}
L^N_\nu(t) &=& \int_0^{r_{\rm max}} 4\pi r^2 n_\nu(r,t) dr,\\
L^E_\nu(t) &=& \int_0^{r_{\rm max}} \int_0^{E_{\rm max}} 4\pi r^2 E_\nu \frac{dn_\nu(r,t)}{dE_\nu} dE_\nu dr, \ \ \\
\frac{dL^N_\nu}{dE_\nu}\left(E_\nu,t\right) &=& \int_0^{r_{\rm max}} 4\pi r^2 \frac{dn_\nu(r,t)}{dE_\nu} dr.
\end{eqnarray}
In these expressions, $r_{\rm max}$ is defined as the radius where the incremental contribution to the neutrino luminosity from a single computational cell, $\Delta L^N_\nu$, falls below $10^{-6}$ of the accumulated $L^N_\nu$ during the outward integration. We calculate the neutrino luminosities and spectra starting from $t_{calc}=-5\times10^9$~s relative to the bounce time. Note that the local neutrino energy loss rates calculated here are not guaranteed to be perfectly identical to those used in the stellar evolution models.

As the core begins to collapse and the density increases, the characteristic neutrino timescale becomes comparable to the dynamical one, invalidating the stationary assumption. Frequent neutrino-matter interactions, such as coherent scattering on nuclei, trap and thermalize neutrinos within the core. Consequently, neutrino absorption becomes significant, and Fermi blocking effects play a crucial role in the emission process. Therefore, for the postcollapse phase, we directly extract the neutrino spectra and luminosities from the radiation-hydrodynamics simulations, which explicitly solve the Boltzmann equations. We estimate these quantities from the neutrino distribution function $f_\nu$:
\begin{eqnarray}
\frac{dL^N_\nu}{dE_\nu}\left(E_\nu,t\right) &=&  4\pi r_i^2 \nonumber \\
&\times&\frac{2\pi c}{\left(2\pi \hbar c \right)^3}\int_0^1 E_\nu^2\cos{\theta_\nu}f_\nu\left(E_\nu,\cos{\theta_\nu}\right)  d\cos{\theta_\nu}, \ \ \ \ \\
L^N_\nu(t) &=& \int_0^{E_{\rm max}} \frac{dL_\nu^N}{dE_\nu}  dE_\nu, \\
L^E_\nu(t) &=& \int_0^{E_{\rm max}} E_\nu\frac{dL_\nu^N}{dE_\nu}  dE_\nu.
\end{eqnarray}
Here, $\theta_\nu$ is the angle between the neutrino propagation direction and the radial vector. We evaluate these quantities at two locations: $r=500$~km and a point near the outer boundary of the computational domain. Before core bounce, we adopt the results from the location with the larger luminosity; after core bounce, we consistently use the values at 500~km. To maintain causality across different radii, we redefine the proper time for neutrinos as $t_\nu = t_{\rm hydro} - r/c$. The detailed procedure is summarized in Appendix~\ref{appendix_neutrino_time}. In all subsequent figures related to neutrinos, we use this redefined time $t_\nu$ without further notification.

Even during the core collapse, $\bar{\nu}_e$ and $\nu_x$ are treated in a postprocessing manner until the central density reaches $\rho_c \gtrsim 10^{13}~{\rm g/cm^3}$. This is because $\beta^-$ decay—a key emission process for $\bar{\nu}_e$—is not included in the radiation-hydrodynamics simulations, which would otherwise lead to an underestimation of their luminosity. Most $\bar{\nu}_e$ are emitted in the outer part of the core ($\rho\sim10^{11.5}~{\rm g/cm^3}$), which is relatively optically thin, meaning that their feedback on the collapse dynamics is negligible. For these reasons, the postprocessing treatment remains valid and reasonable \citep{Kato2017}. We also note that $\beta^-$ decay never contributes to $\bar{\nu}_e$ emission after core bounce, as the shock wave completely disintegrates the heavy nuclei and positron captures become the dominant emission channel.

\section{Neutrino Emission from Presupernova to Early Supernova Phases} \label{ch3}

\begin{figure*}[ht]
    \centering
\includegraphics[width=\textwidth,clip]{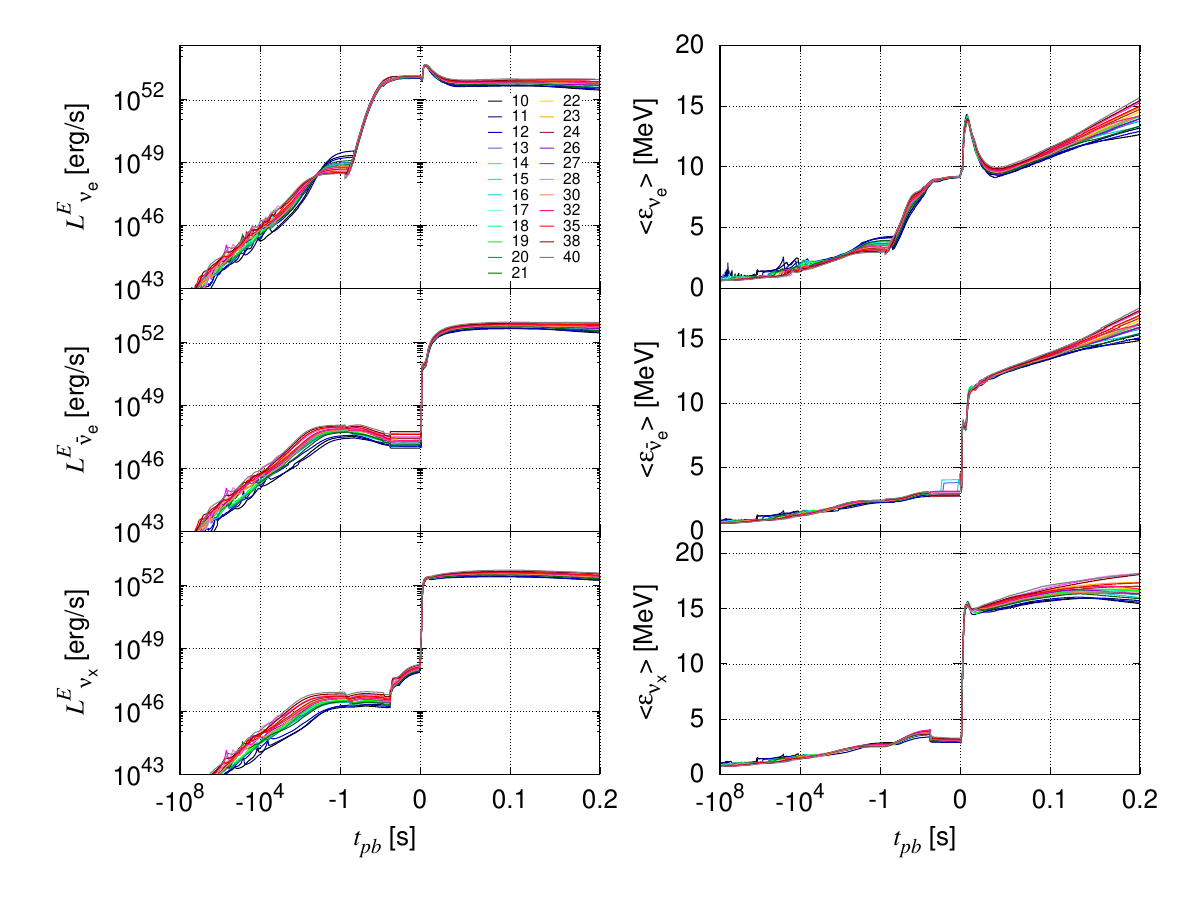}
    \caption{Time evolution of neutrino luminosities (left) and average energies (right) of $\nu_e$, $\bar{\nu}_e$ and $\nu_x$ from top to bottom. Colors distinguish the progenitor models.}  
    \label{fig:numberlum}
\end{figure*}

\begin{figure*}[ht]
    \centering
\includegraphics[width=17cm,clip]{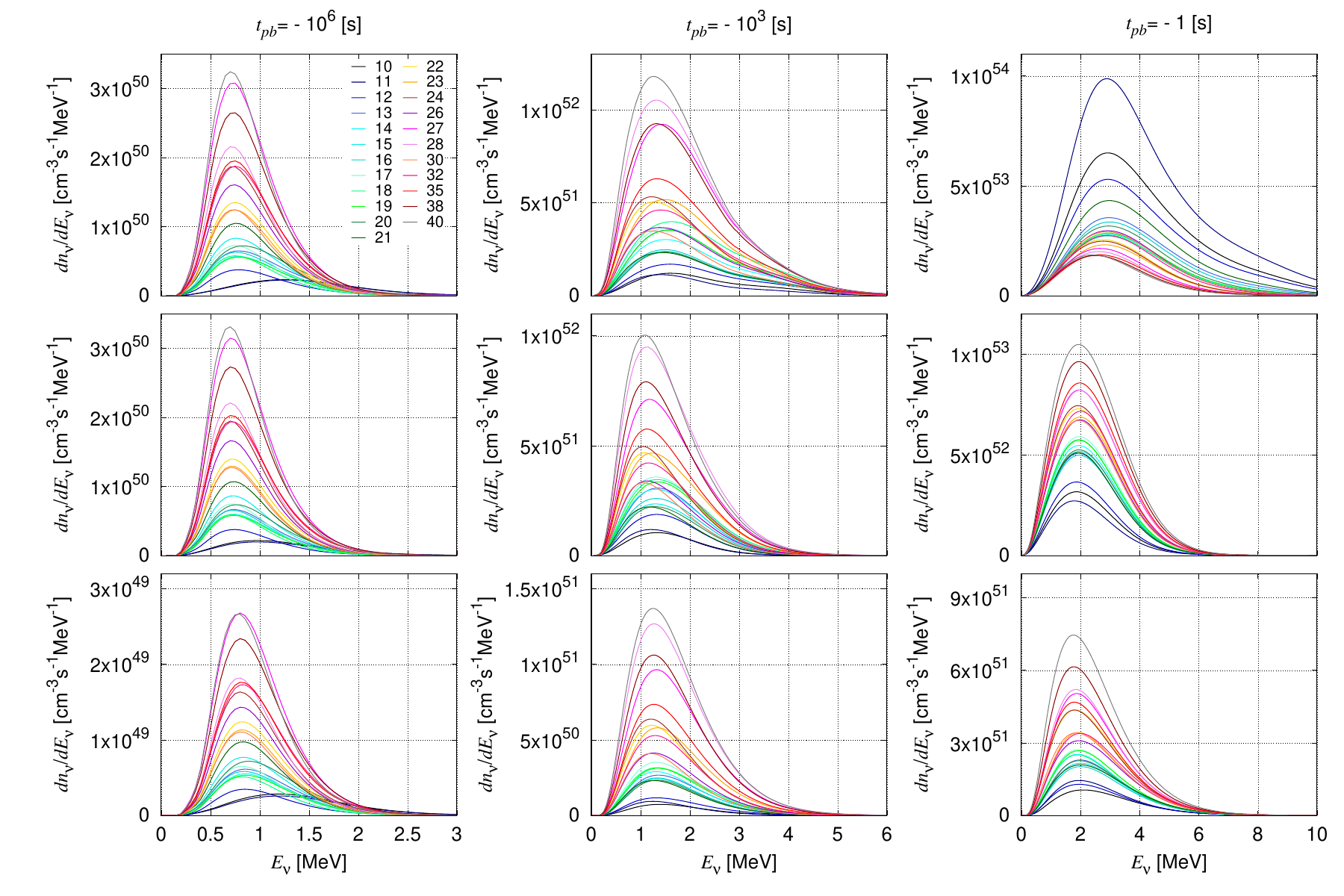}
\includegraphics[width=17cm,clip]{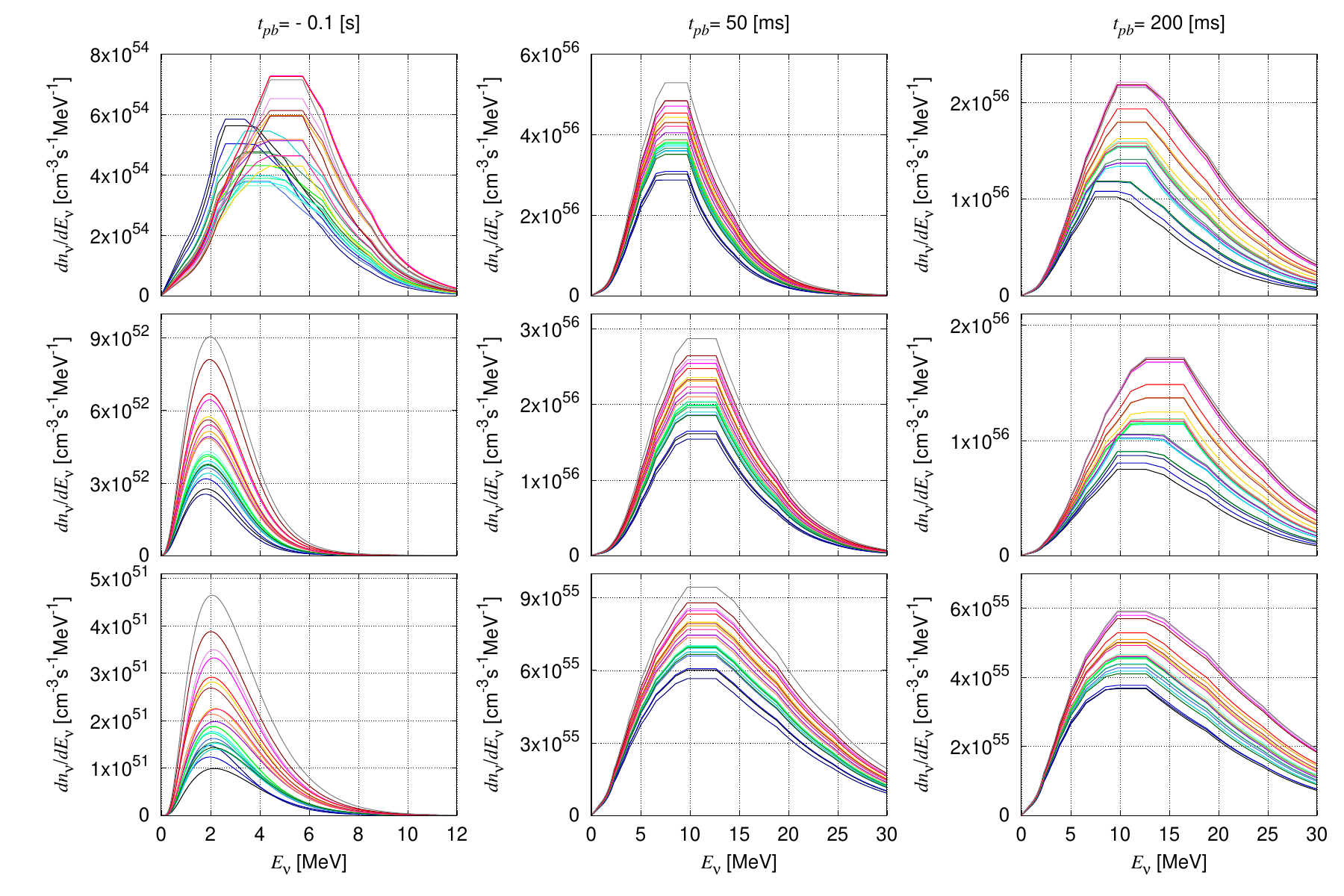}
    \caption{Neutrino spectrum for $\nu_e$, $\bar{\nu}_e$ and $\nu_x$ from top to bottom. We show six profiles at $t_{pb}\sim -10^6$~s, $-10^3$~s, $-1$~s, $-0.1$~s, 50~ms and 200~ms. Colors distinguish the progenitor models.}
    \label{fig:numberspe}
\end{figure*}

In this section, we detail the long-term evolution of neutrino luminosities and spectra from the preSN through the early SN phase. First, we provide an overview of the progenitor and SN models that serve as the foundation for our neutrino emission calculations (Section~\ref{sub:BG}). Based on these models, we then summarize the characteristics of the neutrino luminosities and spectra across each evolutionary stage (Section~\ref{sub:neutrino}). Because preSN neutrino emissions have not yet been systematically investigated in this context, we place particular emphasis on analyzing this phase. Although differences in numerical implementations and input physics between the MESA and HOSHI codes lead to some variations in the resulting stellar evolution models, their overall qualitative trends remain consistent. Therefore, in this section, we primarily base our discussion on the HOSHI models, while the MESA models are comprehensively summarized in Appendix~\ref{appendix_hydro} and \ref{appendix_mesa}.

It is important to note that while $M_{\rm ZAMS}$ is a critical input parameter for numerical simulations, it is not an ideal physical metric for characterizing SN progenitors. Previous studies have demonstrated that the final core structure does not map monotonically to $M_{\rm ZAMS}$. For instance, \citet{OConnor2011} and \citet{Sukhbold2016} showed that the compactness parameter exhibits a complex, nonmonotonic dependence on the initial mass. This complexity is further exacerbated by significant uncertainties in mass-loss rates and binary interactions \citep{Smith2014, Sana2012}. These factors can drastically alter late-stage evolution, meaning that stars with identical $M_{\rm ZAMS}$ may develop divergent preSN structures depending on their mass-loss history or binary status \citep{Podsiadlowski1992,Eldridge2008}. Consequently, we emphasize that throughout this paper, $M_{\rm ZAMS}$ is used solely as a convenient label to distinguish between models, rather than as a predictive physical parameter for our discussions.

\subsection{Overall Features of Background Models} \label{sub:BG}

Before addressing the main results regarding neutrino emission, we briefly overview the background stellar and SN simulations, which provide essential context for the subsequent discussions. We refer the reader to Appendix~\ref{appendix_hydro} for complete details of these models.

We introduce our background models by examining the radial profiles of $\rho$, $T$, $Y_e$ and $\mu_e$ in Figure~\ref{fig:progenitor_radial_prof}.
We select snapshots at $t_{pb} \sim -10^6$, $-10^3$, and $-1$~s to represent the progenitor phase, $t_{pb} \sim -0.1$~s for the core collapse phase and $t_{pb} \sim 50$ and $200$~ms for the early SN phase. Here, time is measured relative to core bounce ($t_{pb}$), which means that the preSN and SN phases correspond to $t_{pb} < 0$ and $t_{pb} > 0$, respectively.

During the progenitor phase, the star evolves by maintaining hydrostatic equilibrium while undergoing successive stages of core and shell burning. This process progressively increases the central temperature and density (Figure~\ref{fig:rhoc_tc}). At all epochs, the density profile is primarily dictated by the core size. For example, the $40\,M_\odot$ model, which has the largest core, exhibits a relatively low central density but maintains a high density extending into the outer regions. In contrast, models with smaller cores, such as the $11\,M_\odot$ model, are characterized by high central densities that rapidly decline outward. The distributions of $Y_e$ and $\mu_e$ naturally follow from this density structure. Because $\mu_e \propto \rho^{1/3}$, it rises with increasing density, which in turn accelerates electron capture. Consequently, higher-density regions generally exhibit lower $Y_e$ values.

The temperature distribution, however, behaves differently from the other physical quantities, reflecting the combined effects of nuclear burning, convection, and neutrino cooling. At $t_{pb} \sim -10^6$~s, the profiles exhibit significant diversity because, as noted previously, each model is captured at a slightly different evolutionary stage. For $t_{pb} \gtrsim -10^3$~s, as the iron core forms, these stage-dependent differences gradually diminish. Generally, progenitors with more massive cores reach higher temperatures to support themselves against self-gravity. At the onset of core collapse ($t_{pb} \sim -1$~s), the temperature profiles in the core become almost identical. This convergence occurs because the onset of collapse in the HOSHI models is strictly defined by a specific central temperature. As mentioned in Section~\ref{subsec:setup}, this definition is somewhat arbitrary, and altering it could eliminate features such as the apparent reversal of the temperature ordering among the models—a behavior we do not observe in the MESA models. Nevertheless, as demonstrated in Section~\ref{ch4}, the total number of neutrinos integrated over a fixed duration remains robust and completely independent of this threshold definition.

Once core collapse commences ($t_{pb} \sim -1$~s), the core undergoes a nearly universal evolution regardless of the progenitor model, as shown in the bottom-left panels of Figure~\ref{fig:progenitor_radial_prof} (see also $\rho_c \gtrsim 10^{11}$~g~cm$^{-3}$ in Figures~\ref{fig:rhoc_tc} and \ref{fig:progenitor_timeevo}). Iron cores at the onset of collapse share fundamentally similar structures as they approach the Chandrasekhar mass limit. Although slight thermal differences arise due to neutrino emission—reflecting variations in initial entropy that affect the nuclear composition—these differences are largely washed out during the collapse. Ultimately, at core bounce ($t_{pb} = 0$~s), all thermodynamic profiles except temperature converge to nearly identical states, a consequence known as Mazurek's law.

Following core bounce, a shock wave forms and propagates outward. Confined to strictly one-dimensional calculations, neutrino heating behind the shock are insufficient to trigger a successful explosion; the shock wave stalls at a radius of approximately 150~km and gradually recedes across all models (Figure~\ref{fig:shock_evo}). During the very early postbounce phase ($\lesssim 50$~ms), the radial profiles of most thermodynamic quantities remain largely independent of the progenitor model. This initial universality persists as long as the accreting material originates from the innermost iron core, which shares a nearly identical density structure across all progenitor models. As evolution progresses into the sustained accretion phase ($\sim 50$--200~ms), the accreting material transitions to the outer core region. At this stage, distinct progenitor-dependent density gradients drive varying mass accretion rates. These divergent accretion flows directly dictate the compressive heating of the PNS, causing the temperature profiles to become highly sensitive to the progenitor's specific internal structure (see the bottom-right panels of Figure~\ref{fig:progenitor_radial_prof}).

\subsection{Neutrino Luminosities and Spectra} \label{sub:neutrino}

A primary outcome of this study is to provide, for the first time, of comprehensive neutrino luminosities and spectra spanning the preSN phase to the early SN phase. Figure~\ref{fig:numberlum} presents the time evolution of the neutrino luminosities (left panels) and the average energies (right panels) for $\nu_e$, $\bar{\nu}_e$, and $\nu_x$, from top to bottom. We also display the neutrino energy spectra at several selected time steps in Figure~\ref{fig:numberspe}. We divide our discussion of these results into three distinct evolutionary phases: the progenitor phase (Section~\ref{subsec:preSN_lumi}), core collapse (Section~\ref{subsec:collapse_lumi}), and the early SN phase (Section~\ref{subsec:SN_lumi}).

\subsubsection{Progenitor Phase} \label{subsec:preSN_lumi}

Here, we focus on the quasi-stationary progenitor phase prior to the onset of core collapse ($t_{pb} \lesssim -1$~s).

\begin{figure*}[ht]
    \centering
\includegraphics[width=\textwidth,clip]{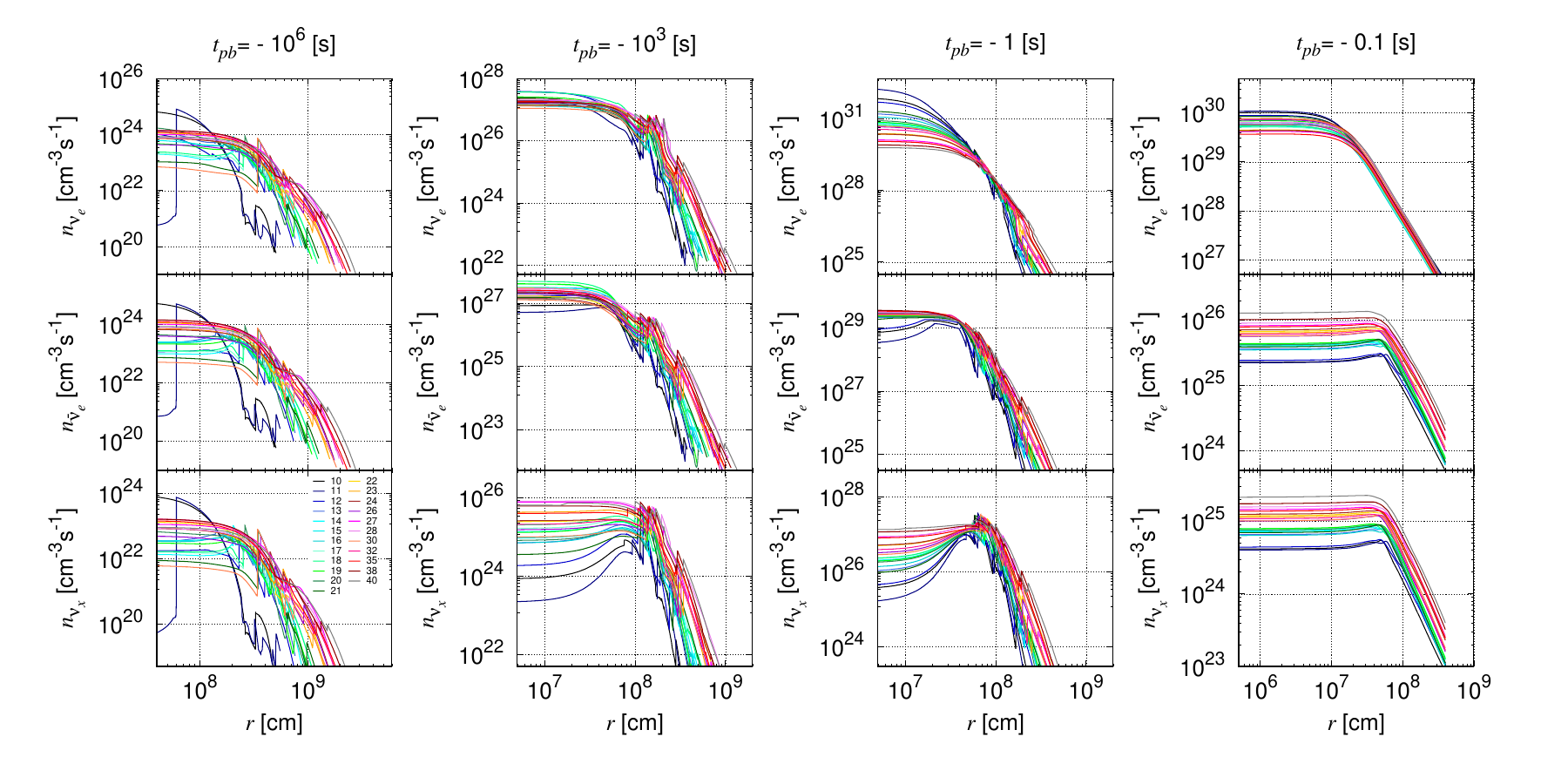}
    \caption{Radial profiles of neutrino number densities for $\nu_e$ (top), $\bar{\nu}_e$ (middle) and $\nu_x$ (bottom) in the preSN phase. We show the profiles at $t_{pb}\sim -10^6$, $-10^3$, $-1$ and $-0.1$~s from left to right. Colors distinguish the progenitor models.}  
    \label{fig:radial_prof_neutrino}
\end{figure*}

\begin{figure*}[ht]
    \centering
\includegraphics[width=\textwidth,clip]{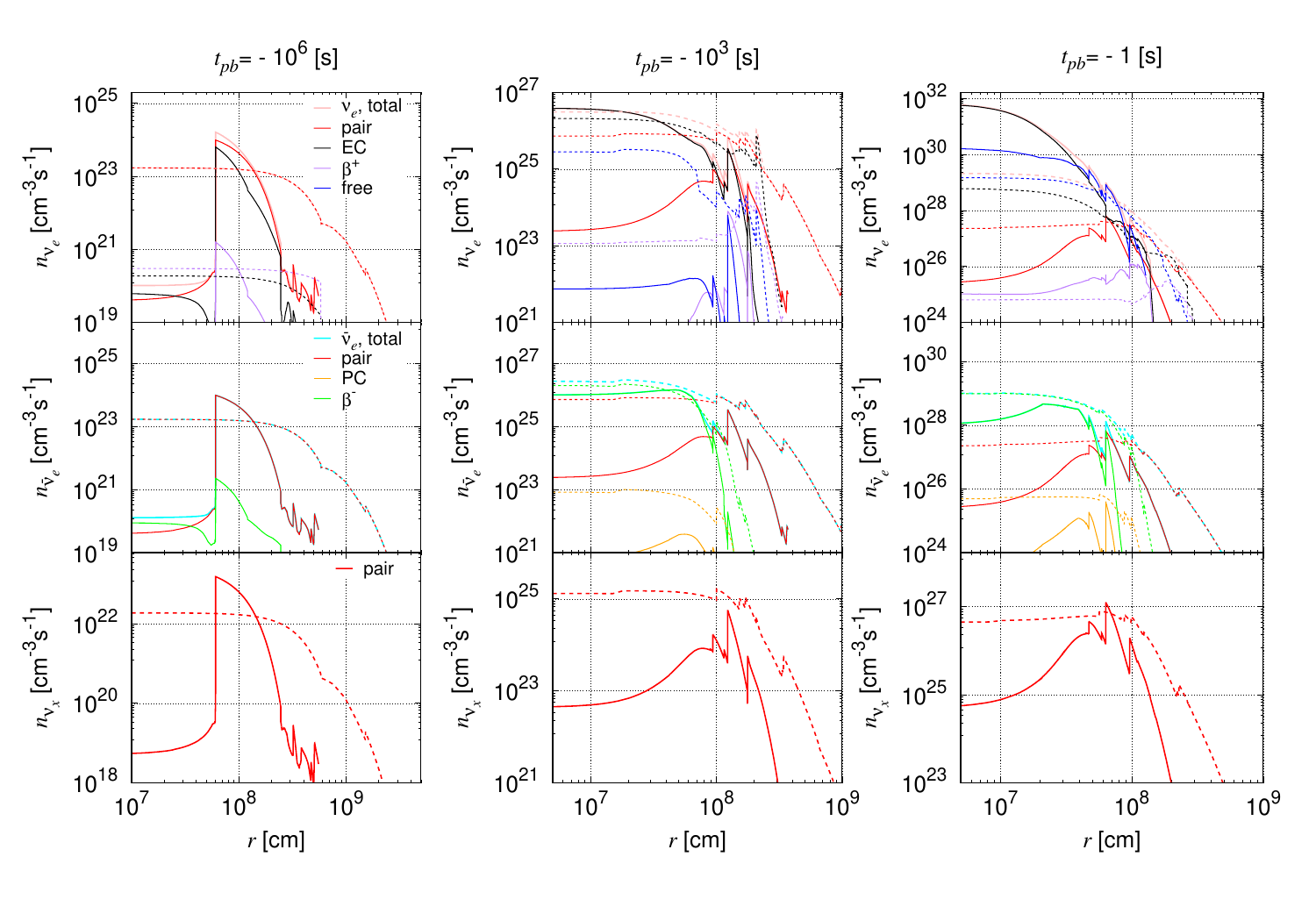}
    \caption{Radial profiles of neutrino emissivities via dominant neutrino processes for the 11 (solid) and 40~$M_\odot$ progenitors (dotted). Colors distinguish the neutrino emission process.}  
    \label{fig:radial_prof_eachmass}
\end{figure*}

As shown in the left panels of Figure~\ref{fig:numberlum}, the precollapse neutrino luminosities generally increase monotonically across all stellar models, although they exhibit temporary increases and decreases associated with specific nuclear burning episodes. However, at any given $t_{pb}$, the luminosities vary by roughly one to two orders of magnitude depending on the neutrino flavor and the evolutionary phase. We observe that the 40 $M_\odot$ model consistently emits more $\bar{\nu}_e$ and $\nu_x$ across all progenitor phases. In contrast, the hierarchy among the models temporarily reverses for $\nu_e$ in the period of $-10 \lesssim t_{pb} \lesssim -1$~s, where the 11 $M_\odot$ model exhibits the highest $\nu_e$ luminosity. These trends can be understood by examining the radial distributions of the neutrino emissivities and identifying the dominant emission processes, as illustrated in Figures~\ref{fig:radial_prof_neutrino} and \ref{fig:radial_prof_eachmass}, respectively. In the latter figure, we present the 11 $M_\odot$ and 40 $M_\odot$ models as representative cases because they bracket the extreme behaviors in luminosities across all flavors.

First, we examine $\nu_x$, which exhibits the simplest emission dynamics because these neutrinos are produced exclusively via pair. Because this emissivity is highly sensitive to temperature, models with extended high-temperature cores (e.g. the 40 $M_\odot$ model) maintain high neutrino emissivity over a much broader volume at all epochs. In contrast, in models with smaller cores, such as the 11 $M_\odot$ model, the high-temperature regions are highly centrally concentrated. With the exception of a brief period at $t_{pb} \sim -10^6$~s when some specific burning episodes temporarily elevate the temperature, the overall core temperature in these compact models remains relatively low. Furthermore, more compact core models exhibit higher electron chemical potentials ($\mu_e$). Because high degeneracy strongly suppresses positron production, pair emission is correspondingly inhibited—for example, in the central region of the 11 $M_\odot$ model at $t_{pb} \sim -10^3$ and $-1$~s. Therefore, we conclude that the $\nu_x$ luminosity correlates positively with the size of the progenitor's core.

Next, we turn our attention to $\bar{\nu}_e$. Up until $t_{pb} \lesssim -10^3$~s, pair is the dominant emission channel, similar to $\nu_x$, resulting in a comparable model-dependence trend. In later phases ($t_{pb} \sim -10^3$ and $-1$~s), although $\beta^-$ decay overtakes pair in the central region, it does not produce a large flux of $\bar{\nu}_e$ due to the strong Pauli blocking induced by the high $\mu_e$. Additionally, because density and temperature decrease rapidly outward, nuclei suitable for $\beta^-$ decay become scarce, tightly confining the emission to the innermost core. Consequently, models with more extended high-$T$ and low-$\mu_e$ cores yield a higher overall $\bar{\nu}_e$ emission. Therefore, as with $\nu_x$, we conclude that the $\bar{\nu}_e$ luminosity is primarily governed by the size of the progenitor's core.

Finally, during the early stages, $\nu_e$ are also primarily emitted via pair, mirroring the other flavors. However, as $\mu_e$ increases, EC become more frequent and rapidly dominate the $\nu_e$ emission. At $t_{pb} \sim -10^3$~s, the $\mu_e$ in the outer regions remains relatively low, meaning that the total $\nu_e$ emission is still largely dictated by pair. Thus, models with extended high-$T$ cores still exhibit higher $\nu_e$ luminosities at this stage. However, as the region dominated by EC expands over time, the trend completely reverses immediately prior to core bounce: models with compact cores and extremely high central $\mu_e$ exhibit the highest $\nu_e$ luminosities. As noted in the previous section, it is worth remembering that the precise timing of this reversal just before collapse is somewhat sensitive to how the onset of collapse is defined.

The right panels of Figure~\ref{fig:numberlum} display the time evolution of the average neutrino energies. We encourage the reader to examine the energy spectra in Figure~\ref{fig:numberspe} in conjunction with these results. Similar to the luminosities, the average energies monotonically increase over time for all flavors. In addition to this smooth secular increase, several sharp features are visible, particularly in the average energy of $\nu_e$ (upper-right panel of Figure~\ref{fig:numberlum}). These peaks are not caused by an abrupt global change in the composition, but by a temporary enhancement of particular nuclear weak channels that emit relatively energetic $\nu_e$. For example, in the 10~$M_\odot$ model, the feature around 
$t_{\rm pb}\sim -5\times10^7$~s is associated with Ne-shell burning, which enhances EC and $\beta^+$ decay of intermediate-mass nuclei such as S, P, and Cl isotopes. The peak therefore reflects a temporary change in the dominant nuclear weak-emission channel, rather than the duration of the shell-burning episode itself.

The average energies of $\bar{\nu}_e$ and $\nu_x$ show very little dependence on the progenitor models. Because pair dominates the emission for these flavors, the characteristic neutrino energy is dictated almost entirely by the local temperature where the emission peaks. Comparing the highest core temperatures at each time point reveals only a 10\%--20\% variation across the progenitor models (see Figure~\ref{fig:progenitor_radial_prof}), which cleanly translates into the small model dependence observed in the average energies of $\bar{\nu}_e$ and $\nu_x$. For $\nu_e$, on the other hand, $\mu_e$ plays a critical role, as the capture of higher-energy electrons produces correspondingly higher-energy neutrinos. Therefore, compact core models with high $\mu_e$ naturally have the higher $\nu_e$ average energies. As seen in Figure~\ref{fig:progenitor_radial_prof}, $\mu_e$ can vary by a factor of a few depending on the progenitor model, which explains why the average energy of $\nu_e$ exhibits a much stronger model dependence than those of $\bar{\nu}_e$ and $\nu_x$.

A final remark is warranted regarding the neutrino emission in the earliest phases ($t_{pb} \lesssim -10^5$~s). In our previous study \citep{Kato2017}, we neglected neutrino emission from nuclei in nonNSE regions, assuming that the core almost entirely transitions to NSE just prior to collapse and that NSE emissions vastly dominate \citep{Kato2020a}. However, the situation is fundamentally different during these very early phases. We find that EC on certain specific nuclei possess large positive $Q$-values, emitting neutrinos with energies close to typical detector thresholds even within nonNSE regions. For example, focusing on the region with a temperature peak ($T \sim 0.28$~MeV) at $r \sim 6 \times 10^7$~cm in the 11 $M_\odot$ model at $t_{pb} \sim -10^6$~s, EC on iron-group elements such as $^{51}$Cr, $^{53}$Mn, and $^{56}$Fe act as the primary emission processes for $\nu_e$.

\subsubsection{Core Collapse Phase} \label{subsec:collapse_lumi}

\begin{figure*}[ht]
    \centering
    \includegraphics[width=17cm]{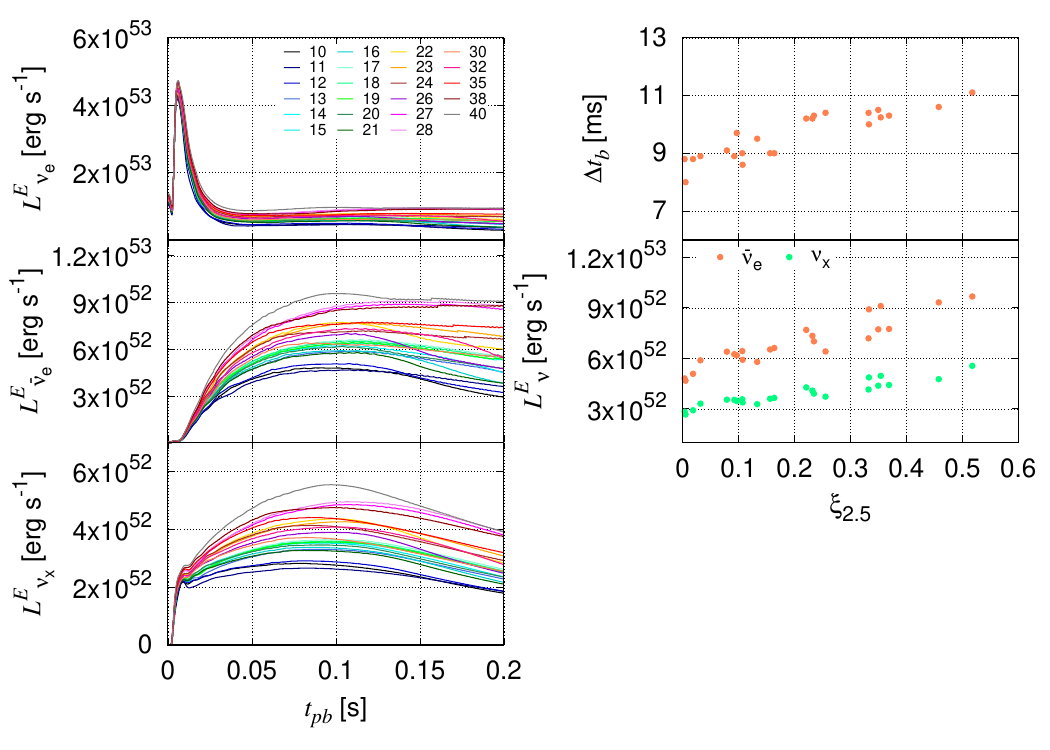}
    \caption{Left panels show SN neutrino lightcurves of $\nu_e$, $\bar{\nu}_e$ and $\nu_x$ from top to bottom. Right panels show the correlation between $\xi_{2.5}$ and neutrino-burst width (right top) or peak luminosities (right bottom).}
    \label{fig:burstwidth}
\end{figure*}

During the core collapse phase, as density rapidly increases, $\nu_e$ become trapped at a specific density threshold ($\rho_{\text{trap}} \sim 10^{11}$--$10^{12}\,\text{g~cm}^{-3}$). The emitted $\nu_e$ diffuse outward from a specific radius known as the neutrinosphere, a process that effectively masks any variations in the deep internal temperature structure.  Consequently, the resulting $\nu_e$ luminosity is largely determined by the universal properties of the collapsing degenerate core. In contrast, pair plays a crucial role in the emission of $\bar{\nu}_e$ and $\nu_x$, resulting in a significantly stronger temperature dependence compared to $\nu_e$. Because this process requires the presence of positrons, it is heavily suppressed in the highly degenerate central core. Instead, this emission tends to occur in the outer core regions, where $\mu_e$ is lower. Since these outer layers are more susceptible to the specific structural and thermal histories of the progenitors, the emission of $\bar{\nu}_e$ and $\nu_x$ clearly manifests a strong model dependence. These results are entirely consistent with our previous study \citep{Kato2017}.

We must also comment on several minor discontinuities observed in Figure~\ref{fig:numberlum}, particularly in the interval of $-1 \lesssim t_{pb} \lesssim 0$~s. These artifacts arise from three primary computational causes. First, the transition from the quasi-static stellar evolution calculation to the fully dynamical core-collapse simulation inevitably introduces small numerical transients. Although the central hydrodynamical quantities are connected smoothly, the mapping is not completely identical at all radii because the EOS is different. Moreover, while the same weak-rate set is adopted in both phases, its numerical implementation differs slightly. The small mismatch in the luminosity slope around $t_{\rm pb}\sim -1$~s is therefore likely caused by these numerical differences and input physics. Second, the location at which the neutrino luminosity is evaluated changes; as described in Section~\ref{lumcalc}, we adopt the larger of the luminosities calculated at $500\,\text{km}$ or near the outer boundary of the computational domain. This specific discontinuity appears only in a few models, such as the $13\,M_\odot$ and $15\,M_\odot$ cases shown in the middle-right panel. Finally, for $\bar{\nu}_e$ and $\nu_x$, the neutrino emission rates in regions with densities below $10^{13}\,\text{g~cm}^{-3}$ are calculated via a postprocessing approach. This numerical transition creates the small feature visible at $t_{pb} \sim 10^{-2}$~s for all models. We emphasize that these discontinuities are negligible in magnitude, occur over very short timescales, and have no significant impact on the subsequent physical discussions.

\subsubsection{Early Supernova Phase} \label{subsec:SN_lumi}

The flavor-dependent characteristics observed during the collapse are further amplified in the postbounce phase. The $\nu_e$ luminosity remains robust across different progenitor models, with variations limited to approximately $20\%$ throughout both the neutronization burst and the early accretion phase. This stability arises because the early $\nu_e$ emission is largely regulated by the self-similar collapse of the universal inner iron core. In contrast, the luminosities of $\bar{\nu}_e$ and $\nu_x$ exhibit clear time-dependent divergences. Since their primary production mechanisms—positron capture on neutrons for $\bar{\nu}_e$ and pair for $\nu_x$—rely heavily on the thermal generation of abundant positrons, they are highly sensitive to the temperature of the PNS mantle. During the very early burst phase ($\lesssim 50$~ms), these emissions emerge from the outer core and strictly reflect the progenitor's temperature structure. As the evolution transitions into the accretion phase ($\sim 50$--$200$~ms), the variations across models amplify by factors of a few. During this stage, the temperature of the PNS mantle becomes directly governed by the compressive heating driven by the mass accretion rate. Consequently, the diverse emissions of $\bar{\nu}_e$ and $\nu_x$ naturally reflect the distinct density gradients of the outer core across different progenitor models.

\citet{Choi2025} recently reported a correlation between the compactness parameter and various quantities characterizing SN neutrino emission. Motivated by this, we investigate whether similar features are observable in our simulations. Specifically, we examine the width of the $\nu_e$ neutronization burst and the peak luminosities of $\bar{\nu}_e$ and $\nu_x$. Figure~\ref{fig:burstwidth} presents the luminosity lightcurves for $\nu_e$, $\bar{\nu}_e$, and $\nu_x$ (from top to bottom) in the left panels, while the right panels show the correlations of the burst width (top right) and peak neutrino luminosities (bottom right) with the compactness parameter. As is evident from the figure, our results generally support their findings. In particular, the peak luminosities of $\bar{\nu}_e$ and $\nu_x$ exhibit a robust correlation with the compactness parameter ($\xi_{2.5}$). This strong dependence naturally reflects the underlying physical processes. Progenitors with higher compactness possess shallower density gradients in their outer cores, leading to heavier mass accretion onto the PNS. This sustained accretion drives the strong compressive heating and extreme temperatures required to boost the $\bar{\nu}_e$ and $\nu_x$ emissions.

\section{Correlation between Stellar Key Parameters and Neutrino Emission} \label{ch4}

In this section, we explore the possibility of inferring the internal structure of SN progenitors based on neutrino observations. Before delving into detailed observational prospects, we first investigate the physical quantities that are theoretically linked to neutrino emission. As discussed in the previous section, we successfully reproduced the correlations for postbounce neutrino emission reported in the previous studies, even when employing our independent computational codes and distinct input physics. In contrast, such correlations have not yet been systematically explored for preSN neutrino emission. Therefore, it is essential to identify the key physical quantities that govern this early phase in a manner that is robust against variations in specific progenitor models.

First, we define the critical parameters used to characterize the progenitor models in Section~\ref{subsec:key_parameter}. As we will demonstrate in Section~\ref{ch5}, even with next-generation neutrino detectors, we expect to observe only a few hundred preSN neutrino events in total. This limited statistics makes detailed time-resolved analyses highly challenging. Consequently, we rely on time-integrated neutrino quantities: the total number ($N$) and the total energy of emitted neutrinos ($E$). The correlations between these key stellar parameters and the integrated quantities ($N$ and $E$) are discussed separately for the preSN and SN phases in Sections~\ref{subsec:preSN} and \ref{subsec:SN}, respectively.

\subsection{Stellar Key Parameters} \label{subsec:key_parameter}

\begin{figure*}[htpb]
    \centering
\includegraphics[width=15cm,clip]{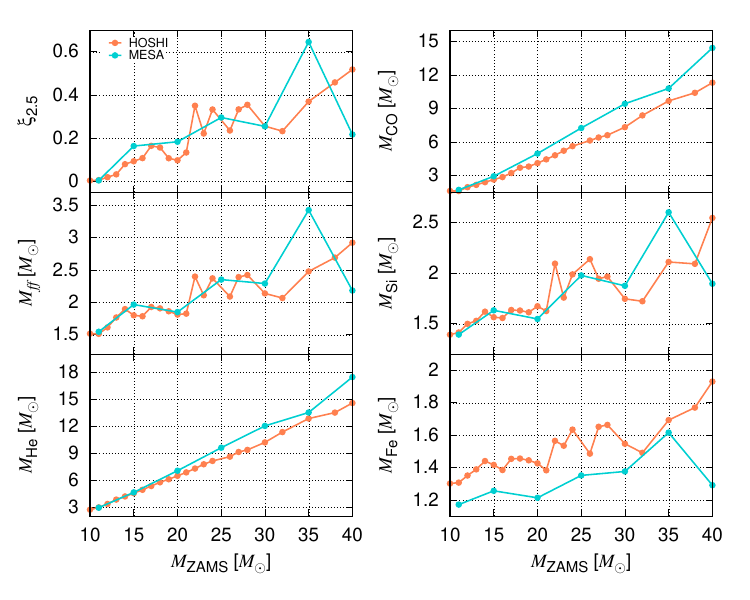}
    \caption{The key parameters for our progenitor models. We estimate them at the onset of the core collapse. Colors distinguish the stellar code.}
    \label{fig:keyphys}
\end{figure*}

We first highlight the key parameters used to characterize the SN progenitors. Following the classification scheme proposed by \citet{Takahashi2023}, we focus on two broad categories: parameters related to the density structure (density parameters) and those related to the chemical stratification (composition parameters). The specific parameters examined within each category are as follows:
\begin{enumerate}
\item{Density parameters: $\xi_{2.5}$, $M_{\mathrm{ff}}$}
\item{Composition parameters: $M_{\mathrm{He}}$, $M_{\mathrm{CO}}$, $M_{\mathrm{Si}}$, $M_{\mathrm{Fe}}$}
\end{enumerate}

The first category includes the compactness parameter ($\xi_{2.5}$) and the freefall mass ($M_{\mathrm{ff}}$). The compactness parameter is defined as
\begin{eqnarray}
\xi_{2.5} = \left .\frac{M(r)/M_\odot}{r/1000~{\rm km}}\right|_{M=2.5M_\odot},
\end{eqnarray}
which serves as a robust indicator of how much matter is concentrated in the central core. Its relationship with both stellar explodability and SN neutrino luminosity has been extensively discussed in the literature \citep[e.g.][]{OConnor2011, Ugliano2012, Nakamura2015, Ertl2016, Sukhbold2016}. Alongside compactness, the freefall mass ($M_{\mathrm{ff}}$) exhibits a significant correlation with the parameters that characterize stellar evolution \citep{Takahashi2023}. We define $M_{\mathrm{ff}}$ as the mass coordinate where the freefall time ($\tau_{\mathrm{ff}}$) equals 1~s:
\begin{eqnarray}
\tau_{\mathrm{ff}}=\frac{\pi}{2\sqrt{2}}\sqrt{\frac{r^3}{GM(r)}}=1~{\rm s}.    
\end{eqnarray}

The parameters in the second category represent the mass coordinates at the boundaries between distinct chemical layers. We define $M_{\mathrm{He}}$, $M_{\mathrm{CO}}$, and $M_{\mathrm{Si}}$ as the mass coordinates at the base of the regions where the mass fractions $X_{\mathrm{H}}$, $X_{\mathrm{He}}$, and $X_{\mathrm{O}}$ exceed 0.1, respectively. $M_{\mathrm{Fe}}$ is defined as the outermost mass coordinate that satisfies $X_{\mathrm{Fe}} > 0.1$ and $X_{\mathrm{Si}} < 0.01$.

\begin{figure*}[ht]
    \centering
    \includegraphics[width=\textwidth]{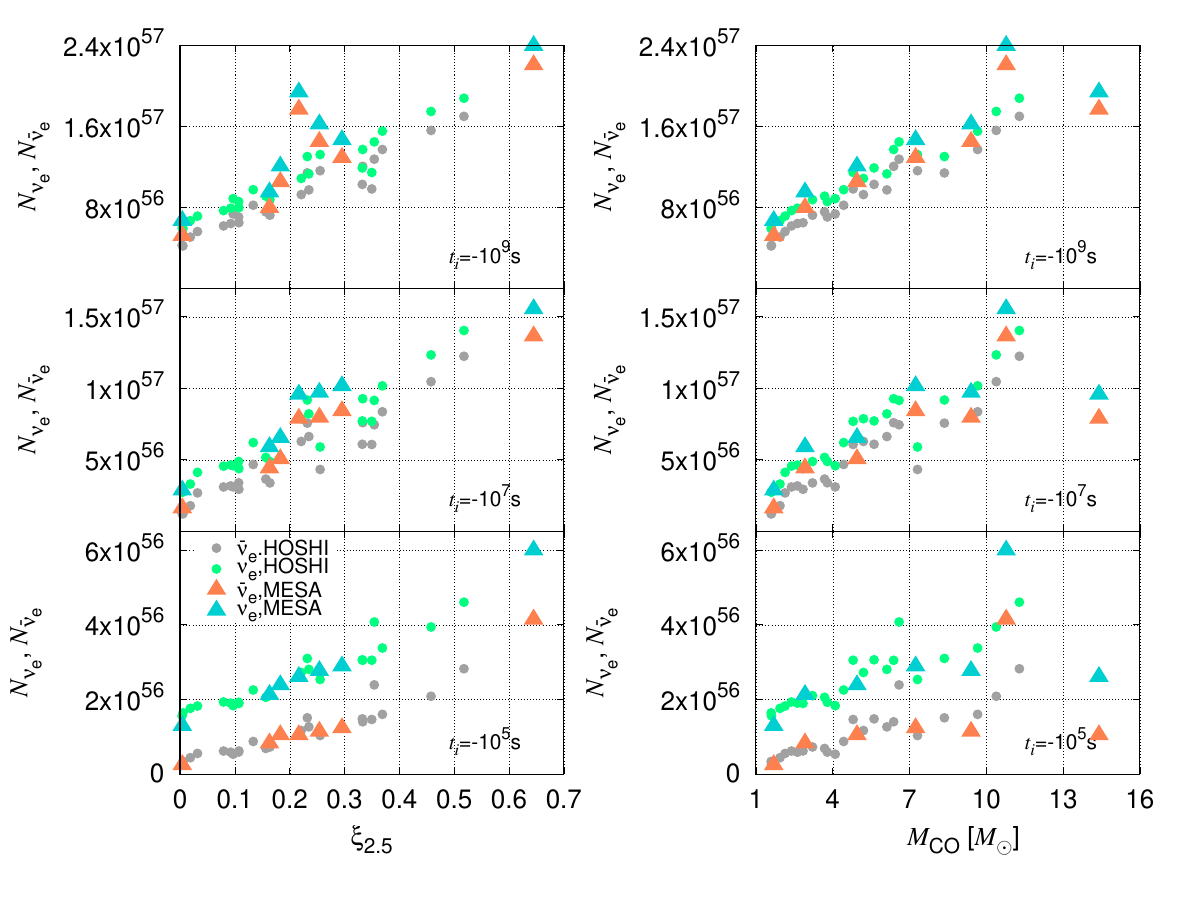}
    \caption{Relationship between integrated neutrino number and $\xi_{2.5}$ (left) or $M_{\rm CO}$ (right) in the preSN phase. We take three initial times for the integration $t_i=-10^9$, $-10^7$ and $-10^5$~s from top to bottom. Dot types distinguish the stellar evolution codes.}
    \label{fig:correlation_preSN}
\end{figure*}

Figure~\ref{fig:keyphys} illustrates these parameters at the onset of core collapse for our suite of progenitor models. We observe two distinct types of distributions with respect to the initial progenitor mass ($M_{\rm ZAMS}$). The first exhibits a linear dependence, characteristic of $M_{\mathrm{He}}$ and $M_{\mathrm{CO}}$. The second displays a nonmonotonic jagged dependence, as seen with $\xi_{2.5}$. Although the exact distribution profiles vary depending on the stellar evolution code employed, the existence of these two distinct behavioral types remains consistent. In the subsequent analysis, we select $M_{\mathrm{CO}}$ and $\xi_{2.5}$ as the representative parameters for each distribution type.
\begin{enumerate}
\item {\textbf{$M_{\mathrm{CO}}$} \\
Hydrogen and helium burning are known to be stable in massive stars; their properties are largely dictated by convection efficiency and mass-loss rates \citep[e.g.][]{Woosley2002, Sukhbold2014}. While the convection efficiency is comparable between our two sets of stellar models, the MESA models do not incorporate mass loss. Without mass loss, the stellar envelopes persist longer, providing an extended supply of fuel for nuclear burning. Consequently, the core tends to grow more massive, particularly in heavier progenitors. Additionally, we note that the relationship between $M_{\mathrm{CO}}$ and $M_{\mathrm{ZAMS}}$ can be significantly altered by binary interactions and uncertainties in the reaction rate of $^{12}\mathrm{C}(\alpha,\gamma)^{16}\mathrm{O}$ \citep[e.g.][]{Schneider2020, deBoer2017}.}
\item{\textbf{$\xi_{2.5}$} \\
Several studies have argued that this jagged mass-dependence is driven by the dynamics of carbon core and shell burning \citep{Sukhbold2014, Laplace2025}. In their models, carbon core burning proceeds radiatively for the progenitors above $\sim 20~M_\odot$. Once core carbon burning ceases, neutrino cooling causes the core to contract and increase pressure, forcing the carbon-burning shell to migrate inward. If this carbon shell settles at a small radius, subsequent oxygen burning cannot ignite effectively in the core. This suppresses neutrino cooling, leading to a less degenerate, higher-$T$ core, which ultimately results in relatively high compactness. However, because the carbon shell is stratified into multiple layers, the specific layer that ignites depends sensitively on the progenitor mass, naturally producing a jagged distribution. These trends established during carbon-shell burning propagate into the subsequent oxygen-shell burning and are carried forward through all final burning stages. Consequently, the mass coordinates of elements synthesized in these late phases—such as $M_{\mathrm{ff}}$, $M_{\mathrm{Si}}$, and $M_{\mathrm{Fe}}$—exhibit a similarly jagged mass dependence.

Compactness is also highly sensitive to the treatment of mass loss and mixing. As previously mentioned, when the envelope is stripped through mass loss, the fuel supply is curtailed, yielding a smaller iron core. If the core is smaller, the 2.5~$M_\odot$ reference coordinate shifts relatively farther outward in radius, thereby reducing the compactness parameter. A similar logic applies to mixing: when convection and overshooting operate more efficiently, the core tends to grow larger, which conversely leads to a decrease in compactness.}
\end{enumerate}

It is crucial to specify the epoch at which these parameters are evaluated. In this study, we evaluate them at the onset of core collapse. While this evaluation time could technically be chosen arbitrarily, our ultimate goal is to retrieve progenitor structure information from future neutrino observations using theoretical models. Therefore, evaluating these parameters at the onset of collapse—a well-understood epoch frequently used as a benchmark in theoretical studies—is the most physically appropriate choice. Additionally, as noted earlier, the exact definition of the onset of core collapse differs slightly between models. However, because both $\xi_{2.5}$ and $M_{\mathrm{CO}}$ remain nearly constant in the moments immediately preceding collapse, these definitional differences do not affect the evaluation of these key parameters.

Finally, we also investigated the relationship between neutrino emission and the quantities in the third category (thermal parameters) defined by \citet{Takahashi2023}, but found no meaningful correlations.

\subsection{PreSN Phase} \label{subsec:preSN}

The total number ($N$) and total energy ($E$) of neutrinos emitted over a given time interval defined as
\begin{eqnarray}
N\left(t_i\right) &=& \int_{t_i}^{t_f} L_\nu^N\left(t\right) dt, \\
E\left(t_i\right) &=& \int_{t_i}^{t_f} L_\nu^E\left(t\right) dt,
\end{eqnarray}
where we fix the final time $t_f$ to the moment of core bounce ($t_{pb}=0$~s). To investigate the dependence on the integration range, we test three different initial times: $t_i=-10^9$, $-10^7$, and $-10^5$~s.

Figure~\ref{fig:correlation_preSN} illustrates the relationship between the integrated neutrino numbers and the key parameters $\xi_{2.5}$ (left) and $M_{\mathrm{CO}}$ (right). The dependence of the correlation strength on $t_i$ exhibits a contrasting behavior for $\xi_{2.5}$ versus $M_{\mathrm{CO}}$. The correlation between $\xi_{2.5}$ and $N$ strengthens for later starting times (i.e., larger $t_i$), whereas the correlation with $M_{\mathrm{CO}}$ weakens. This divergence is primarily driven by progenitors with extended cores. For instance, in the upper-left panel ($t_i = -10^9$~s), the 40~$M_\odot$ MESA model appears as a significant outlier near $\xi_{2.5} \sim 0.2$. In the lower-right panel ($t_i = -10^5$~s), the 35~$M_\odot$ model similarly deviates from the main trend at $M_{\mathrm{CO}} \sim 10~M_\odot$. We emphasize that these trends are robust and entirely independent of the specific stellar evolution code employed. Identical behaviors are also observed for the integrated energy $E$.

\begin{figure*}[htbp]
    \centering
    \includegraphics[width=8cm]{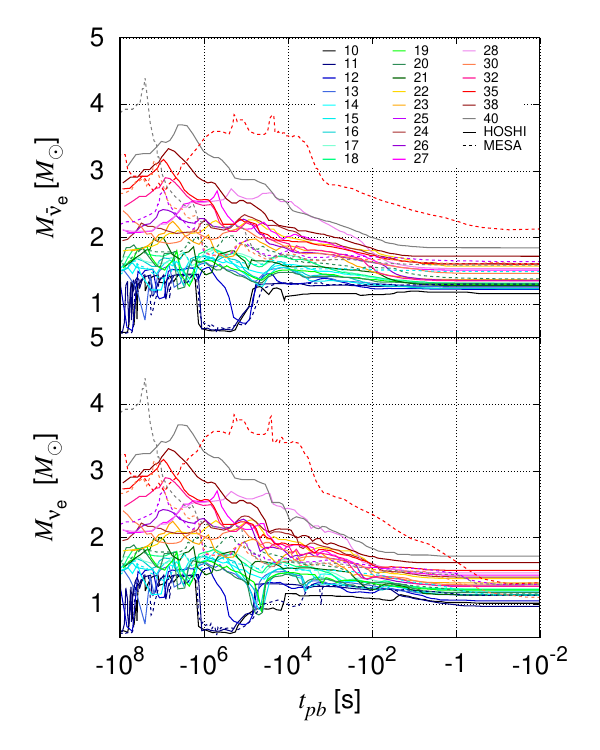}
    \includegraphics[width=8cm]{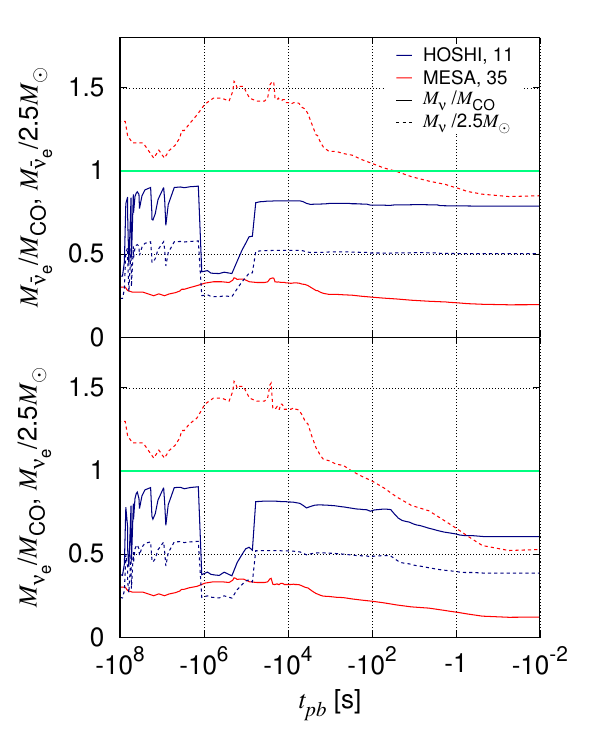}
    \caption{Time evolution of $M_{\bar{\nu}_e}$ (top) and $M_{{\nu}_e}$ (bottom) in the left panels. The right panels show the time evolution of $M_\nu/M_{\rm CO}$ and $M_\nu/2.5M_\odot$ for two representative models: the HOSHI 11~$M_\odot$ model, which has both a small compactness and a small $M_{\rm CO}$, and the MESA 35~$M_\odot$ model, which exhibits the opposite trend. Colors distinguish the progenitor models.}
    \label{fig:Mnu_evo}
\end{figure*}

To understand this $t_i$ dependence, we must identify the precise regions from which neutrinos are emitted. We introduce two new metrics, $M_{\nu_e}$ and $M_{\bar{\nu}_e}$, defined as the mass coordinates enclosing 95\% of the total neutrino luminosity when integrating the local emission outward from the center. The left panels of Figure~\ref{fig:Mnu_evo} displays the time evolution of these emission boundaries. The different behavior of the correlations with $M_{\rm CO}$ and $\xi_{2.5}$ can be understood as a consequence of the different physical information carried by these quantities. The CO core mass should not be interpreted as the outer boundary of the neutrino-emitting region. Rather, $M_{\rm CO}$ acts as a proxy for the long-term evolutionary history of the progenitor, including the size and masses of the cores and shells. When the neutrino number is integrated from an early time, such as $t_i=-10^9$~s, the accumulated signal includes contributions from these extended burning stages. It therefore retains memory of the overall CO core evolution, which explains the stronger correlation with $M_{\rm CO}$. By contrast, when the integration is restricted to the final $\sim 10^5$~s before core collapse, the neutrino emission is dominated by the hot and dense inner core. In this late phase, $M_{\nu_e}$ and $M_{\bar{\nu}_e}$ become much smaller than $M_{\rm CO}$, particularly for progenitors with extended CO cores, while they are comparable to or smaller than the mass scale used in the definition of $\xi_{2.5}$. Thus, $M_{\rm CO}$ becomes too global a quantity to characterize the region that actually dominates the late-time neutrino emission. The compactness parameter, on the other hand, is more directly connected to the inner-core structure and the degree of contraction, and therefore shows a stronger correlation with the neutrino number for later starting times.

The right panels of Figure~\ref{fig:Mnu_evo} illustrate this point more explicitly for two representative models. We show the ratios $M_{\nu}/M_{\rm CO}$ and $M_{\nu}/2.5M_\odot$ for the HOSHI 11~$M_\odot$ model and the MESA 35~$M_\odot$ model, which represent opposite trends in the structural parameters. The HOSHI 11~$M_\odot$ model has a relatively small CO core ($M_{\rm CO}\sim1.60~M_\odot$) and low compactness ($\xi_{2.5}\sim5.32\times10^{-3}$), whereas the MESA 35~$M_\odot$ model has an extended CO core ($M_{\rm CO}\sim10.8~M_\odot$) but a comparatively different inner-core compactness ($\xi_{2.5}\sim0.645$). For the HOSHI 11~$M_\odot$ model, $M_{\nu}$ remains comparable to both $M_{\rm CO}$ and $2.5\,M_\odot$ over a significant part of the evolution, indicating that the characteristic scales of the CO core and the inner core are not widely separated. In this case, the distinction between the two structural indicators is relatively modest. In contrast, for the MESA 35~$M_\odot$ model, $M_{\nu}/M_{\rm CO}$ becomes much smaller than unity just before the collapse, while $M_{\nu}/2.5M_\odot$ remains of order unity or smaller. This demonstrates that, in extended-core progenitors, the late-time neutrino-emitting region occupies only the inner part of the much larger CO core. For these reasons, the weakening of the $M_{\rm CO}$ correlation at late times does not imply that $M_{\rm CO}$ is unimportant for the preSN evolution; rather, it reflects the fact that the late-time neutrino signal is controlled by the inner-core conditions instead of the global CO core size.

We also investigated whether the correlation with the preSN neutrino emission persists when evaluating the compactness parameter at different mass coordinates, specifically $1.5\,M_\odot$ or $2.0\,M_\odot$. The detailed results of this parameter study are presented in Appendix~\ref{appendix_compactness}. The correlation noticeably weakens when compactness is defined at these smaller mass coordinates, particularly for the most massive progenitors. For this reason, we adopt $\xi_{2.5}$ as the standard definition for all subsequent discussions.

\begin{figure*}[htbp]
    \centering
    \includegraphics[width=\textwidth]{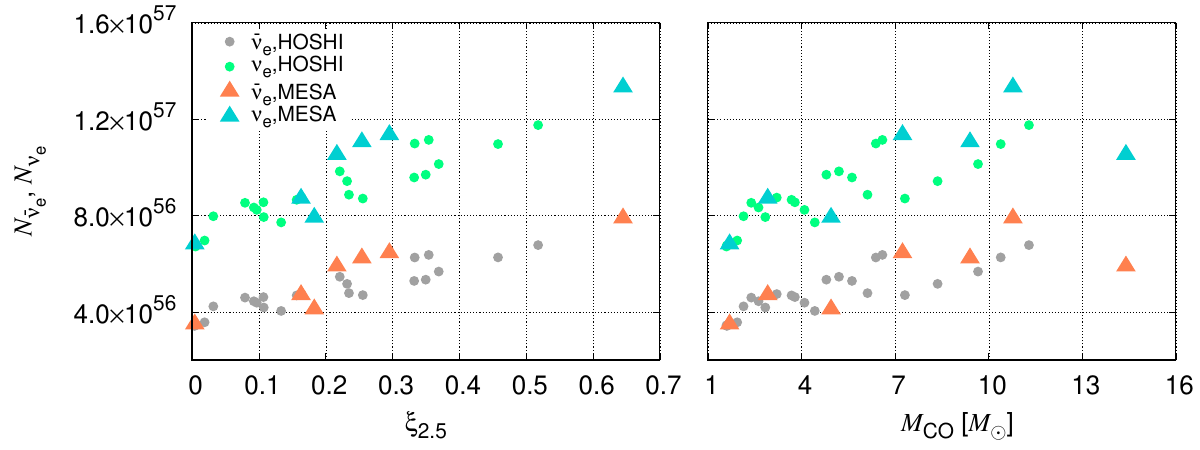}
    \caption{Relationship between integrated neutrino number and $\xi_{2.5}$ (left) or $M_{\rm CO}$ (right) in the early SN phase. Dot types distinguish the stellar evolution code.}
    \label{fig:correlation_SN}
\end{figure*}

\subsection{Early SN Phase} \label{subsec:SN}

We now turn our investigation to the correlations during the early SN phase, presented in Figure~\ref{fig:correlation_SN}. The total number of emitted neutrinos, $N$, is calculated by time-integrating the neutrino luminosity from $t_{pb} = 0$ to $200$~ms. As shown in Figure~\ref{fig:correlation_SN}, a strong correlation with $\xi_{2.5}$ persists regardless of the specific stellar evolution code. The initial burst ($\lesssim 50$~ms) is governed by the universal inner core. However, the subsequent accretion phase heavily dominates the 200~ms integrated emission, driving this consistent correlation across different codes. During this extended period, the mass accretion rate and the resulting thermal structure of the PNS are directly dictated by the density gradients of the outer core, which $\xi_{2.5}$ elegantly characterizes. For $M_{\mathrm{CO}}$, the correlation with the integrated neutrino emission is noticeably weaker than that with $\xi_{2.5}$, perfectly mirroring the trends observed just prior to core collapse.

Isolating this early SN phase offers a distinct observational advantage: it provides a clean window into the inner core structure, largely unaffected by the complex hydrodynamic instabilities that emerge after the shock stalls.  Because the neutrino light curve can be resolved with high temporal precision for a nearby event, adopting an analysis strategy focused on these specific early time windows will be essential for extracting accurate progenitor properties.

\section{Neutrino Detection} \label{ch5}

In the previous section, we identified strong correlations between the total number of emitted neutrinos and the key parameters characterizing the progenitor structure. In this section, we discuss whether these correlations remain observable under realistic detection conditions. To this end, we evaluate the expected neutrino event rates and initial detection (alert) times in Section~\ref{subsubsec:alart}, followed by the total number of detected events for several current and future detectors in Section~\ref{subsubsec:event}. We then investigate the correlations between these observable quantities and the key progenitor parameters in Section~\ref{subsec:obs_correlation}.

\subsection{Methods for Estimation of Neutrino Events} \label{subsec:method_obs}

\subsubsection{Neutrino Detectors} \label{subsubsec:detector_info}

\begin{table}[htbp]
\centering
 \caption{The detector parameters assumed in this paper. It should be noted that we use the full sizes of detectors. The numbers of background rates are taken from \citet{Simpson2019} for SK, \citet{Asakura2016} for KamLAND and \citet{Abusleme2024} for JUNO. For HK, we scale the SK background rates by the fiducial volume ratio. For DUNE, we adopt a background rate of $\sim 200$ events $\text{day}^{-1}$, assuming effective removal of spallation events and dominance of solar neutrino interactions \citep{Capozzi2018}.} 
\begin{tabular}{cccc}
 \hline\hline
 Detector&  Volume  & Energy Threshold $(E_{th})$ & BG \\ 
         &   (kt)   &   (MeV)                     &   (${\rm day}^{-1}$) \\ \hline
 SK & 32 & 5.3 & 25 \\
 HK & 260 & 5.3 & 200 \\
 KamLAND & 1 & 1.8 & 0.1 \\
 JUNO & 20 & 1.8 & 20 \\
 DUNE & 70 & 8 & 200 \\ \hline\hline
 \label{Tab:detector}
\end{tabular}
\end{table}

We estimate the number of observed neutrino events at five representative neutrino detectors: SK, HK, KamLAND, JUNO, and DUNE. Their assumed performance parameters are summarized in Table~\ref{Tab:detector}. We consider the primary detection channels: inverse beta decay (IBD) for $\bar{\nu}_e$, and charged-current interactions with $^{40}\mathrm{Ar}$ for $\nu_e$. Throughout this analysis, we adopt optimistic performance parameters to explore the maximum potential of these facilities. Although a restricted fiducial volume is typically utilized to mitigate background noise depending on the specific observation target, we employ the full inner volume for all detectors. For the water Cherenkov detectors (SK and HK), we assume the implementation of the delayed coincidence technique for IBD events, utilizing gadolinium for efficient neutron tagging \citep{Beacom2004}. This tagging is used to identify IBD events and to suppress single-event backgrounds. It also effectively lowers the usable energy threshold for the IBD sample. In the case of DUNE, the primary background sources at energies relevant for preSN neutrinos ($E_\nu \lesssim$ 15~MeV) are cosmic-ray muon spallation products and solar neutrinos. Because recent studies demonstrate that spallation backgrounds can be effectively suppressed to negligible levels through sophisticated tagging and cut strategies \citep{Zhu2018}, the irreducible background is primarily driven by solar $^{8}\mathrm{B}$ neutrinos. Based on the event rate estimations for these solar neutrinos in \citet{Capozzi2018}, we assign an optimistic background rate of $\sim$200 events per day for the full 70~kt detector volume. To contextualize our results for realistic nearby progenitor candidates like Betelgeuse, we scale our calculations to reference distances of 200~pc and 1~kpc.

In addition to these primary channels, we also estimate the contribution from neutrino--electron elastic scattering (ES) for SK, HK, and JUNO. This channel is sensitive to all neutrino flavors and can in principle provide directional information, but its event sample is qualitatively different from the IBD sample, because ES events appear as single electron-recoil events and are therefore more sensitive to detector-specific single-event backgrounds, such as solar neutrinos, radioactive backgrounds, and spallation products. Therefore, we do not include ES events for estimating the alert time. Instead, we compute the number of ES events after the IBD alert time. We apply a recoil-electron visible-energy threshold of $E_{\rm th,e}=3.5~{\rm MeV}$ and assume unit detection efficiency above this threshold as an optimistic upper-limit estimate. A realistic ES analysis would require detector-specific efficiencies, angular responses, and time-dependent background models, which are beyond the scope of this work.

Other flavor-insensitive channels may also provide complementary information. In particular, coherent elastic neutrino--nucleus scattering (CE$\nu$NS) is sensitive to all neutrino flavors and has been experimentally observed \citep[e.g.,][]{Akimov2017}. Because CE$\nu$NS can be detected with low nuclear-recoil thresholds, dark-matter direct-detection experiments have been discussed as possible detectors of both preSN neutrinos and the postbounce supernova burst 
\citep[e.g.,][]{Duba2008,Lang2016, Raj2020, Pattavina2021, Ko2023}. However, the target masses of current dark-matter detectors are much smaller than those of the neutrino detectors considered here. We therefore do not include CE$\nu$NS in the present event-rate analysis.

\begin{figure*}
    \centering
    \includegraphics[width=8cm]{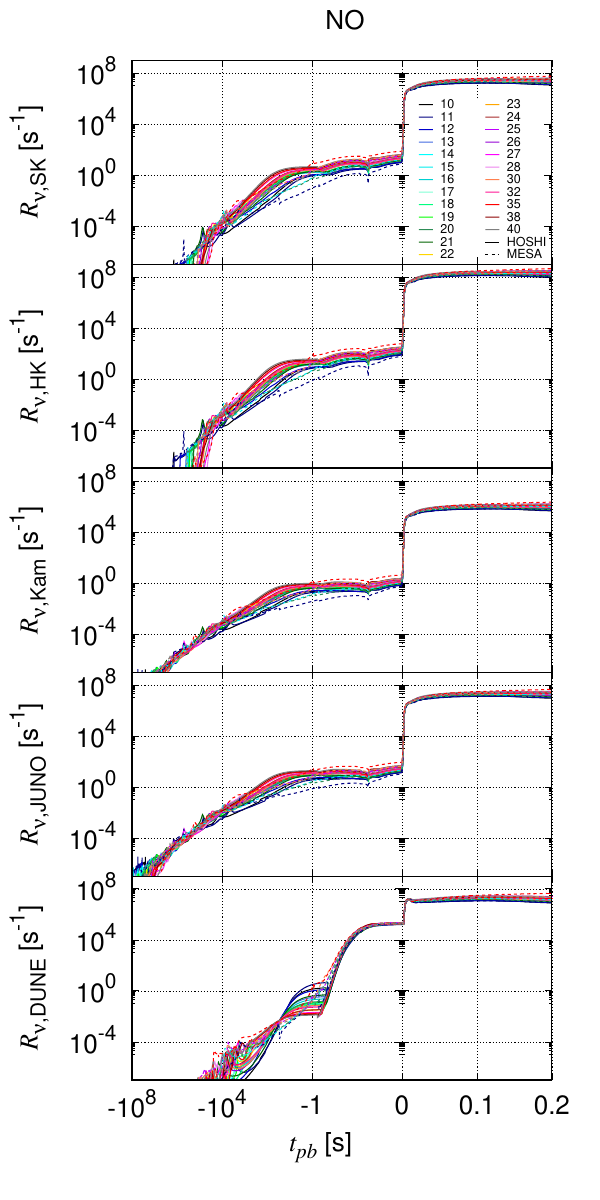}
    \includegraphics[width=8cm]{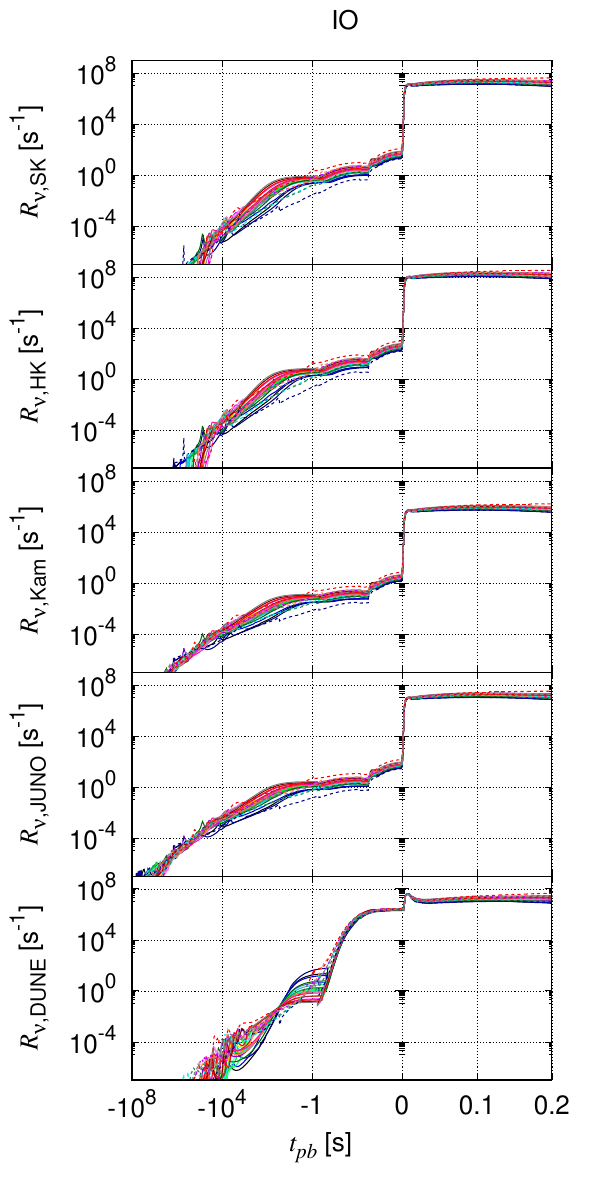}
    \caption{Time evolution of event rates for IBD at SK, HK, KamLAND, JUNO, and for Ar reactions at DUNE from top to bottom. Left and right panels show the cases with NO and IO, respectively. We assume that the progenitor is located at 200~pc.}
    \label{fig:eventrate}
\end{figure*}

\begin{figure*}
    \centering
    \includegraphics[width=8cm]{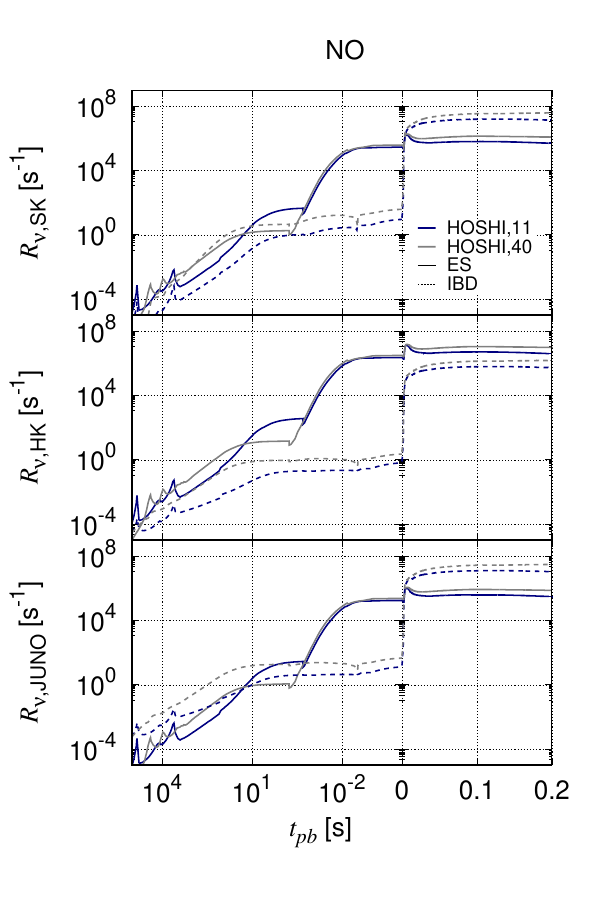}
    \includegraphics[width=8cm]{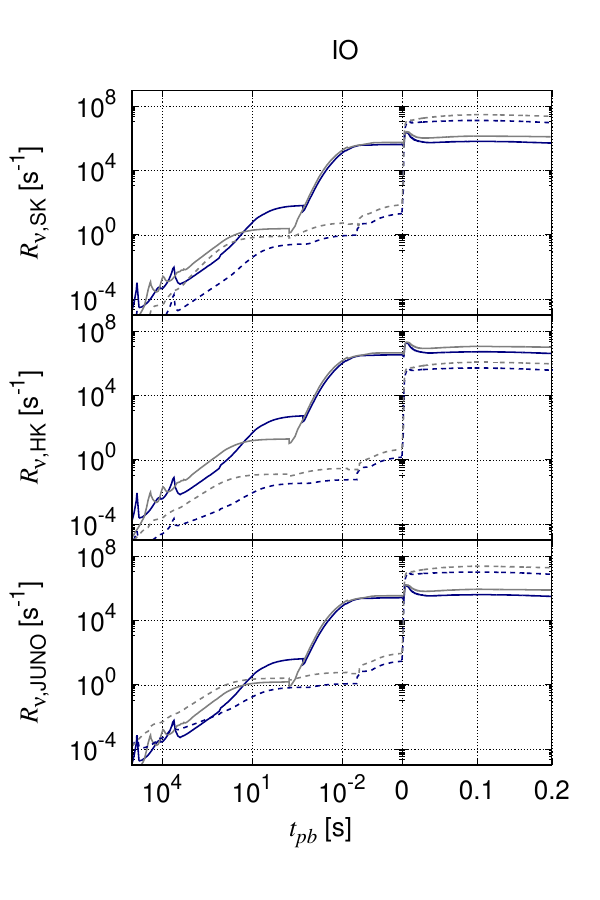}
    \caption{Time evolution of the ES and IBD event rates for the HOSHI 11~$M_\odot$ and 40~$M_\odot$ models in SK, HK, and JUNO. Solid and dashed lines show the ES and IBD channels, respectively.  Left and right panels show the cases with NO and IO, respectively. We assume that the progenitor is located at 200~pc.}
    \label{fig:eventrate_elesca_IBD}
\end{figure*}

\subsubsection{Neutrino Oscillation}

As neutrinos propagate from the progenitor to Earth, they undergo flavor conversion.  We account for both vacuum oscillations and the Mikheyev-Smirnov-Wolfenstein (MSW) effect in the adiabatic limit, modeling the oscillated spectra as follows:
\begin{eqnarray}
\left(\frac{dL^N_{\nu_e}}{dE_\nu}\right)_\text{osc} = p \left(\frac{dL^N_{\nu_e}}{dE_\nu}\right)_0 + (1-p) \left(\frac{dL^N_{\nu_x}}{dE_\nu}\right)_0, \nonumber \\
\left(\frac{dL^N_{\bar{\nu}_e}}{dE_\nu}\right)_\text{osc} = p^\prime \left(\frac{dL^N_{\bar{\nu}_e}}{dE_\nu}\right)_0 + (1-p^\prime) \left(\frac{dL^N_{\bar{\nu}_x}}{dE_\nu}\right)_0.
\end{eqnarray}
Here, the subscript 0 denotes the original unoscillated spectra evaluated at the source. The terms $p$ and $p^\prime$ represent the survival probabilities for $\nu_e$ and $\bar{\nu}_e$, respectively. For our calculations, we adopt $p = 0.0234$ (0.301) and $p^\prime = 0.676$ (0.0240) for the normal (inverted) mass ordering. For simplicity, we currently neglect the effects of collective neutrino oscillations induced by neutrino-neutrino self-interactions \citep[e.g.][]{duan2010}.

\subsection{Results for Neutrino Detection} \label{subsec:obs_results}

\subsubsection{Neutrino Event Rates}

Figure~\ref{fig:eventrate} displays the expected neutrino event rates in each detector as a function of time. The panels correspond to SK, HK, KamLAND, JUNO, and DUNE from top to bottom. The left and right columns show the results for the normal ordering (NO) and inverted ordering (IO) cases, respectively. Focusing initially on $\bar{\nu}_e$ detection under the NO scenario, we observe distinct temporal trends driven by detector energy thresholds. For $t_{pb} \lesssim -500$~s, the average energy of preSN neutrinos is generally below the detection thresholds of all facilities. As stellar evolution proceeds, this average energy gradually increases (see the right panels of Figure~\ref{fig:numberlum}). Because liquid scintillator detectors possess a significantly lower-energy threshold of $1.8$~MeV, they begin to register higher event rates at much earlier epochs. In contrast, after core bounce, the average neutrino energy becomes sufficiently high that the impact of these thresholds diminishes. The event rate then scales almost entirely with the detector volume, naturally resulting in the highest expected rates for the massive HK detector.

The $\nu_e$ detection in DUNE exhibits a fundamentally different behavior. Although the energy threshold for the CC of argon is significantly higher than that for IBD, $\nu_e$ are primarily produced via EC. This physical mechanism produces systematically higher-energy neutrinos, causing the average $\nu_e$ energy to rise earlier than that of $\bar{\nu}_e$. As a result, during the collapse phase ($t_{pb} \lesssim -1$~s), DUNE occasionally registers a higher $\nu_e$ event rate than the corresponding $\bar{\nu}_e$ rates observed in the other detectors.

Comparing the two mass orderings reveals expected shifts in the observed flavor distributions. In the NO scenario, the survival probability for $\bar{\nu}_e$ ($p^\prime$) remains relatively high, whereas the IO scenario strongly enhances the survival probability for $\nu_e$ ($p$). Because of these well-understood oscillation dynamics, shifting from the NO to the IO case naturally decreases the $\bar{\nu}_e$ event rates across all IBD-sensitive detectors, while simultaneously boosting the $\nu_e$ detection rate in DUNE. These overarching trends are entirely consistent with the previous literature \citep[e.g.][]{Kato2020b}.

Time evolution of ES events are shown in Figure~\ref{fig:eventrate_elesca_IBD}. We pick up two representative progenitors, the HOSHI 11~$M_\odot$ and 40~$M_\odot$ models. Here we show the results for SK, HK, and JUNO, adopting the same recoil-electron threshold for the ES channel as described above. The relative importance of ES and IBD depends on both the detector properties and the neutrino mass ordering. In the NO case, the 11~$M_\odot$ model, or the degenerate core model, exhibits a particularly large $\nu_e$ flux. As a result, for water Cherenkov detectors such as SK and HK, whose large target volumes compensate for the recoil-electron threshold, the ES event rate can exceed the IBD event rate over most of preSN phase. In contrast, for JUNO the IBD rate is initially larger, and the ES rate becomes comparable to or larger than the IBD rate only at later times, when the $\nu_e$ luminosity rapidly increases toward core collapse.

The same qualitative behavior is found for the IO case, but the difference is more pronounced. Because the $\nu_e$ flux at Earth is larger, the ES channel is enhanced relative to IBD. Consequently, the ES rate exceeds the IBD rate earlier and more clearly, particularly for the low-mass progenitor. These results show that ES can provide a useful complementary channel to IBD, especially in models with strong $\nu_e$ emission.

\subsubsection{First Detection Time} \label{subsubsec:alart}

\begin{figure}
    \centering
    \includegraphics[width=8cm]{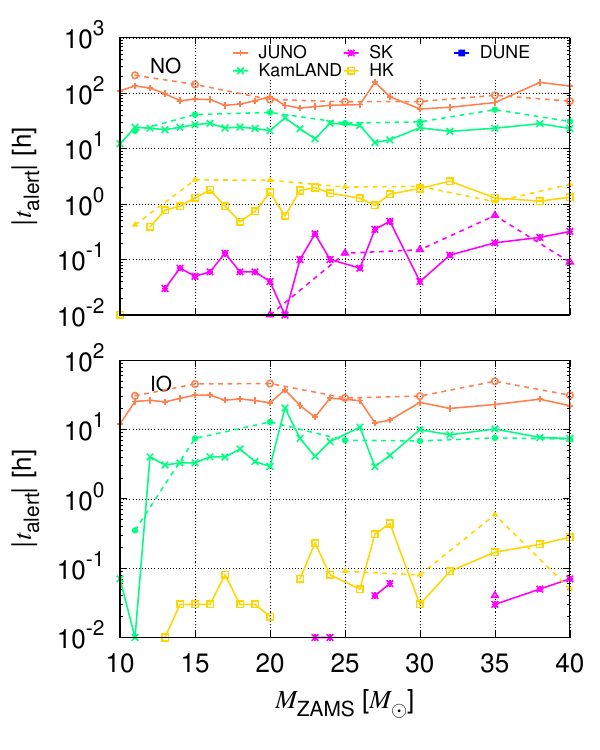}
    \caption{The time of the first detection $t_{\rm alert}$ in the case of NO (top) and IO (bottom), respectively. 
    Solid and dotted lines denote the HOSHI and MESA models, respectively. We assume that the distance to progenitors is $d=200$ pc and $(T_w, R_{\rm false})=(24~{\rm h}, 10/{\rm yr})$. Colors distinguish neutrino detectors. }
    \label{fig:begintime}
\end{figure}

Estimating the total number of emitted neutrinos from actual observations requires determining the exact time of the first detection. This moment simultaneously corresponds to issuing an early alert for the impending SN explosion. This is one of the most critical practical roles of preSN neutrino monitoring. To estimate this first detection time, we employ a statistical analysis based on the false alarm rate (FAR), following the methodologies of \citet{Simpson2019} and \citet{Kato2020b}. The detection logic relies on monitoring the event rate within a continuous sliding time window of width $T_w$. In this analysis, we test two distinct time windows: 12 and 24~hr. We apply this FAR analysis only to the IBD channel, for which prompt--delayed coincidence provides a clean event sample and background estimates are available. The ES channel is not included in the alert criterion because its single-event backgrounds require detector-specific modeling. Nevertheless, if ES events can be reliably identified in future analyses, this channel may also contribute to preSN alerts, because the ES event rate can exceed the IBD event rate at earlier times for some progenitor models (see Fig.~\ref{fig:eventrate_elesca_IBD}).

Let $R_{\mathrm{BG}}$ denote the constant background event rate for a specific detector, with values for each facility summarized in Table~\ref{Tab:detector}. The expected number of background events within a given window is simply $B = R_{\mathrm{BG}} \times T_w$. The critical detection threshold, $N_{\mathrm{th}}$, is determined to ensure that the frequency of false positives triggered by background fluctuations does not exceed a predefined FAR. We define $N_{\mathrm{th}}$ as the smallest integer satisfying the following condition:
\begin{eqnarray}
\sum_{k=N_{\text{th}}}^{\infty} P(k|B) \le R_{\rm false} \times T_w,
\end{eqnarray}
where the left-hand side represents the probability of observing $N_{\text{th}}$ or more background events in a single window, assuming Poisson statistics $P(k|B)$ for $k$ events. We consider two FAR scenarios: $R_{\rm false}=1$/yr and $10$/yr.

Using this derived threshold, we can calculate the alert time, $t_{\mathrm{alert}}$. The expected number of signal events, $S(t)$, accumulated in the window ending at time $t$ is given by:
\begin{eqnarray}
S(t) = \int_{t-T_w}^{t} R_{\nu}(t') \, dt'. \label{eq:alart_time}
\end{eqnarray}
A successful detection is declared at the earliest time $t_{\mathrm{alert}}$ when the total expected number of events (signal plus background) exceeds the required threshold:
\begin{eqnarray}
S(t_{\text{alert}}) + B \ge N_{\text{th}}.
\end{eqnarray}

As detailed in Appendix~\ref{appendix_alert}, the derived alert time is highly sensitive to the choices of $T_w$ and $R_{\mathrm{false}}$, and the optimal combination varies across different detectors. For liquid scintillator detectors characterized by low background rates, longer $T_w$ values prove more suitable because they allow the cumulative neutrino signal to safely dominate over the background. As dictated by its definition, higher values of $R_{\mathrm{false}}$ naturally favor earlier alerts regardless of the detector type. Throughout the remainder of this study, we adopt $T_w = 24$~hr and $R_{\mathrm{false}} = 10$/yr as our standard parameters.

Figure~\ref{fig:begintime} summarizes the estimated alert times, displaying the NO and IO cases in the top and bottom panels, respectively. JUNO consistently provides the earliest alert times, entirely independent of the stellar evolution code, progenitor mass, or neutrino mass ordering. KamLAND, another liquid scintillator detector, follows closely behind. Because water Cherenkov detectors (SK and HK) possess higher energy thresholds, their first detection times are significantly delayed, rendering them less optimal for early warning purposes. In DUNE, the combination of a high background rate and a high energy threshold poses severe challenges for issuing an early preSN alert.

\begin{figure*}[t]
    \centering
    \includegraphics[width=\textwidth]{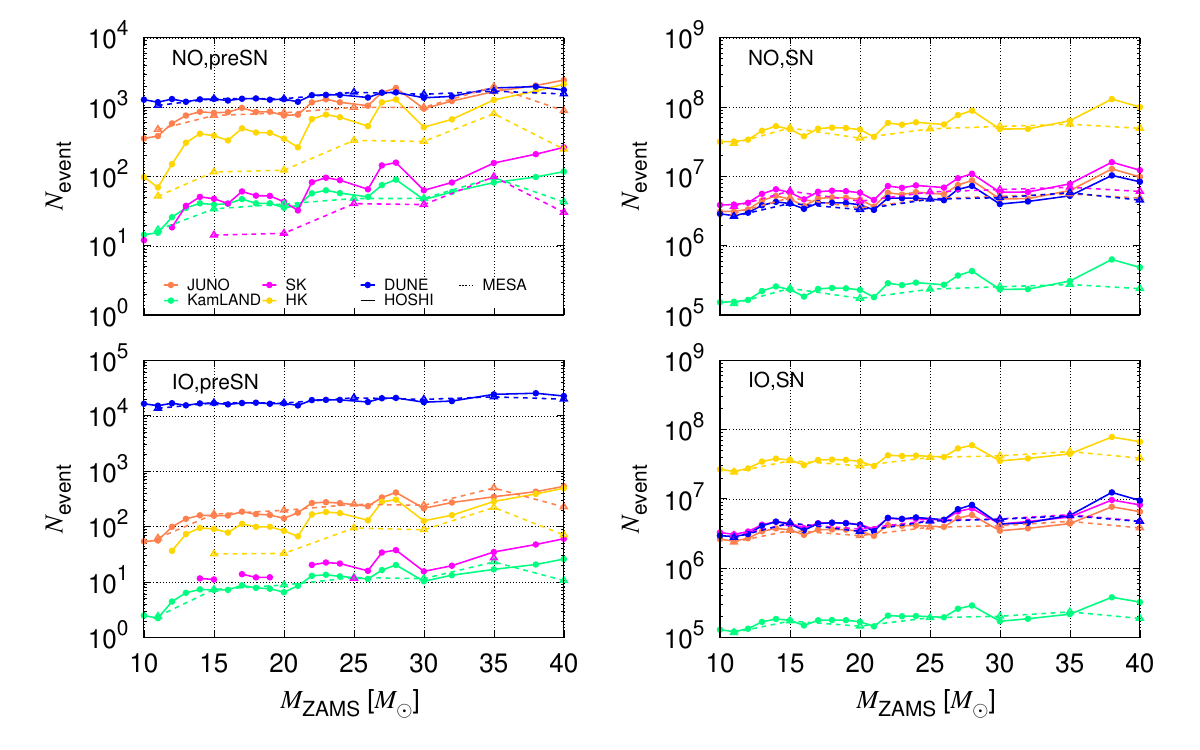}
    \caption{Total number of IBD events in the preSN (left) and SN phases (right). Top and bottom panels show the NO and IO cases, respectively.
    We assume that the progenitor is located at 200~pc and $(T_w, R_{\rm false})= (24~{\rm h},10/{\rm yr})$.
    Colors distinguish neutrino detectors.}
    \label{fig:totevent}
\end{figure*}

Crucially, the alert time does not exhibit a clear dependence on the progenitor model. Comparing the alert times for JUNO under the NO scenario, the $27~M_\odot$ model and the $11~M_\odot$ model yield the earliest detection times for the HOSHI and MESA progenitor sets, respectively. This stochasticity arises from evaluating the event rate over a fixed time window of $T_w = 24$~hr. This approach makes the alert trigger highly sensitive to the specific timing of internal evolutionary phases and the transient peaks in their associated neutrino emission. For example, in the HOSHI $27~M_\odot$ model, oxygen-core burning occurs approximately 150~hr prior to the bounce ($t_{pb} \sim -150$~hr), perfectly coinciding with the alert time. This advanced burning stage violently alters the central core state and causes a transient enhancement in the neutrino luminosity. A similar localized enhancement occurs in the MESA $11~M_\odot$ model, where off-center oxygen burning triggers around $t_{pb} \sim -200$~hr.

\subsubsection{Total Number of Neutrino Events} \label{subsubsec:event}

\begin{figure*}
    \centering
    \includegraphics[width=15cm]{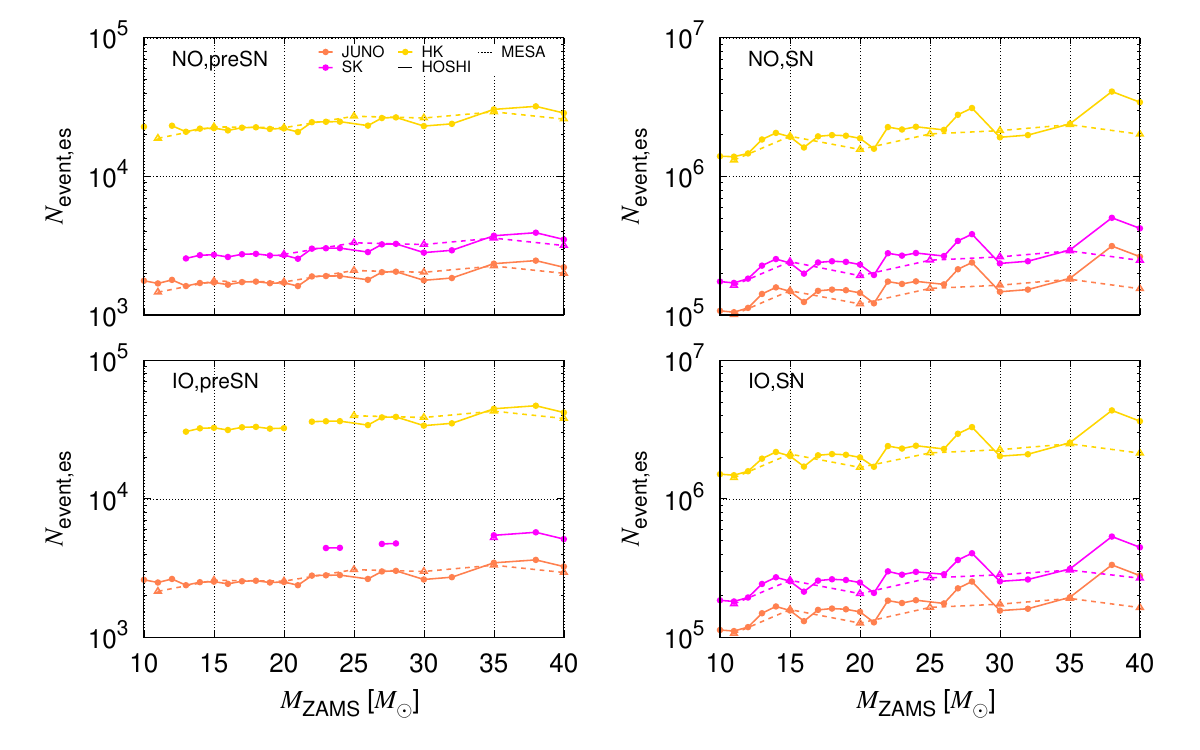}
    \caption{The same as Fig.~\ref{fig:totevent} but for ES events. }
    \label{fig:totevent_elesca}
\end{figure*}

We next evaluate the total number of expected neutrino events, dividing our analysis into the preSN and SN phases. Throughout this calculation, we assume a standard progenitor distance of 200~pc. For the preSN phase, the total event count is obtained by time-integrating the expected event rate $R_\nu(t)$ for each detector from the alert time ($t_{\mathrm{alert}}$) up to the moment of core bounce ($t_{pb}=0$).

The results regarding IBD events for the preSN phase are presented in the left panels of Figure~\ref{fig:totevent}, with the NO and IO cases shown in the top and bottom rows, respectively. For $\bar{\nu}_e$ detection during the preSN phase, JUNO consistently records the highest number of events across all progenitor masses. This dominance highlights that the duration of the detectable window—dictated by the low energy threshold—plays a more critical role than detector volume. Because the NO scenario yields a higher survival probability for $\bar{\nu}_e$ ($p^\prime$), the accumulated event counts are naturally larger compared to the IO case.

The $\nu_e$ detection in DUNE presents a contrasting scenario. Although the initial detection occurs almost immediately before core bounce, the instantaneous event rate rises up exceptionally high ($R_\nu(t) \sim 10^4$--$10^6$~s$^{-1}$). This intense flux allows DUNE to accumulate approximately $10^3$--$10^4$ events within a very brief window, strongly dependent on the chosen mass ordering\footnote{While the formal alert time for DUNE often coincides with $t_{pb} = 0$~s, the integration encompasses the neutrino emission and background counts within the preceding window $T_w$. The number of detected preSN neutrinos thus remains strictly finite (see Equation~\ref{eq:alart_time}).}.

Turning to the SN phase, literature conventionally scales neutrino event estimates to a distance of 10~kpc (the approximate distance to the Galactic center). Because we deliberately adopt a distance of 200~pc to emphasize the continuity between the preSN and SN phases, the event counts presented in the right panels of Figure~\ref{fig:totevent} are substantially larger than standard reference values. 

It is worth noting that if an SN were to occur at a distance of approximately 200~pc, SK would detect approximately $10^6-10^7$ neutrino events. This flux intensity would exceed the capacity of the current data acquisition system. Therefore, we emphasize that specific countermeasures, such as modifications to the electronics or buffer systems, are currently being implemented to handle such an extremely high event rate \citep{Mori2024}.

We also estimate the total number of ES events accumulated after the IBD-based alert time shown in Figure~\ref{fig:totevent_elesca}. The model dependence is much weaker than in the IBD channel as well as Ar reaction at DUNE. This is because ES is sensitive to all neutrino flavors, and the accumulated ES events are dominated by the rapidly increasing $\nu_e$ emission immediately before core bounce. As a result, differences in the IBD alert time, the adopted energy threshold, and the detailed precollapse evolution are largely washed out by the intense late-time $\nu_e$ emission. The total number of ES events is therefore controlled mainly by the detector size once the recoil-electron threshold is fixed.

The ES channel therefore provides information complementary to IBD. While IBD primarily constrains the $\bar{\nu}_e$ component, ES measures a cross-section-weighted contribution from all flavors. If the $\bar{\nu}_e$ contribution is constrained by the IBD sample, the ES events can in principle provide information on the non$\bar{\nu}_e$ component, especially the $\nu_e$ emission associated with EC in the final collapse phase.

\subsubsection{Source Pointing}
The ES channel can also provide directional information because the recoil electron is preferentially emitted in the incoming neutrino direction. However, in the preSN phase the accumulated ES statistics are limited, and our simple estimate indicates that a pointing accuracy better than $\sim10^\circ$ is typically achieved only very close to bounce. Thus, preSN ES events may provide a useful directional hint for nearby progenitors, but they are unlikely to determine the progenitor position much earlier than the postbounce neutrino burst.

\subsection{Observability of Correlations with Stellar Key Parameters} \label{subsec:obs_correlation}

\begin{figure*}
    \centering
    \includegraphics[width=15cm]{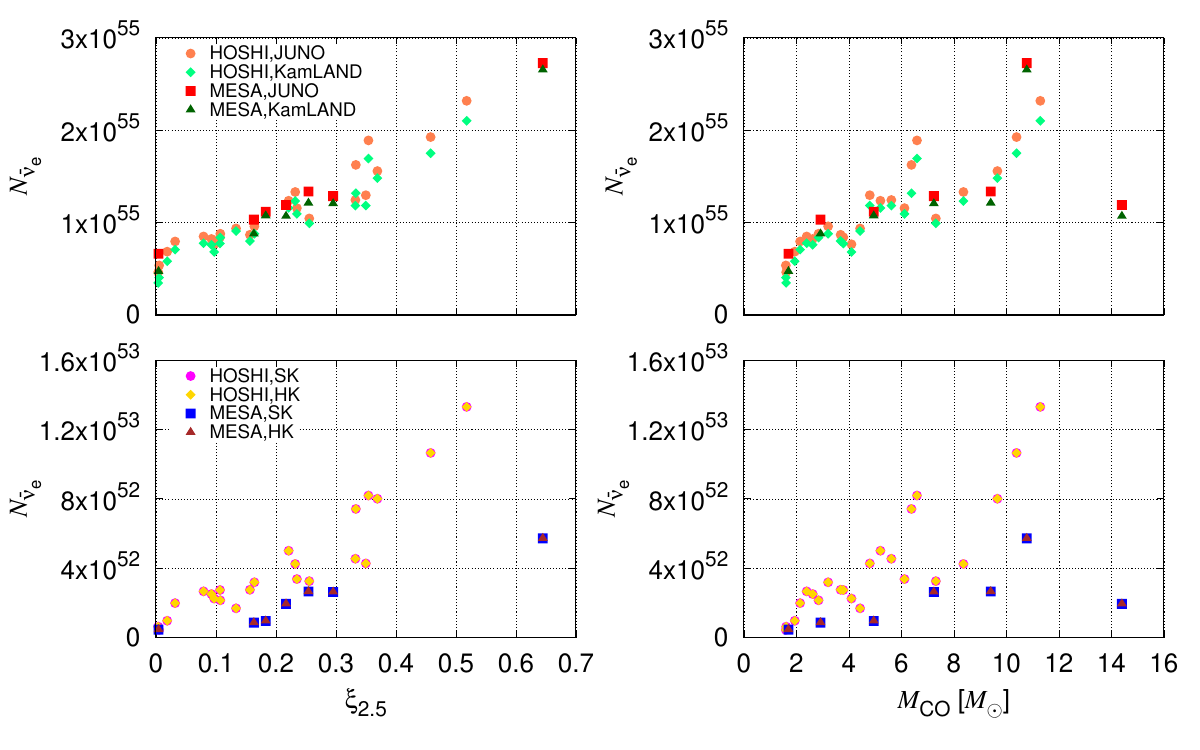}
    \caption{Relationship between integrated neutrino number and $\xi_{2.5}$ (left) or $M_{\rm CO}$ (right) in the preSN phase. In these figures, we assume that the progenitor is located at 200~pc and the mass ordering is normal. Colors and point types distinguish detectors and stellar evolution codes, respectively. }
    \label{fig:correlation_preSN_period}
\end{figure*}

We discuss how the theoretical correlations between key progenitor parameters and time-integrated neutrino emission, identified in Section~\ref{ch4}, manifest in actual observations using specific detectors. In our previous theoretical analysis, integrating the total number of neutrinos, $N(t_i)$, from a defined initial time $t_i$ up to core bounce revealed distinct trends: correlations with $\xi_{2.5}$-type parameters emerge when considering only the final evolutionary phase ($t_i \sim -10^5$~s), while correlations with $M_{\mathrm{CO}}$-type parameters require a much longer integration window ($t_i \sim -10^9$~s). Here, we reexamine these relationships by incorporating three critical observational factors: neutrino oscillations, detector energy thresholds, and realistic alert times. Our primary objective is to verify whether these correlations survive under practical detection conditions.

Figure~\ref{fig:correlation_preSN_period} illustrates these observationally filtered correlations for $\bar{\nu}_e$ during the preSN phase. The left and right panels display the dependence on $\xi_{2.5}$ and $M_{\mathrm{CO}}$, respectively, comparing liquid scintillator detectors (top) with water Cherenkov detectors (bottom). The $\bar{\nu}_e$ information is reconstructed from the IBD events. Strikingly, the theoretical correlations remain largely intact even after applying these three observational constraints. Because the estimated alert times for all detectors generally fall within the window of $t_{pb} \sim -140$ to $0$~hr, they naturally align with the integration timescale ($t_i \sim -10^5$~s) where the neutrino emission originally exhibited a strong correlation with $\xi_{2.5}$. This alignment ensures that the tight relationship with $\xi_{2.5}$ survives under realistic detection scenarios. Furthermore, liquid scintillator detectors with lower-energy thresholds capture a larger fraction of the early lower-energy flux, thereby exhibiting more pronounced correlations than their water Cherenkov counterparts. Although a more rigorous treatment of detection uncertainties is ultimately required—particularly because our current analysis assumes perfect reconstruction of neutrino number luminosities from the observed data—these findings strongly suggest that future preSN neutrino detections can provide direct, albeit rough, observational constraints on the core compactness $\xi_{2.5}$.

\begin{figure*}
    \centering
    \includegraphics[width=16cm]{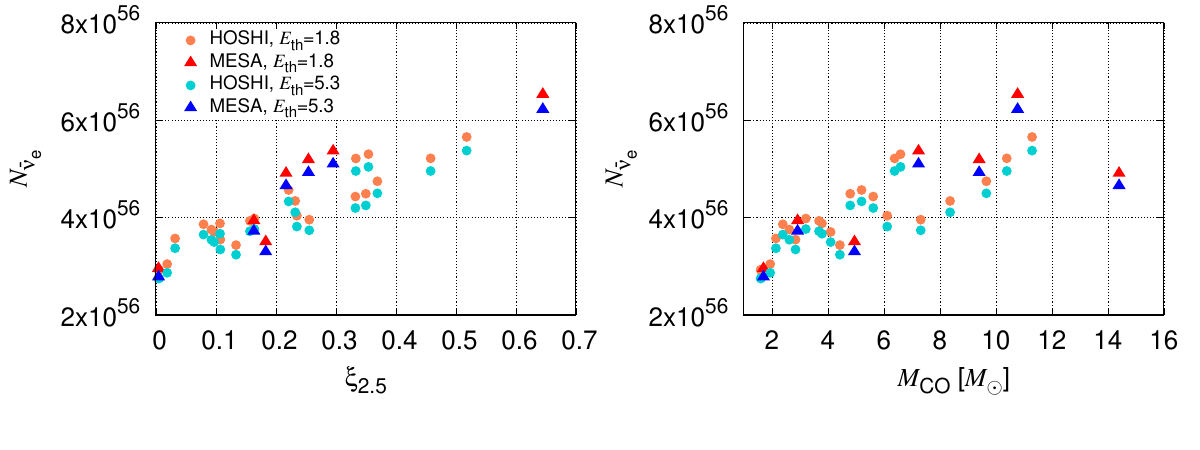}
    \caption{Relationship between integrated neutrino number and $\xi_{2.5}$ (left) or $M_{\rm CO}$ (right) in the SN phase. In these figures, we assume that the progenitor is located at 200~pc and the mass ordering is normal. Colors and point types distinguish detectors and stellar evolution codes. }
    \label{fig:correlation_SN_period}
\end{figure*}

For neutrinos emitted during the SN phase, the detection process is completely independent of an alert time trigger. Because background rates become negligible during the intense SN burst and detector volume acts merely as a uniform scaling factor for the event counts, the functional differences between detectors naturally reduce to their energy thresholds. We thus focus solely on how these energy thresholds affect the observable correlations. Figure~\ref{fig:correlation_SN_period} presents these results. Because our integration is strictly limited to the early postbounce phase ($\lesssim$ 200~ms), the observable correlations show virtually no dependence on the underlying stellar evolution code. This remarkable insensitivity to the specific evolutionary code represents a major advantage of targeting the early SN phase as we mentioned earlier. Therefore, we can cleanly extract physical correlations with parameters like $\xi_{2.5}$ and $M_{\mathrm{CO}}$ without severe contamination from model-dependent uncertainties.

We move on to the non$\bar{\nu}_e$ component. As mentioned above, the ES channel is sensitive to all neutrino flavors. Therefore, once the $\bar{\nu}_e$ component is constrained by IBD, the remaining ES signal can in principle provide information on $\nu_e$, $\nu_x$, and $\bar{\nu}_x$. However, the ES cross section depends on the neutrino flavor. Thus, the observable quantity is not the simple sum of the fluences of these flavors, but a cross-section-weighted sum. We define the effective neutrino number relevant for ES as:
\begin{eqnarray}
\Phi_{\rm eff} = \sum_{\alpha}
\int_{t_{\rm alert}}^{0} dt
\int_{E_\nu(E_{th,e})}^{E_{\nu,{\rm max}}} dE_\nu \nonumber \\
\ \ \ \ \ \times \left(\frac{dL_{\alpha}^N}{dE_\nu}\right)_{\rm osc}\,
\sigma_{\alpha e}(E_\nu;E_{th,e}),
\end{eqnarray}
where $\alpha=\nu_e,\nu_x,\bar{\nu}_x$, and the energy spectrum after the oscillation is taken into account, $\left(dL_\alpha^N/dE_\nu\right)_{\rm osc}$. The ES cross section $\sigma_{\alpha e}$ is 
\begin{eqnarray}
\sigma_{\alpha e}(E_\nu;E_{th,e}) =
\int_{E_{th,e}}^{E_{e,{\rm max}}(E_\nu)}
\frac{d\sigma_{\alpha e}}{dE_e}(E_\nu,E_e)\,dE_e ,
\end{eqnarray}
with $E_{e, \rm{max}}(E_\nu)=2E_\nu^2/(m_e+2E_\nu)$.

\begin{figure*}
    \centering
    \includegraphics[width=15cm]{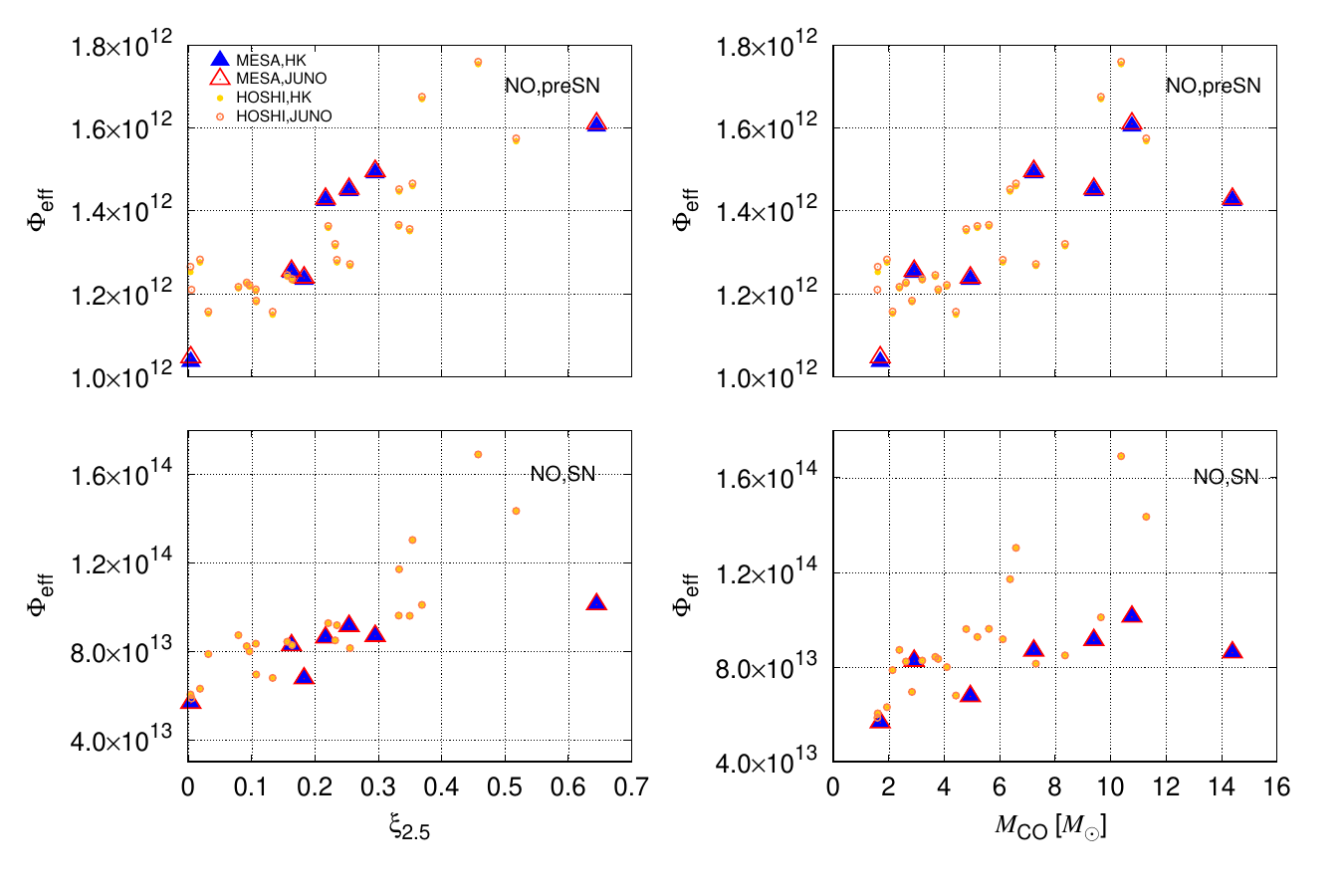}
    \caption{Relationship between effective flux of $\nu_e$+$\nu_x$ and $\xi_{2.5}$ (left) or $M_{\rm CO}$ (right). The results in the preSN and SN phases are shown in top and bottom panels, respectively. In these figures, we assume that the progenitor is located at 200~pc and the mass ordering is normal. Colors and point types distinguish detectors and stellar evolution codes, respectively.}
    \label{fig:correlation_elesca_flux}
\end{figure*}

Figure~\ref{fig:correlation_elesca_flux} shows the correlation between $\Phi_{\rm eff}$ and either $\xi_{2.5}$ or $M_{\rm CO}$. In the preSN phase, the effective number for non$\bar{\nu}_e$ flavors shows a clearer trend with $M_{\rm CO}$ than the IBD-reconstructed $\bar{\nu}_e$ number. This behavior is not simply due to the total number of emitted $\nu_e$. We find that the model-to-model scatter is reduced when only neutrinos above the effective ES threshold are included. The corresponding neutrino-energy threshold for $E_{th,e}=3.5~{\rm MeV}$ is $E_\nu^{\rm min}\sim3.74~{\rm MeV}$. In all models, the energy peak of the $\nu_e$ spectrum lies below this threshold and introduces additional scatter in the total $\nu_e$ number. By contrast, the above-threshold component, which is more directly relevant for ES detection, exhibits a weaker model dependence and therefore shows a clearer relation with $M_{\rm CO}$.

The ES quantity should therefore be interpreted as a threshold- and cross-section-weighted non$\bar{\nu}_e$ fluence, rather than as a simple measure of the total $\nu_e+\nu_x+\bar{\nu}_x$ emission. 
Since IBD primarily probes $\bar{\nu}_e$, while ES is sensitive to $\nu_e$, $\nu_x$, and $\bar{\nu}_x$, the two channels provide complementary information. The clearer $M_{\rm CO}$ trend in $\Phi_{\rm eff}$ suggests that the high-energy, ES-accessible part of the non$\bar{\nu}_e$ emission retains information on the CO core evolution, although this relation is indirect and depends on the adopted recoil-electron threshold.

In the early SN phase, followed here up to $t_{\rm pb}=200$~s, the correlation with $\xi_{2.5}$ is weaker than that found for the IBD-reconstructed number of $\bar{\nu}_e$ discussed above. This does not mean that the individual ES-weighted flavor components are uncorrelated with compactness. Rather, each flavor component shows a positive trend, but with different levels of scatter. Because the total ES quantity is a threshold- and cross-section-weighted sum of $\nu_e$, $\nu_x$, and $\bar{\nu}_x$, these different scatters partially wash out a single tight compactness relation. For $M_{\rm CO}$, the qualitative behavior is similar to the IBD case. The ES-weighted fluence shows only a moderate positive trend with substantial scatter, indicating that the neutrino emission during the first 200~ms after bounce is not determined solely by $M_{\rm CO}$, but is also affected by the early postbounce hydrodynamics and neutrino transport.

DUNE provide another important channel through charged-current interactions of $\nu_e$ on argon, especially for the final prebounce stage, the neutronization burst, and the early postbounce phase \citep[e.g.,][]{Abi2021}. A combined analysis of IBD, ES, and argon charged-current events is particularly powerful for reconstructing flavor-dependent supernova neutrino spectra, and such multidetector approaches have been investigated for the postbounce supernova signal \citep[e.g.,][]{Nagakura2021a, nagakura2021b, Huang2023}. In the present work, however, we do not include DUNE in the correlation analysis. As we discussed in the previous section, the detectable $\nu_e$ signal in DUNE is expected to become significant mainly very close to core bounce and the total number of events is almost independent of progenitor models. Ongoing studies of low-energy physics in DUNE aim to improve the sensitivity to MeV-scale signals, including supernova and solar neutrinos \citep[e.g.,][]{Caratelli2022, Paulucci2022}. If such developments enable a lower-threshold and more stable detection of preSN $\nu_e$ events, a combined flavor-resolved analysis with IBD, ES, and argon charged-current events will become an important future extension of this work.

Extending the analysis into later accretion phases using multidimensional simulations would likely reveal much larger discrepancies tied to the progenitor structure. nonlinear hydrodynamic instabilities, such as SASI and neutrino-driven convection, dominate this later phase. These multidimensional processes act as powerful amplifiers, highly sensitive to subtle differences in the density profiles of progenitors that vary significantly among stellar evolution codes \citep[e.g.][]{Nagakura2021}. We therefore expect the later evolutionary phases to exhibit highly code-dependent trends, and we leave the detailed verification of these long-term multidimensional effects for future work.

\subsection{Detection Prospects for a Progenitor at 1~kpc}

\begin{figure*}
    \centering
    \includegraphics[width=18cm]{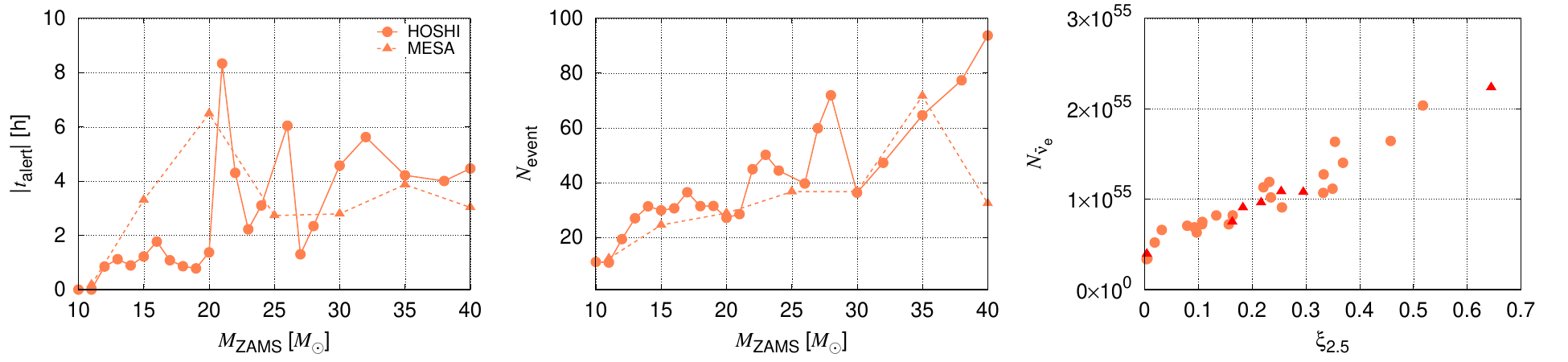}
    \caption{Alert time (left) and total numbers of neutrino events (middle) in the preSN phase for JUNO. The correlation between $\xi_{2.5}$ and $N_{\bar{\nu}_e}$ is shown in the right panel. In these figures, we assume that the progenitor is located at 1~kpc and neutrino follows the NO. We take $(T_w, R_{\rm false})=(24~{\rm h}, 10/{\rm yr})$.}
    \label{fig:correlation_SN_period_1kpc}
\end{figure*}

Our discussion has so far focused on a nearby distance of 200~pc. It is crucial to evaluate the detectability of preSN neutrinos from more distant progenitors because of the low Galactic SN rate. In this final section, we extend our analysis to a progenitor located at 1~kpc.

Figure~\ref{fig:correlation_SN_period_1kpc} presents the alert time (left), the total number of neutrino events (middle), and the correlation between $\xi_{2.5}$ and the total number of preSN neutrinos (right). For these calculations, we assume the normal ordering (NO) and standard alert parameters of $T_w = 24$~hr and $R_{\mathrm{false}} = 10$/yr. Even at this extended distance of 1~kpc, JUNO retains the capability to issue an early warning 0--8~hr prior to core bounce, strongly depending on the progenitor mass. Within this 1~kpc cosmic neighborhood, approximately 40 candidate progenitor stars could potentially be monitored \citep{Mukhopadhyay2020}.

The expected total number of events scales with the progenitor mass, allowing JUNO to register approximately 10--100 events. After incorporating the three critical observational factors—neutrino oscillations, energy thresholds, and alert times—the physical correlation with $\xi_{2.5}$ persists even at this extended 1~kpc baseline. However, for scenarios yielding low detection statistics (around 10 events), accurately reconstructing the total intrinsic neutrino emission from the observed signal becomes inherently challenging. More sophisticated statistical reconstruction techniques and detailed detector simulations will be essential in future work to fully unlock the potential of these distant preSN observations.

\section{Synergistic Strategy for Probing Progenitor Structures via PreSN and SN Neutrino Observations} \label{ch6}

In this paper, we have comprehensively investigated the neutrino emission spanning the preSN stages through the early SN phase.  Past studies heavily discuss the relationship between SN neutrinos and their progenitor stars; those analyses inherently suffer from uncertainties tied to the SN explosion mechanism, including shock propagation and complex neutrino transport. To circumvent these limitations, our study isolates the correlation between preSN neutrinos and progenitor properties. Observing these preSN neutrinos offers an entirely independent means of probing the progenitor core, allowing direct validation of competing progenitor models. A defining feature of our approach is the deliberate focus on the early SN phase, where the obscuring effects of the explosion mechanism remain minimal.

As detailed in Section~\ref{ch5}, detecting preSN neutrinos requires the source to be in the vicinity of Earth, implying a decidedly low expected event rate. A fortuitous detection of even a single nearby SN would thus be an invaluable, once-in-a-lifetime scientific milestone, making it vital to understand exactly what physical information can be extracted from such a rare signal. Our findings demonstrate a strong correlation between the time-integrated number of preSN neutrinos and key progenitor parameters, particularly $M_{\mathrm{CO}}$ and $\xi_{2.5}$. These relationships guarantee that actual preSN neutrino observations can place direct empirical constraints on these parameters.

Observations of subsequent SN neutrinos independently yield information on $\xi_{2.5}$. Extracting consistent values for $\xi_{2.5}$ from both the preSN and SN phases would place more precise constraints on the progenitor's structure. A discrepancy between the observational results with the two phases would immediately expose underlying contradictions within either the stellar evolution or SN models, forcing a critical reassessment of current theoretical frameworks. Deriving such independent constraints across distinct evolutionary epochs provides an exceptionally powerful diagnostic tool, even in the event of a single detection. Extending this methodology to multidimensional SN simulations and analyzing later epochs (e.g., the accretion and PNS cooling phases beyond 200~ms) will help map the progenitor structure outside the iron core. We reserve this crucial extension for future work.

Our current analysis relies on specific numerical parameters and input physics intrinsic to the HOSHI and MESA codes. Introducing alternative physical treatments, such as modifying convective mixing parameters or incorporating magnetic fields and stellar rotation, might shift the baseline correlations. Such theoretical shifts would naturally alter the inferred values of $\xi_{2.5}$ and $M_{\mathrm{CO}}$ derived from actual observations. Even with these model dependencies, assuming a specific theoretical framework allows these combined preSN and early SN neutrino signatures to provide meaningful, albeit approximate, constraints on the core properties. Ultimately, the synergistic methodology presented here offers valuable clues for identifying which input physics and numerical parameters are better suited to approximate the deep internal structure of an observed progenitor star.

\section{Discussion} \label{ch7}

Before summarizing our results, we discuss several uncertainties and limitations that should be considered when interpreting our findings. Since our analysis connects preSN neutrino emission to progenitor structure and observational event numbers, the results are affected by uncertainties in the neutrino-emission physics (Section~\ref{sec:weak_rate_uncertainties}), stellar evolution modeling (Section~\ref{sec:stellar_evolution_uncertainties}), and detector response (Section~\ref{sec:observational_uncertainties}). Below we summarize the main sources of uncertainty relevant to this study.

\subsection{Uncertainties in Neutrino Emission and Propagation}
\label{sec:weak_rate_uncertainties}

Theoretical predictions of preSN neutrino emission depend sensitively on the adopted nuclear weak rates. The weak-rate sets used in this work cover sd-shell nuclei, pf-shell nuclei, and heavier nuclei that appear under NSE-like conditions. Nevertheless, these rates remain an important source of uncertainty. Recent studies have updated weak rates for selected sd-shell nuclei and Urca pairs, including Ne, F, Na, Mg, and Si-region nuclei, and have improved the treatment of forbidden transitions and excited-state contributions 
\citep[e.g.,][]{MartinezPinedo2014, Suzuki2016, Suzuki2019, Kirsebom2019, Langanke2021, Suzuki2024, Sharma2025}. In the pf-shell and iron-group region, ECs on nuclei play a central role in deleptonization and collapse dynamics 
\citep[e.g.,][]{Langanke2000, Langanke2001, Langanke2003, Sullivan2016}. For neutron-rich and heavier nuclei, shell-model calculations, finite-temperature quasiparticle random-phase approximation (TQRPA/QRPA) methods, and energy-density-functional approaches continue to improve the treatment of ECs, $\beta$ decays, forbidden transitions, and thermal unblocking effects \citep[e.g.,][]{ Sarriguren2013,Dzhioev2020, Giraud2022, Ravlic2025}. These finite-temperature nuclear effects may affect neutrino spectra and energy-loss rates from hot nuclei in the preSN environments \citep[e.g.,][]{Dzhioev2023,Dzhioev2024,Dzhioev2025}. 

These uncertainties can affect the absolute luminosities, spectra, and sharp time-dependent features such as shell-burning episodes. For example, sd-shell nuclei can be important during O/Ne/Si shell burning and in low-mass progenitors, whereas pf-shell and iron-group nuclei become increasingly important in the dense inner core just before collapse. Weak-rate uncertainties are also relevant during collapse and in the early postbounce phase. After the onset of collapse, ECs on heavy nuclei control deleptonization, the homologous-core mass, and the neutronization burst. The predicted neutrino emission can therefore depend on the consistency between the nuclear composition supplied by the EOS and the adopted weak-interaction rates, as well as on the treatment of shell effects, neutron-rich nuclei, and single-nucleus approximations under NSE conditions \citep[e.g.,][]{Furusawa2017, Nagakura2019}. The global correlations discussed in this paper should therefore be interpreted within the adopted weak-rate framework. A systematic investigation of the impact of modern weak-rate updates on preSN neutrino signals will be an important subject of future work.

We also note that our treatment of neutrino oscillations is simplified. We adopt an MSW-based prescription, while the dense neutrino environment after the post bounce may allow additional collective effects \citep[e.g.,][]{duan2010}, including fast flavor conversion \cite[e.g.,][]{sawyer2005}, and collisional flavor instability \citep[e.g.,][]{johns2021}. These effects are driven by neutrino--neutrino interactions and depend sensitively on the flavor-dependent angular distributions, matter density, and collision terms. Recent core-collapse supernova studies have investigated the occurrence and impact of fast and collisional flavor instabilities using phonomenological \citep[e.g.,][]{Ehring2023, Ehring2023b, Mori2025, Wang2025a,Wang2026a,Akaho2025a,Akaho2026} or quantum-kinetic approaches \citep[e.g.,][]{zaizen2022, Xiong2024, Nagakura2024}. These studies suggest that flavor conversion can modify the flavor ratios, spectral shapes, neutrino heating, and the observable neutrino signal, although the outcome depends strongly on the local neutrino angular distributions and hydrodynamical conditions. A self-consistent treatment of flavor conversion, including collective effects coupled to neutrino transport, is beyond the scope of the present study and is left for future work.

\subsection{Uncertainties in Progenitor Models}
\label{sec:stellar_evolution_uncertainties}

The second class of uncertainties comes from stellar evolution modeling. Our progenitor models are one-dimensional and therefore rely on simplified prescriptions for mass loss, convection, convective boundary mixing, semiconvection, and advanced-stage shell evolution. Mass-loss rates remain uncertain and can affect the final core structure, and evolutionary path of massive stars \citep[e.g.,][]{Smith2014, Renzo2017a}. The treatment of internal mixing also strongly affects the size of burning shells, the CO core mass, and the final iron-core structure. Furthermore, uncertainties in nuclear reaction rates, especially ($^{12}\mathrm{C}(\alpha,\gamma)^{16}\mathrm{O}$), can modify the carbon-to-oxygen ratio after helium burning and thereby influence the subsequent burning history and final core structure 
\citep[e.g.,][]{deBoer2017}. In this work we also do not include stellar rotation, magnetic fields, or binary interactions, although these effects can substantially modify the evolution of massive stars \citep[e.g.,][]{Heger2005, Podsiadlowski1992, Eldridge2008, Sana2012, Schneider2020}.

A particularly important uncertainty in advanced burning stages is the treatment of shell interactions and shell mergers. Although such phenomena can appear in one-dimensional stellar evolution calculations, their detailed development involves multidimensional hydrodynamic processes, including turbulent convection, convective entrainment, and convective-reactive mixing. These processes are necessarily simplified in one-dimensional models, so the occurrence, duration, and impact of shell mergers remain uncertain \citep[e.g.,][]{Sukhbold2014, Sukhbold2018, Collins2018, Couch2015, Mueller2017, Yadav2020, Fields2020, Rizzuti2024a}.

Shell mergers can mix material across shell boundaries, modifying the local composition, nuclear energy generation, thermal structure, and nucleosynthesis \citep[e.g.,][]{Ritter2017, Yadav2020, Rizzuti2024a, Roberti2025a}. They can also generate large-scale nonspherical perturbations and alter quantities relevant for collapse, such as the density structure outside the iron core, compactness, and explodability, although the sign and magnitude of these effects are model dependent \citep[e.g.,][]{Couch2015, Mueller2017, Collins2018, Davis2019, Yadav2020, Rizzuti2024a}.

These effects may in turn modify the preSN neutrino signal. Changes in temperature can affect thermal neutrino processes, while changes in composition can alter the nuclei that dominate EC and $\beta$ decay. Thus, shell mergers may either enhance or suppress the predicted neutrino luminosities, spectra, and average energies depending on their timing, location, and nuclear composition. A quantitative assessment requires multidimensional hydrodynamic shell-burning simulations, or improved one-dimensional prescriptions calibrated by such simulations, combined with neutrino-emission postprocessing. 
We leave this for future work. However, these hydrodynamical effects influence both the progenitor structure and the neutrino emission, and therefore do not alter the basic picture that the neutrino emission retains information on the progenitor properties. We thus expect that the correlations discussed in this paper would not be qualitatively changed, although their quantitative values may depend on the hydrodynamical treatment.

\subsection{Uncertainties in Observational Estimates}
\label{sec:observational_uncertainties}

The third class of uncertainties concerns the conversion from neutrino emission to observable event numbers. In this work, we adopt optimistic detector assumptions in order to evaluate the maximum potential of current and future facilities. For example, we use the full inner detector volume and assume idealized detection efficiencies above the adopted energy thresholds. In reality, detection efficiencies depend on visible energy, event topology, reconstruction quality, fiducial-volume cuts, and detector-specific analysis selections. Detailed detector simulations including energy-dependent efficiencies, energy resolution, angular resolution, and site-dependent backgrounds are required for realistic event predictions.

For the IBD channel, delayed coincidence provides a relatively clean event sample, especially in Gd-loaded water Cherenkov detectors and liquid scintillator detectors. 
This allows us to apply a FAR-based alert criterion to the IBD channel. 
However, the actual alert performance depends on detector-specific backgrounds, neutron-tagging efficiency, analysis thresholds, and the adopted time window. 
For DUNE, solar neutrinos are expected to be the dominant irreducible background at preSN energies, while spallation and other cosmogenic backgrounds require dedicated rejection strategies.

The ES channel is subject to larger observational uncertainties. 
Unlike IBD, ES events appear as single recoil-electron events without a delayed neutron-capture tag, and are therefore more affected by single-event backgrounds such as solar neutrinos, radioactivity, and spallation products. 
Thus, the ES results in this work should be interpreted as optimistic estimates rather than detector-level alert predictions. 
A realistic ES analysis requires detector-specific efficiencies, angular responses, and time-dependent background modeling.

\section{Summary} \label{ch8}

Recent advances in observational techniques for MeV-band neutrinos have enabled not only more detailed observations of SN neutrinos but also the possibility of continuous long-term monitoring of neutrino observation including preSN neutrinos well before SN explosion. Driven by this observational potential, we present the first comprehensive modeling of neutrino luminosities and spectra spanning continuously from several hundred years prior to collapse up to 200~ms postbounce. We performed systematic calculations across 30 massive star models ranging from 10 to 40~$M_\odot$, encompassing the typical mass range of CCSN progenitors. This extensive data set is publicly available to facilitate future theoretical and observational studies.

Our methodology couples one-dimensional stellar evolution models generated by the HOSHI and MESA codes with postcollapse dynamical simulations computed using our CCSN code. Using these integrated hydrodynamic profiles, we calculate the continuous long-term neutrino luminosities and spectra, analyzing their dependence on dominant emission processes and local thermal conditions. We systematically investigate the physical correlations between the time-integrated neutrino emission—during both the preSN and early-SN phases—and key structural parameters, specifically the carbon-oxygen core mass ($M_{\mathrm{CO}}$) and the core compactness ($\xi_{2.5}$). Furthermore, assuming realistic detector configurations, we estimate early alert times and expected detection counts using a FAR statistical approach. These observational estimates incorporate the critical effects of neutrino oscillations (both vacuum and MSW effects), detector-specific energy thresholds, and background rates. We conclude by evaluating how these practical observational constraints impact the visibility of the underlying theoretical correlations. This analysis ultimately demonstrates the synergistic potential of combining preSN and early-SN neutrino detections to probe deep stellar interiors.

Our key findings are summarized as follows:
\begin{enumerate}
\item Neutrino emission mechanisms (Figures~\ref{fig:numberlum} and \ref{fig:numberspe}): PreSN neutrino emission depends on the interplay between local thermal condition and the emitting volume. Progenitors with extended cores produce higher total $\bar{\nu}_e$ and $\nu_x$ luminosities due to their larger hotter regions. Progenitors with small high-$\mu_e$ cores exhibit higher central $\nu_e$ emission rates driven by extreme densities. During the collapse and postbounce phases, $\nu_e$ emission remains largely universal, governed by robust density and electron fraction profiles. Because $\bar{\nu}_e$ and $\nu_x$ emissions are dominated by temperature-sensitive positron processes, they strongly retain the imprint of the progenitor's thermal structure.

\item PreSN phase correlations (Figure \ref{fig:correlation_preSN}): The time-integrated total number of preSN neutrinos correlates strongly with $\xi_{2.5}$ when isolating the emission from the final $\sim 10^5$~s. On the other hand, integrating over a much longer duration ($\sim 10^9$~s) reveals a distinct correlation with $M_{\mathrm{CO}}$, reflecting the stellar evolutionary history from earlier burning stages.

\item Early SN phase correlations (Figure~\ref{fig:correlation_SN}): During the very early burst phase ($\lesssim 50$~ms postbounce), $\nu_e$ emission from the universal inner core remains independent of progenitor model, whereas $\bar{\nu}_e$ and $\nu_x$ emissions from the outer core reflect the progenitor's temperature structure. In the subsequent accretion phase ($\sim 50$--200~ms), progenitor dependencies emerge across all neutrino flavors. Distinct density gradients outside the inner core drive varying mass accretion rates, which directly govern the heating and temperature of the PNS. These physical differences cause the $\bar{\nu}_e$ and $\nu_x$ luminosities to strongly correlate with the core compactness $\xi_{2.5}$.

\item Retrieving preSN correlations (Figure~\ref{fig:correlation_preSN_period}): 
The theoretical correlation between the time-integrated preSN neutrino emission and $\xi_{2.5}$ is largely preserved even when accounting for practical observational constraints, including neutrino oscillations, detector energy thresholds, and alert time. For the IBD channel, this correlation mainly reflects the detectable $\bar{\nu}_e$ emission. We also find that the ES channel can provide complementary information on the non$\bar{\nu}_e$ component, because it is sensitive to a cross-section-weighted sum of all flavors. In particular, the ES-weighted fluence retains a positive correlation with $\xi_{2.5}$ and shows an additional trend with $M_{\rm CO}$ through the high-energy, above-threshold part of the emission. Consequently, future detections of preSN neutrinos in multiple channels have the potential to provide meaningful constraints on the internal structure of the progenitor star prior to its explosion.

\item Synergy of preSN and early SN neutrino observations: Our integrated analysis demonstrates that preSN neutrinos offer an independent probe of progenitor properties, entirely free from the obscuring uncertainties of the SN explosion mechanism. Even in the event of a single SN detection, combining preSN and early-SN neutrino signals can place meaningful constraints on theoretical models or expose underlying discrepancies between stellar evolution and explosion physics. This synergy highlights the critical importance of future multidimensional simulations and broader parameter space explorations.

\end{enumerate}

In conclusion, we have demonstrated that preSN neutrino emission in the late evolutionary stages exhibits clear correlations with the progenitor's core structure. Furthermore, these structural dependencies persist into the early SN phase, driven by the varying mass accretion flows from the outer core. Our results suggest that continuous neutrino monitoring can connect two otherwise distinct observational windows: the quasi-static precollapse evolution and the dynamical early postbounce phase. If a nearby massive star undergoes core collapse, the combination of preSN neutrinos, the burst signal, and multichannel detector information may provide a unique probe of the internal structure and final evolution of the progenitor. In this sense, future observations will not only serve as an early warning of an imminent supernova, but may also provide quantitative constraints on the compactness, core mass, and flavor-dependent neutrino emission of massive stars immediately before collapse.

\section*{Acknowledgements}
We gratefully thank to Koh Takahashi, Yudai Suwa, Ken'ichiro Nakazato, Akira Harada, and Masamichi Zaizen for useful discussions. This work used high performance computing resources provided by Fugaku supercomputer at RIKEN, the Wisteria provided by JCAHPC through the HPCI System Research Project (hp230056, hp230270, hp240041, hp240264). C.K. is supported by JSPS KAKENHI Grant Numbers JP24H02245 and JP25K17395. R.A. is supported by Waseda University Grant for Special Research Projects (Project number: 2026C-742) and JSPS KAKENHI Grant Number JP24K00632. H.N. is supported by JSPS KAKENHI Grant Number JP23K03468. R.H. acknowledges support from the RIKEN Special Postdoctoral Researcher Program for junior scientists.

\section*{Data Availability}
The time series of neutrino luminosities and spectra for all progenitor models presented in this study are publicly available on Zenodo:\dataset[10.5281/zenodo.18886214]{https://doi.org/10.5281/zenodo.18886214}. The Zenodo record includes both a combined archive of the full dataset and per-model archives for users who wish to download data for individual progenitor models. The MESA inlists, scripts, and related output files used to construct the MESA progenitor models analyzed in this work are also publicly available on Zenodo:\dataset[10.5281/zenodo.20822084]{https://doi.org/10.5281/zenodo.20822084}.

\appendix

\setcounter{figure}{0}
\setcounter{table}{0}
\setcounter{equation}{0}
\renewcommand{\thefigure}{\thesection\arabic{figure}}
\renewcommand{\thetable}{\thesection\arabic{table}}
\renewcommand{\theequation}{\thesection\arabic{equation}}

\section{Impact of Timing for Switching Calculations} \label{appendix_collapse}

\begin{figure*}[htbp]
    \centering

    \includegraphics[width=6cm]{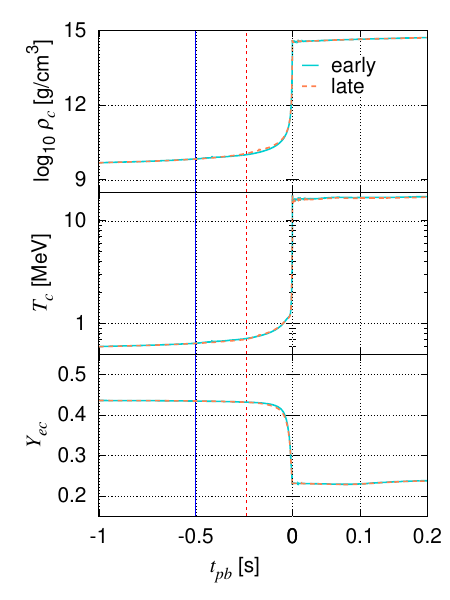}
    \includegraphics[width=11.2cm]{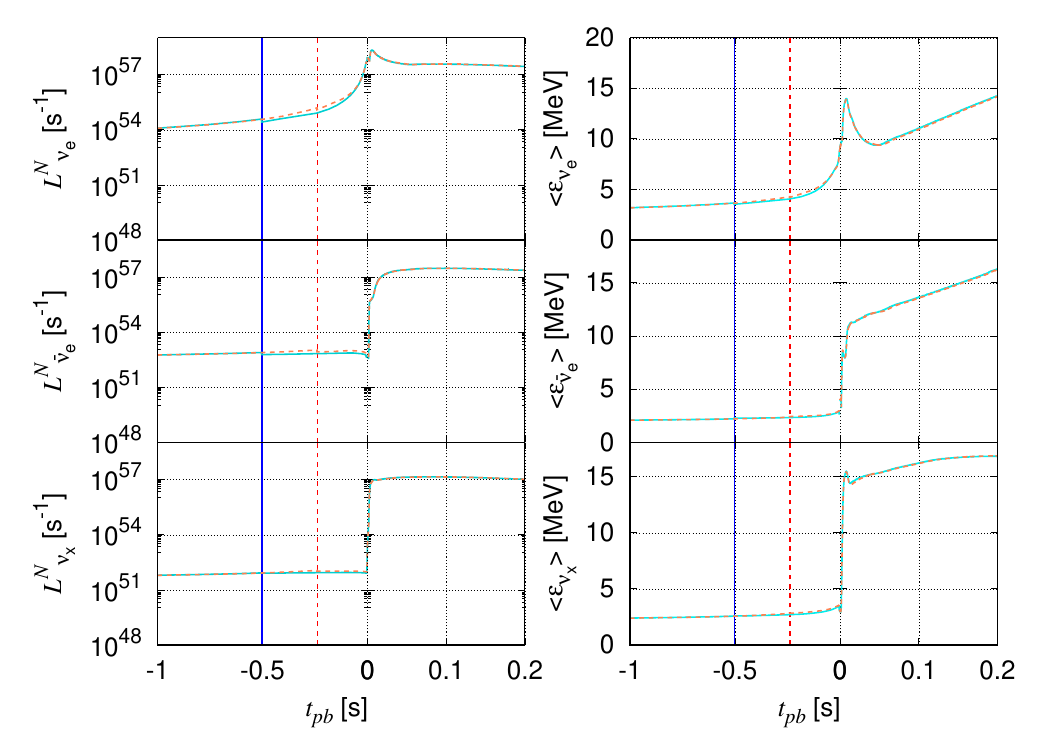}
    \caption{Comparison between early and late timing for switching calculations in 15~$M_\odot$ progenitors. Top and bottom panels show the time evolution of hydro quantities and neutrino properties, respectively. Blue and red vertical lines indicate the timing for switching calculations for early and late cases, respectively.}
    \label{fig:timing}
\end{figure*}

We transition from quasi-steady stellar evolution calculations to dynamical calculations at the onset of collapse. Because these two calculations rely on different hydrodynamical equations, neutrino interaction treatments, and equations of state, the mapping can inherently introduce systematic differences. As discussed in Section~\ref{subsec:setup}, numerical constraints prevent the use of identical onset criteria for mapping the HOSHI and MESA models. To assess the impact of these differing criteria, we compare the evolution of hydrodynamic variables and neutrino luminosities for a 15~$M_\odot$ MESA progenitor using two distinct mapping points. One criterion (adopted for the HOSHI models) maps the star when the central temperature reaches $T_c = 10^{9.9}$~K. The other (the default for MESA models) triggers the transition when more than 10\% of the iron core mass attains an infall velocity exceeding 100~km/s. In the MESA evolutionary track, the velocity-based condition is satisfied earlier ($t_{pb} \sim -0.502$~s, designated as the “early” mapping), while the temperature-based criterion is met later ($t_{pb} \sim -0.239$~s, designated as the “late” mapping).

As shown in Figure~\ref{fig:timing}, we observe only marginal differences in the time evolution of both the hydrodynamic quantities (left panel) and the neutrino properties (center and right panels), even extending into the postbounce phase. These results indicate that the exact timing of the computational transition, within this practical range, does not meaningfully alter the core dynamics or neutrino emissions. We then confirm that our main conclusions are robust against the choice of mapping criteria.

\section{The Modification of Time for Neutrino Emission in the Dynamical Evolution} \label{appendix_neutrino_time}

As described in Section~\ref{lumcalc}, we derive the postbounce neutrino luminosities and spectra from our dynamical simulations. In standard SN simulations, these properties are typically evaluated using distribution functions at a fixed radius of approximately $500\,\mathrm{km}$ (or $1000\,\mathrm{km}$). However, because our study continuously tracks neutrino emission from the collapse phase through $200\,\mathrm{ms}$ postbounce, this conventional approach becomes problematic—especially for progenitors with extended cores. For instance, the $40\,M_\odot$ model possesses a massive iron core extending to roughly 5,000~km prior to collapse. Evaluating the luminosity strictly at $500\,\mathrm{km}$ during the late collapse phase ($\rho_c > 10^{13}\,\mathrm{g\,cm^{-3}}$) would neglect contributions from the outer core, artificially underestimating the total emission. On the other hand, evaluating the emission too close to the outer edge of the computational domain is also problematic, as Boltzmann solvers in CCSN codes are susceptible to numerical artifacts near the outer boundary. To balance these competing issues, we calculate the neutrino luminosity at two locations: at $500\,\mathrm{km}$ and at a radius safely inside the outer boundary. As mentioned in the main text, we dynamically adopt the larger of these two values to ensure the total emission is accurately captured leading up to core bounce.

Implementing this dual-radius evaluation, however, requires careful handling of the time coordinate to preserve causality. Switching the evaluation radius dynamically without accounting for the finite propagation time of neutrinos would introduce artificial discontinuities in the lightcurves. To resolve this, we map all evaluations to a common retarded time, assuming that neutrinos originate from the stellar center and propagate outward at the speed of light. Specifically, letting $t_{\mathrm{hydro}}$ denote the instantaneous simulation time and $r$ the radius where the luminosity is evaluated, we define the proper neutrino time label, $t_\nu$, as:
\begin{eqnarray}
t_\nu = t_{\mathrm{hydro}} - \frac{r}{c}.
\end{eqnarray}
Throughout this study, $t_\nu$ serves as the consistent time coordinate for describing the time evolution of all neutrino properties, including the luminosity.

\section{Hydrodynamic and Thermal Evolution of Progenitor and SN Models} \label{appendix_hydro}

In this appendix, we present the details of hydrodynamic and thermodynamic evolution of the progenitor (Section~\ref{subsec:progenitor_model}) and SN models (Section~\ref{subsec:dynamical_model}). The general evolution of progenitors will be explained using the HOSHI models (solid lines). Finally, the code dependence of these results is discussed by comparing them with the MESA models (dotted lines) in Section~\ref{subsec:code_dependence}.

\begin{figure}[htpb]
    \centering
    \includegraphics[width=9cm,clip]{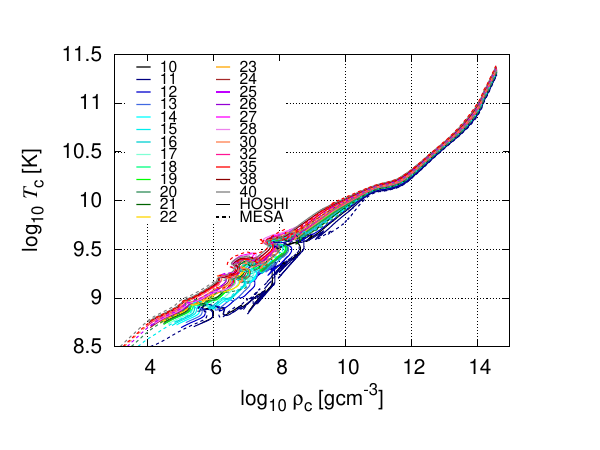}
    \caption{The $\rho_c$-$T_c$ diagram.
    Colors distinguish the progenitor models. The HOSHI and MESA models are described in solid and dotted lines, respectively.}
    \label{fig:rhoc_tc}
\end{figure}

\begin{figure}[htpb]
    \centering
    \includegraphics[width=\columnwidth,clip]{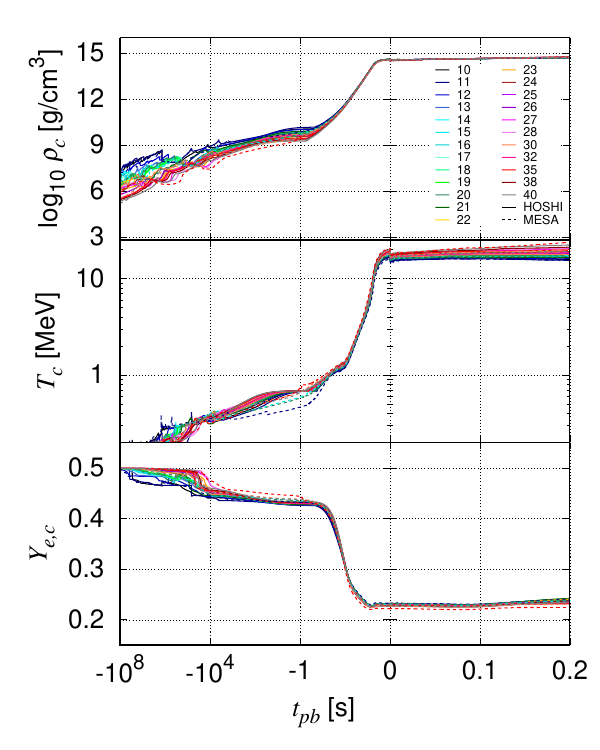}
    \caption{Time evolution of central density, temperature and the electron fraction from top to bottom. The colors distinguish the progenitor models. The HOSHI and MESA models are described in solid and dotted lines, respectively. We take $t_{pb}=0$ at the core bounce and the preSN phase represented by $t_{pb}<0$ and the SN phase represented by $t_{pb}>0$. The temperature evolution in the preSN phase is expanded in a small window.}
    \label{fig:progenitor_timeevo}
\end{figure}

\begin{figure*}[htpb]
    \centering
    \includegraphics[width=\textwidth,clip]{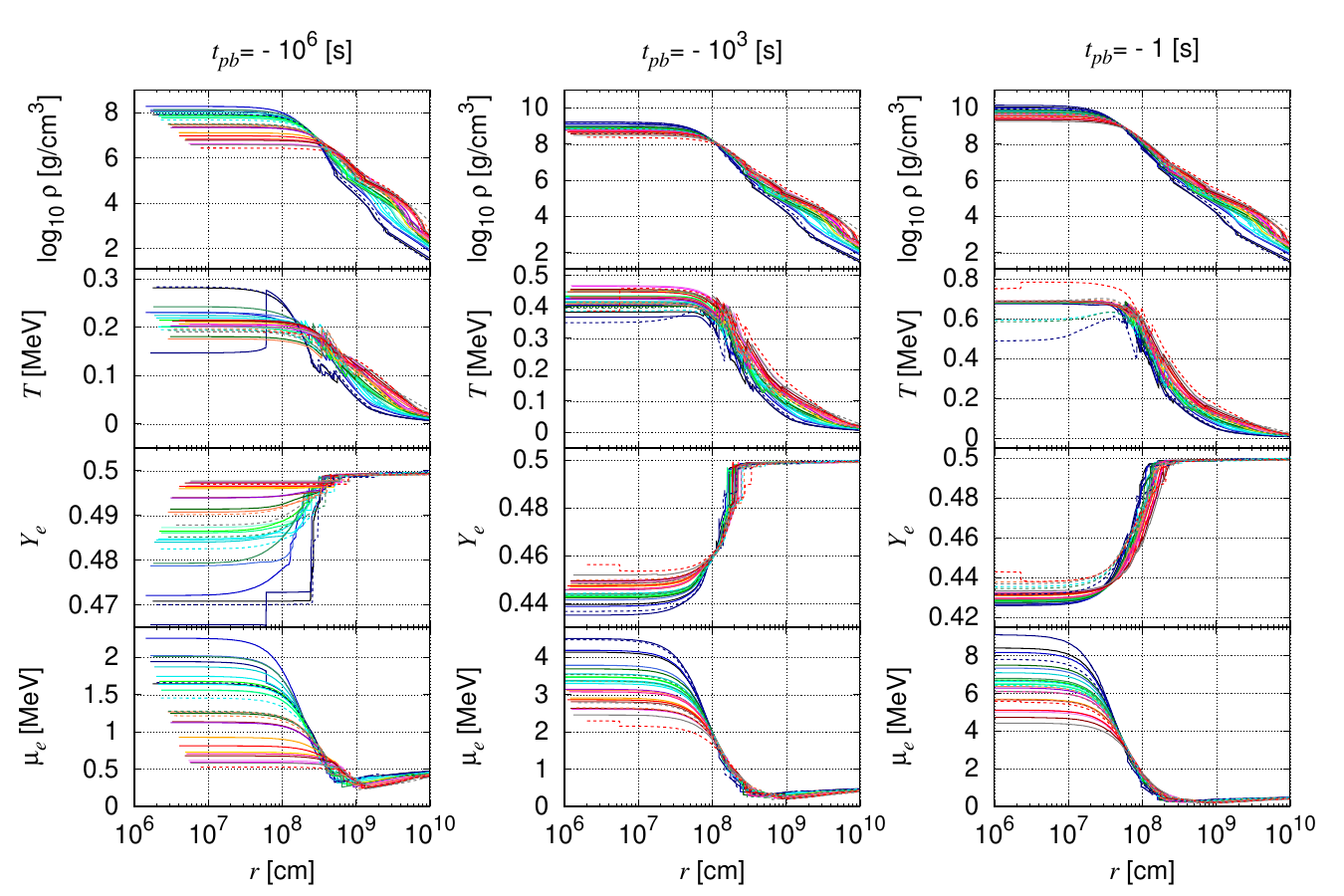}
     \includegraphics[width=\textwidth]{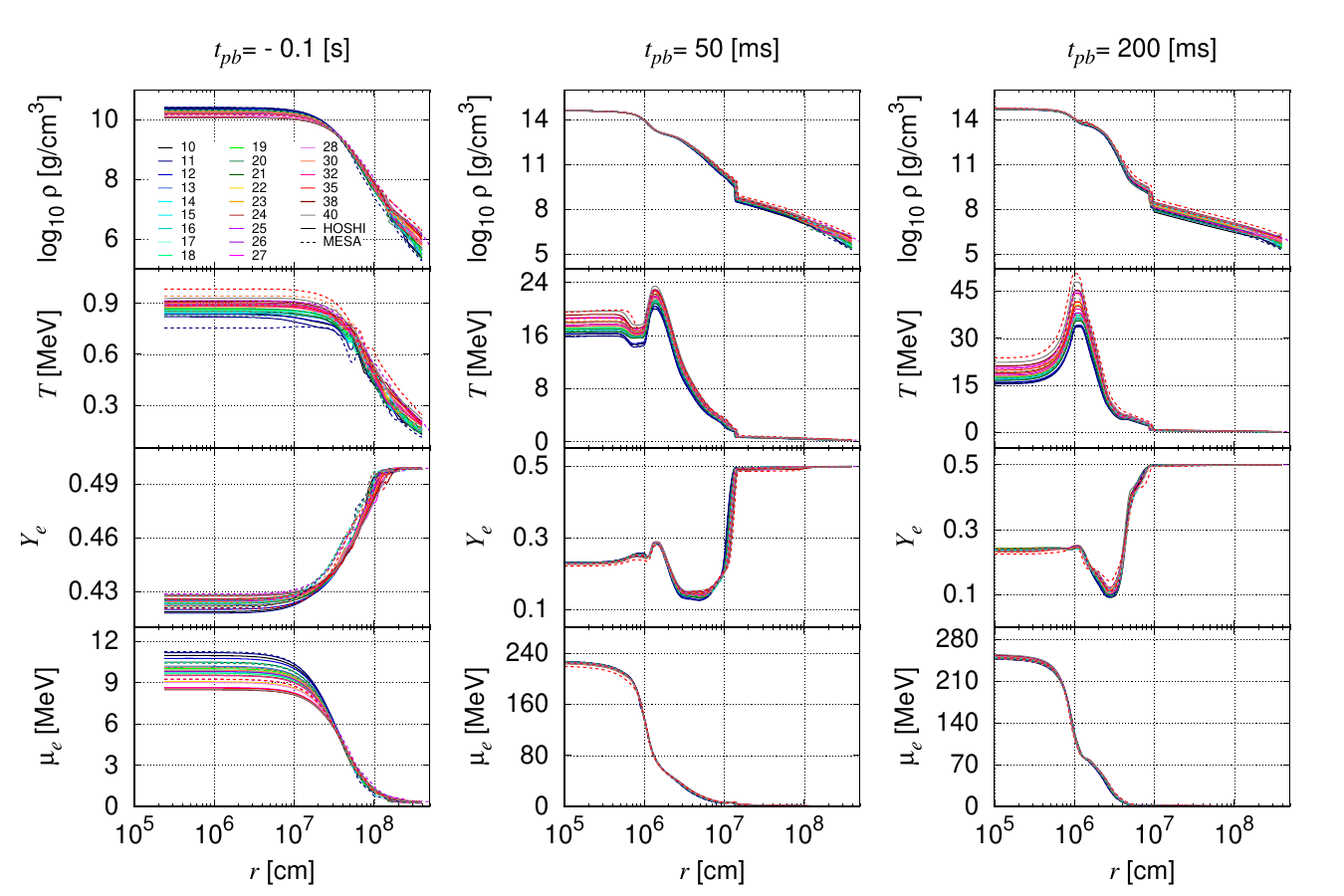}
    \caption{The same as Fig.~\ref{fig:progenitor_radial_prof} but with the MESA models. The HOSHI and MESA models are described in solid and dotted lines, respectively.}
    \label{fig:progenitor_radial_prof_MESA}
\end{figure*}

\subsection{Before Core Collapse} \label{subsec:progenitor_model}

In the mass range of 10--40 $M_\odot$, the central temperature is high enough to ignite carbon despite efficient neutrino cooling, and nuclear burning proceeds through the sequence of C $\rightarrow$ Ne $\rightarrow$ O $\rightarrow$ Si at the core. These burning processes manifest as loop-like structures in the $\rho_c$-$T_c$ diagram (Figure~\ref{fig:rhoc_tc}). However, at the low-mass end (e.g. 10 and 11 $M_\odot$), the core electrons become more degenerate compared to those in heavier progenitors. Consequently, the burning processes become slightly unstable, and these loop-like structures are no longer clearly observed, although nuclear burning continues. Around the core, shell burning generates nuclear energy and occasionally affects the thermal structure of the core, which also produces loop-like features in the diagram. Generally, stars with lighter cores are supported against self-gravity primarily by electron degeneracy pressure, which prevents a rapid rise in temperature. As a result, their evolutionary tracks shift to the right (toward higher densities) until the onset of core collapse at $\rho_c \sim 10^{11}\,\mathrm{g\,cm^{-3}}$.

The time evolution of the central quantities is shown in Figure~\ref{fig:progenitor_timeevo}. Stars with more massive cores achieve substantially higher pressures and temperatures, which makes their nuclear burning phases much shorter despite having a larger reservoir of fuel. As a result, when measured against the time to bounce ($t_{pb}$), lighter progenitors begin their steep increases in density and temperature earlier, while heavier progenitors catch up later in the evolution. For example, O-burning begins at $t_{pb}\sim-10^8$~s for the 11 $M_\odot$ model, whereas for the 40 $M_\odot$ model, it begins at $t_{pb}\sim-10^6$~s.

The radial profiles of the hydrodynamical and thermal quantities are shown in Figure~\ref{fig:progenitor_radial_prof_MESA}. Since a general overview of the radial profile evolution is provided in the main text, we focus here on the details of individual progenitor models. As mentioned above, in the very early phase ($t_{pb}\sim-10^6$~s), the profiles exhibit significant diversity because each model is captured at a slightly different evolutionary stage. For example, the 11 $M_\odot$ model shows high temperatures in the outer layers due to Si burning ignited slightly off-center ($r\sim 5\times10^7$~cm), whereas the central region is effectively cooled by neutrino emission, resulting in a significantly lower central temperature compared to other models. In contrast, the 10 $M_\odot$ model exhibits the highest central temperature due to active core O-burning. Regarding models with more massive cores, taking the 40 $M_\odot$ model as an example, central carbon has just been depleted at this stage. As the evolution progresses further, these differences in burning phases gradually disappear. By $t_{pb} \sim -10^3$~s, with the exception of a few models with the least massive cores, almost all progenitors are actively undergoing core Si-burning. Furthermore, as discussed in the main text, as the evolution approaches core collapse ($t_{pb} \sim -1$~s), the temperature dependence of the core structure becomes almost negligible. Consequently, the distributions of other physical quantities converge to profiles determined almost entirely by the core size.

\begin{figure}[htbp]
    \centering
    \includegraphics[width=8cm]{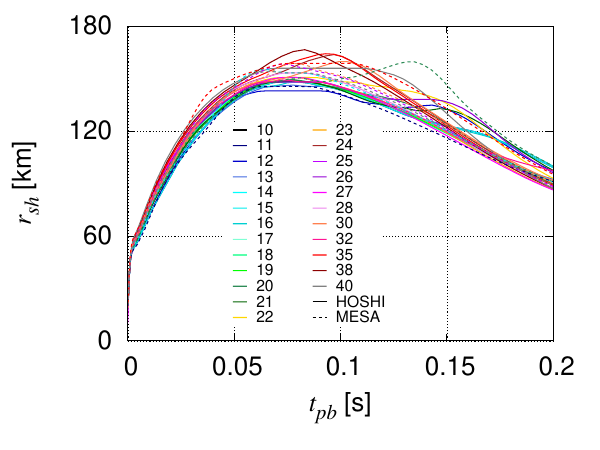}
    \caption{Time evolution of shock radius. The HOSHI and MESA models are described in solid and dotted lines, respectively.}
    \label{fig:shock_evo}
\end{figure}

\subsection{After Core Collapse} \label{subsec:dynamical_model}

Regarding the evolution after the onset of core collapse ($t_{pb}\sim-1$~s), the central quantities show little variation among the models, partly because the same EOS is utilized in this study (Figures~\ref{fig:rhoc_tc} and \ref{fig:progenitor_timeevo}). The only exception is the time evolution of the central temperature, which clearly reflects the initial differences in the core size and thermodynamic state of the progenitors.

The radial profiles of the thermal quantities after core bounce are shown in Figure~\ref{fig:progenitor_radial_prof_MESA}. We select two snapshots at $t_{pb}=50$~ms and 200~ms. We observe a large discontinuity in the density profiles at $r\sim150$~km (100~km) for $t_{pb}=50$~ms ($t_{pb}=200$~ms). This discontinuity corresponds to the position of the shock, which moves inward over time due to the failure of the explosion. The exact evolution of the shock position can be seen in Figure~\ref{fig:shock_evo}. We also find a large temperature peak at $r\sim10$--$15$~km, which roughly corresponds to the outer boundary of the PNS. This peak is generated by heating via matter accretion onto the PNS. The radial distribution of $Y_e$ at $t_{pb} = 50$~ms exhibits a prominent dip at $r \sim 10^6$--$10^7$~cm. This feature demonstrates that the shock passage dissociates heavy nuclei into free nucleons, thereby significantly enhancing the EC rates. By $t_{pb} = 200$ ms, however, this profile evolves: $Y_e$ experiences a slight decline in the inner vicinity of the PNS, while it rises in the outer region immediately behind the shock. This spatial variation arises because the postshock layer corresponds to the gain region, where the absorption of $\nu_e$ and $\bar{\nu}_e$ by nucleons outweighs emission ($n+\nu_e\rightarrow p+e^-$ and $p+\bar{\nu}_e\rightarrow n+e^+$). In contrast, in the deeper layers, neutrino cooling via the inverse reactions ($p+e^-\rightarrow n+\nu_e$ and $n+e^+\rightarrow p+\bar{\nu}_e$) causes $Y_e$ to decrease.

\subsection{Dependence on the Stellar Evolution Code} \label{subsec:code_dependence}

In this section, we compare the time evolution of the central quantities and the radial profiles of the hydrodynamic quantities between the HOSHI and MESA codes. The results for the MESA models are represented by dotted lines in Figures~\ref{fig:rhoc_tc}--\ref{fig:shock_evo}. Overall, the MESA models exhibit evolutionary trends similar to those of the HOSHI models. However, even for progenitors with the same initial ZAMS mass, differences in the thermal evolution emerge due to variations in the incorporated physics and numerical treatments. This fact underscores the limitation of relying solely on the ZAMS mass to characterize SN progenitors.

The most prominent difference between the two codes appears in the central temperature just prior to core collapse ($-100$~s $\lesssim t_{pb} \lesssim -1$~s). As shown in the middle panel of Figure~\ref{fig:progenitor_timeevo}, the central temperature at the onset of collapse is nearly uniform across all HOSHI models due to their specific mapping criteria. In contrast, the MESA models exhibit variations in the central temperature of up to several tens of percent. This dispersion is also evident in the radial distributions shown in the right panel of Figure~\ref{fig:progenitor_radial_prof_MESA}.

Furthermore, the rate of increase in the central temperature immediately prior to core collapse differs significantly between the models generated by the two codes. In the HOSHI models, a gradual deceleration in the temperature rise is observed, whereas the MESA models do not exhibit such a feature. This discrepancy is attributed to differences in how the acceleration term is treated within the respective codes. While the hydrostatic equilibrium equation is generally solved for the evolution leading up to collapse, the MESA models include the acceleration term, effectively solving the full hydrodynamic equations (using the \texttt{change\_v\_flag} option). In contrast, the HOSHI code assumes quasi-static evolution up to the very onset of collapse, estimating the matter velocity from the positional differences of fixed mass coordinates between adjacent time steps in a Lagrangian framework \citep{Takahashi2018}. It is worth noting that the sudden change in the slope of the time evolution for the hydrodynamic quantities, observed when transitioning from the HOSHI code to the dynamic CCSN code, is likely an artifact of this hydrostatic assumption.

The variations in the central temperature immediately prior to core collapse, observed particularly in the MESA models, are not strongly reflected in the central density. In fact, the dispersion in central density is notably smaller. This is because the iron core is predominantly supported by electron degeneracy pressure, which is independent of temperature, despite minor contributions from temperature-dependent thermal pressure. This narrow distribution in density is directly mirrored in $\nu_e$, resulting in a reduced model dependence for $\mu_e$ among the MESA models.

Regarding $Y_e$, its value is fundamentally determined by the extent of EC, making the evolutionary history of density and $\mu_e$ critical. Compared to the HOSHI models, the MESA models generally experience a less pronounced increase in density and $\mu_e$ leading up to core collapse, which results in a higher final $Y_e$. As will be demonstrated in the following section, these structural differences are pivotal in understanding the variations in neutrino luminosity.

Finally, we comment on certain progenitor models that exhibit somewhat peculiar characteristics, such as the 35 $M_\odot$ model. This model possesses an extended core with a relatively low $\rho_c$ and high $T_c$ just prior to collapse. As will be discussed in Section~\ref{subsec:key_parameter}, this is likely because C-shell burning migrates inward, preventing vigorous core O-burning and thereby reducing the efficiency of neutrino cooling. As a result, a compact core with high temperature is formed. 

Even after the core becomes unstable, the postbounce evolution of the MESA models follows the general trends observed in the HOSHI models (Figure~\ref{fig:progenitor_radial_prof_MESA}). Specifically, in accordance with Mazurek's law, the dependence of the physical quantities (other than temperature) on the specific progenitor model gradually diminishes, ultimately leading to nearly identical core structures.

\section{Neutrino Luminosity and Spectra for the MESA Models} \label{appendix_mesa}

\begin{figure*}[t]
    \centering
\includegraphics[width=\textwidth,clip]{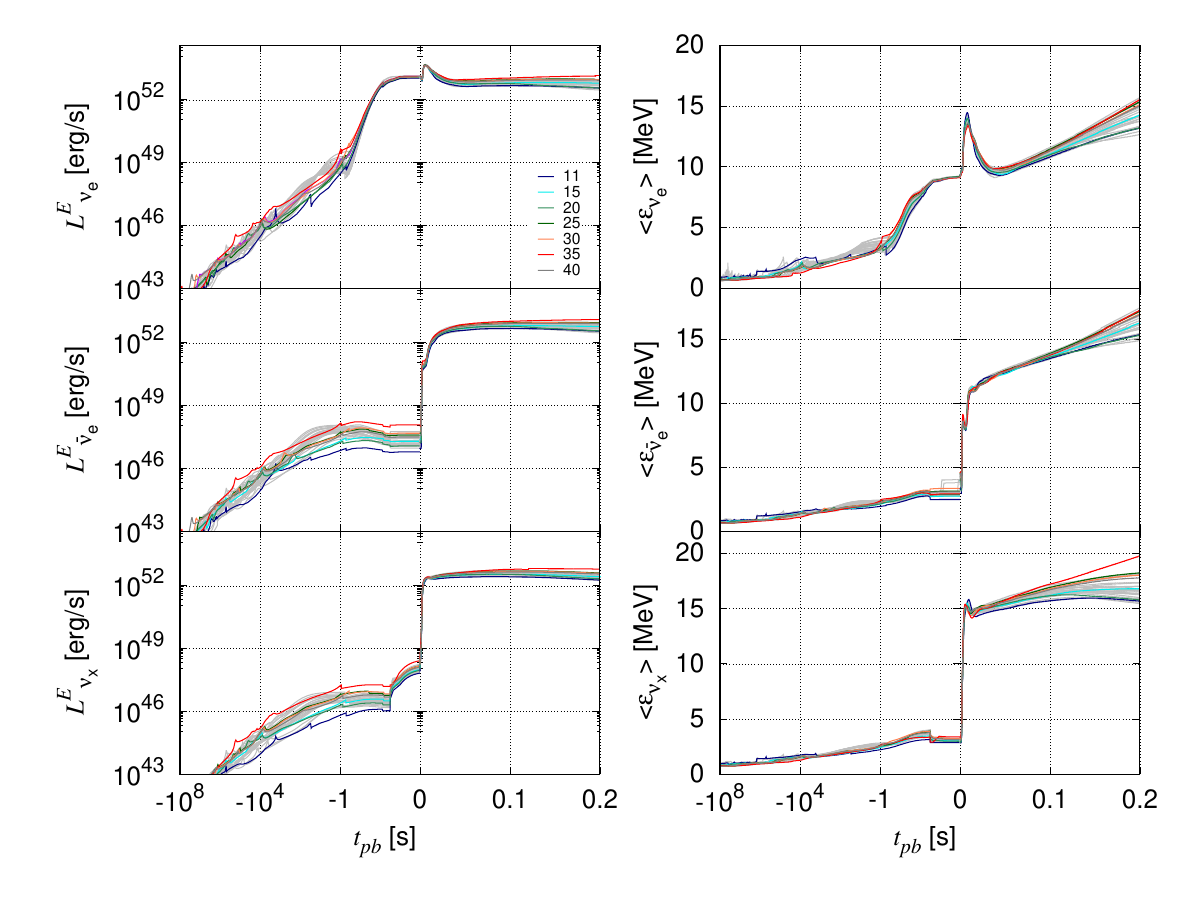}
    \caption{The same as Fig.~\ref{fig:numberlum} but with the MESA models. The HOSHI models are also shown in gray lines.}  
    \label{fig:numberlum_MESA}
\end{figure*}

\begin{figure*}
    \centering
\includegraphics[width=17cm,clip]{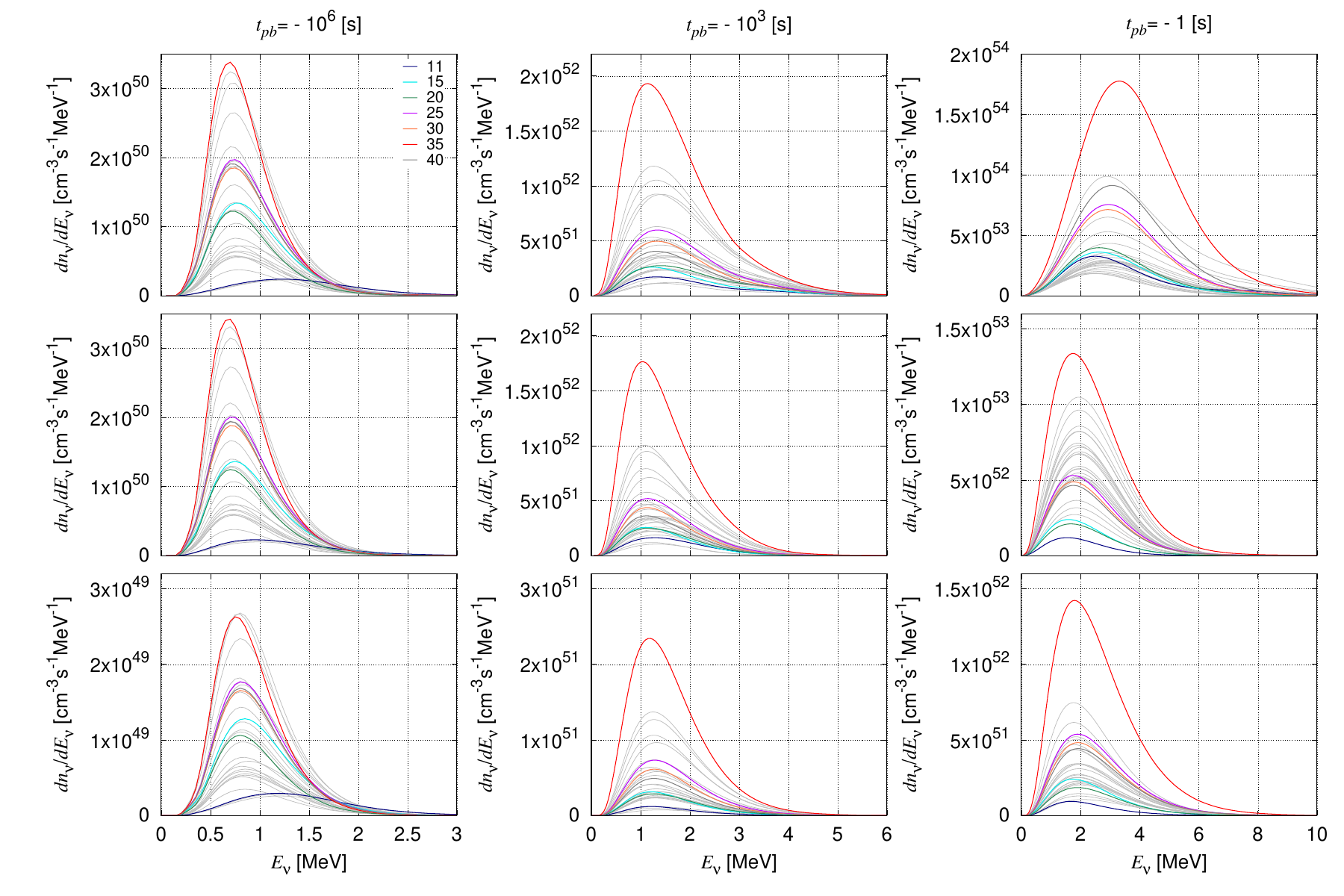}
\includegraphics[width=17cm,clip]{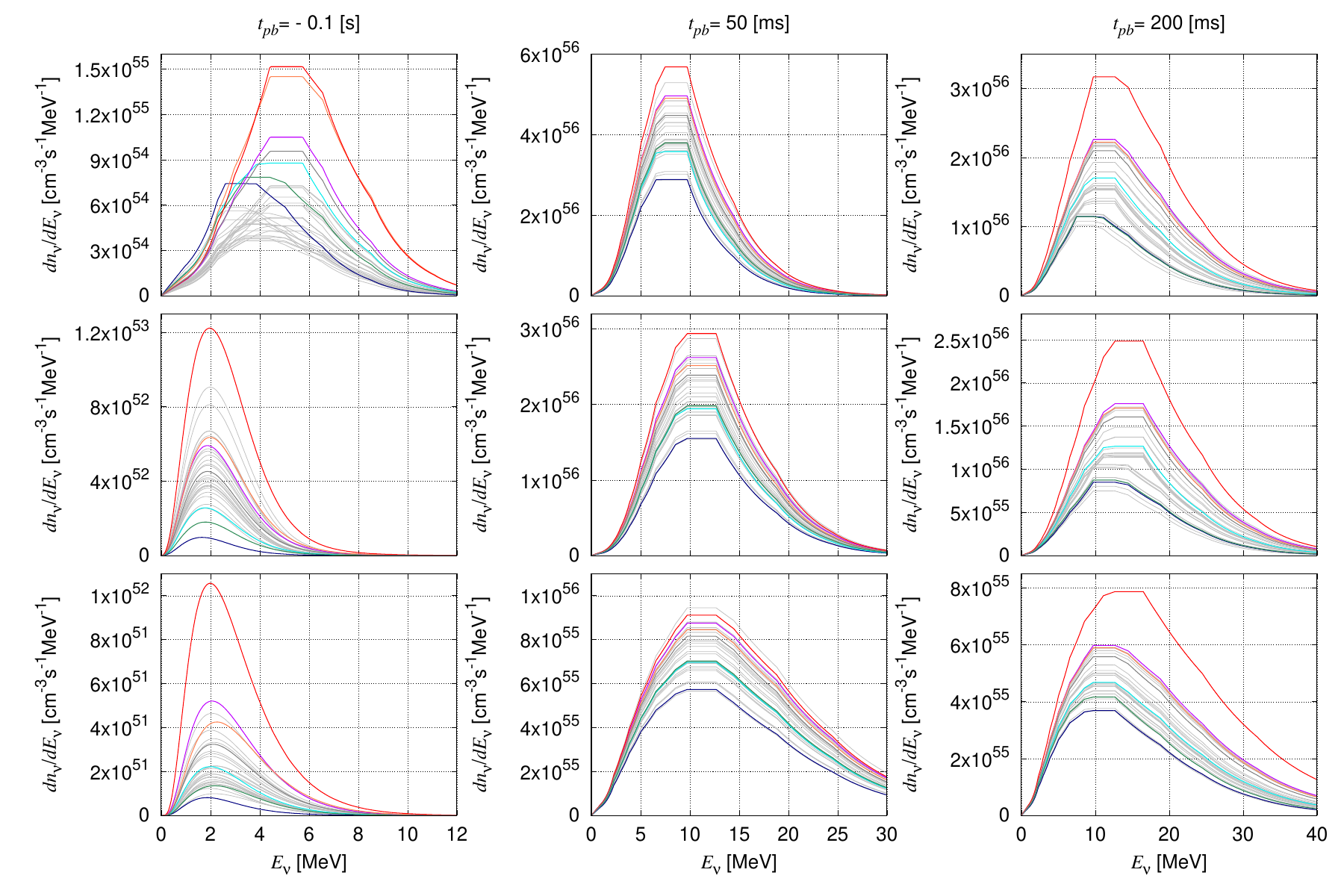}
    \caption{The same as Fig.~\ref{fig:numberspe} but with the MESA models. The HOSHI models are also shown in gray lines.} 
    \label{fig:numberspe_MESA}
\end{figure*}

In this appendix, we detail the properties of neutrino emission for the MESA models, focusing primarily on their deviations from the HOSHI models. The luminosities and spectra are summarized in Figures~\ref{fig:numberlum_MESA} and \ref{fig:numberspe_MESA}, respectively.

\begin{figure*}[t]
    \centering
\includegraphics[width=\textwidth,clip]{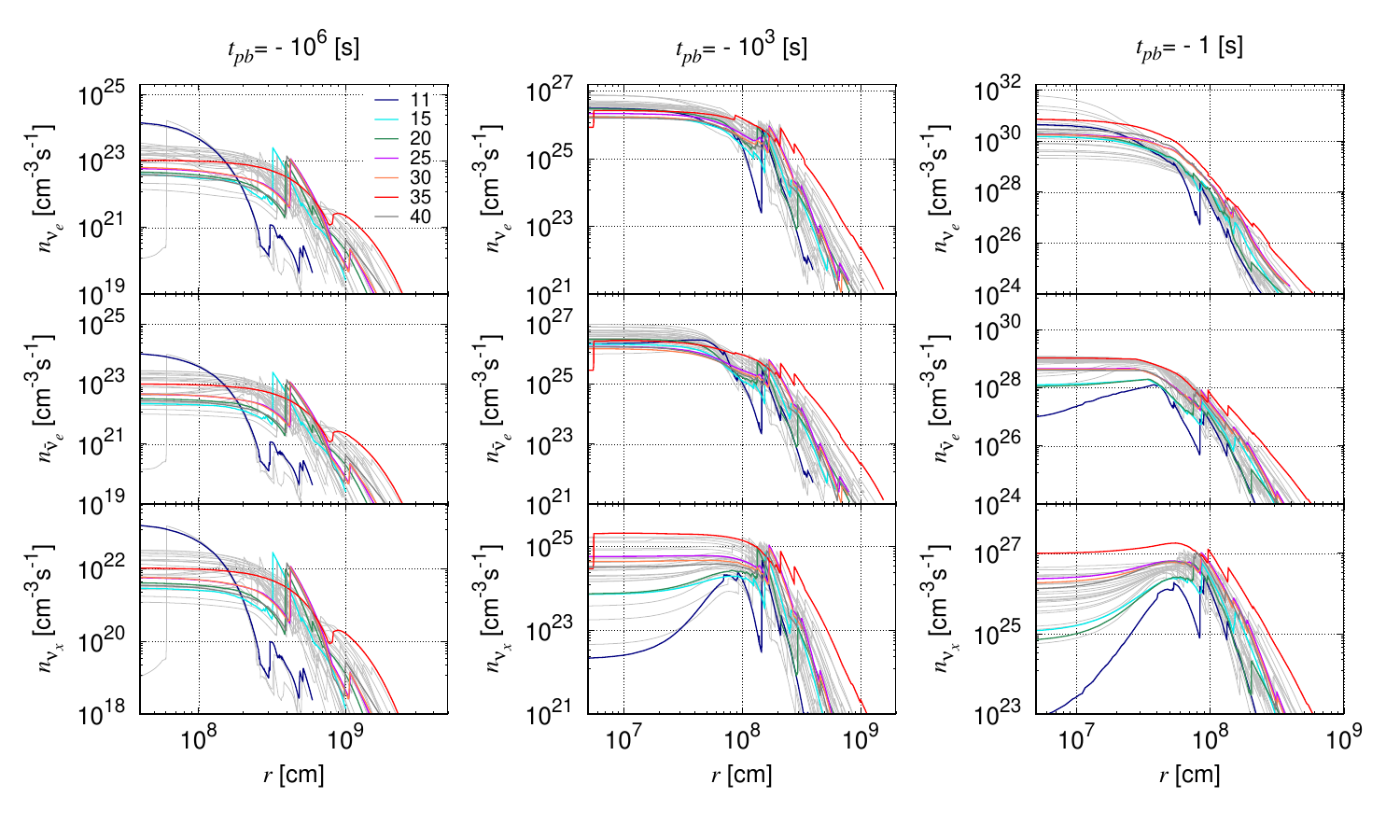}
    \caption{Radial profiles of neutrino number densities for $\nu_e$ (top), $\bar{\nu}_e$ (middle) and $\nu_x$ (bottom) in the MESA models. The HOSHI models are also shown in gray lines.}  
    \label{fig:radial_prof_neutrino_MESA}
\end{figure*}

The most significant deviation from the HOSHI results concerns the $\nu_e$ luminosity immediately preceding core collapse. As shown in the top panel of Figure~\ref{fig:numberlum}, the MESA models do not have the characteristic inversion or crossing of $\nu_e$ luminosities observed among the HOSHI models around $t_{pb} \sim -100$ to $-1$~s. Instead, the 35~$M_\odot$ model maintains the highest $\nu_e$ luminosity throughout the progenitor evolution, while the 11~$M_\odot$ model remains the lowest. This disparity is further illustrated by the radial distributions in Figure~\ref{fig:radial_prof_neutrino_MESA} (top right). For instance, at $t_{pb} \sim -1\,\text{s}$, the 35~$M_\odot$ MESA model possesses a highly extended high-$T$ core, where the contribution of pair to the $\nu_e$ emission is substantially larger than in the corresponding HOSHI model. Consequently, its total $\nu_e$ luminosity vastly exceeds that of the 11~$M_\odot$ model. While the 11~$M_\odot$ model still generates substantial $\nu_e$ emission via electron capture in its compact, high-density core, it is outpaced by the sheer volume of pair emission in the 35~$M_\odot$ model.

Crucially, it should be emphasized that these systematic differences between the two sets of stellar evolution codes do not alter our primary conclusion: the time-integrated neutrino emission retains a remarkably strong correlation with the key structural parameters of the SN progenitors (e.g. $\xi_{2.5}$).

Regarding the postbounce evolution, we observe no significant discrepancies between the MESA and HOSHI models. The results are fundamentally consistent with the physical scenario presented in the main text; specifically, the emission of $\nu_e$ exhibits robust universality, showing significantly less model dependence than the other neutrino flavors.

\section{Dependence on the Mass Coordinate of Compactness} \label{appendix_compactness}

\begin{figure*}[ht]
    \centering
    \includegraphics[width=\textwidth]{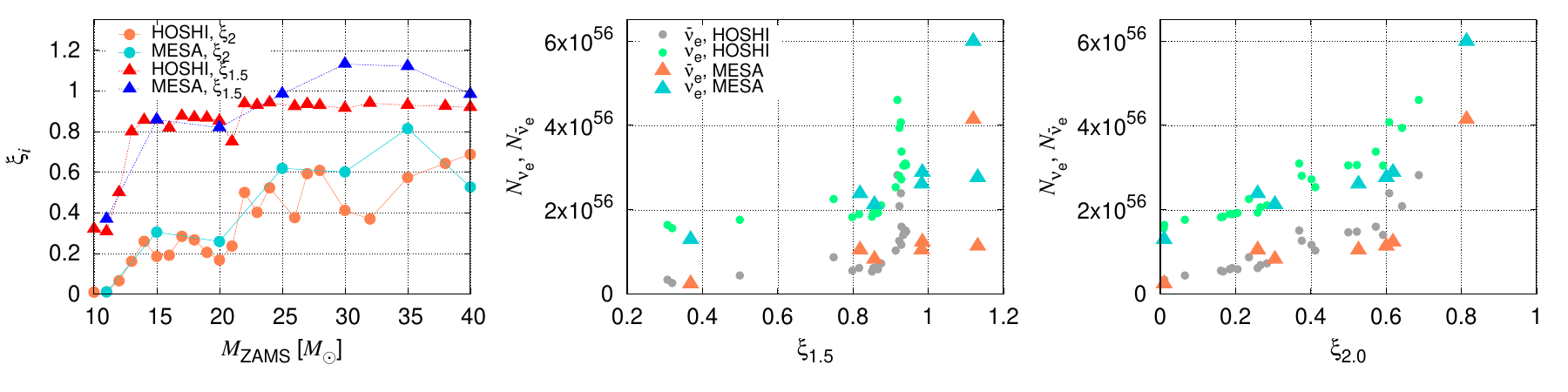}

    \caption{Left panel shows $\xi_{1.5}$ and $\xi_{2}$ for our progenitor models. Middle and right panels show the relationship between the integrated number of neutrinos and $\xi_{1.5}$ and $\xi_{2}$, respectively.}
    \label{fig:compactness_depend}
\end{figure*}

In this section, we investigate how the correlation with the time-integrated preSN neutrino emission changes when the mass coordinate defining the compactness parameter is varied. Specifically, focusing on the phase immediately prior to the explosion (i.e., the start time of integration $t_i = -10^5\,\text{s}$), where a strong correlation was previously observed with $\xi_{2.5}$, we examined the impact of altering the reference mass coordinate from 2.5~$M_\odot$ to $1.5\,M_\odot$ and $2.0\,M_\odot$.

The results are presented in Figure~\ref{fig:compactness_depend}. As evident from this figure, the correlation weakens as the mass coordinate defining the compactness is shifted inward. This behavior can be understood from the relationship with the neutrino emission regions, represented by the mass coordinates $M_{\nu_e}$ and $M_{\bar{\nu}_e}$, as discussed in the main text. As shown in Figure~\ref{fig:Mnu_evo}, particularly in heavier progenitors, the effective mass coordinates $M_{\nu_e}$ and $M_{\bar{\nu}_e}$ are located outside $2.0\,M_\odot$, meaning that they fall outside the region crucial to determine the compactness if defined at lower masses. We interpret this mismatch as the primary cause of the weak correlation. Therefore, we conclude that, in the context of this study, $2.5\,M_\odot$ serves as a superior mass coordinate to determine the compactness parameter.

\section{Dependence of the Time Window and False Alarm Rate} \label{appendix_alert}

\begin{figure}[ht]
    \centering
    \includegraphics[width=\columnwidth]{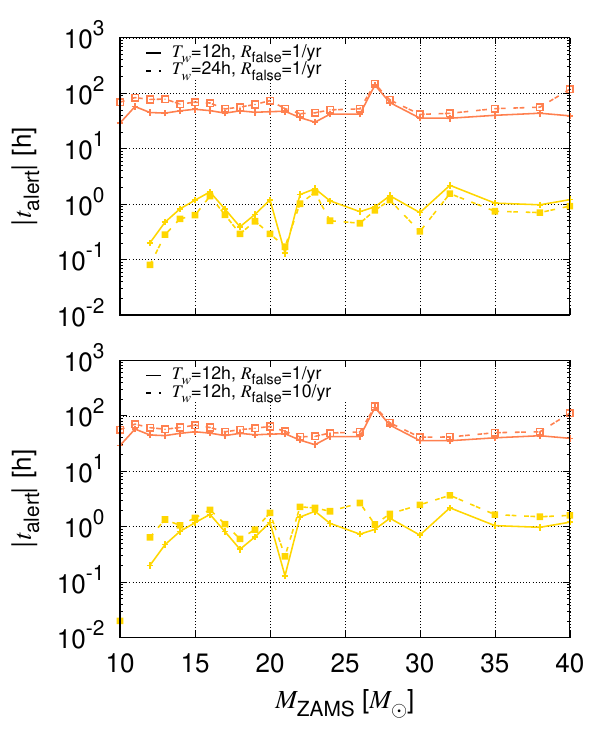}
    \caption{Dependence of the alert time $t_{\rm alert}$ on $T_w$ (top) and $R_{\rm false}$ (bottom). Solid and dotted lines denote the HOSHI and MESA models, respectively. We assume that the distance to progenitors is $d=200$ pc and neutrinos follow the NO. Colors distinguish neutrino detectors. }
    \label{fig:begintime_Tw_Rfalse}
\end{figure}

Figure~\ref{fig:begintime_Tw_Rfalse} shows the dependence of the alert time on $T_w$ (top) and $R_{\text{false}}$ (bottom) for the NO case. Only the results for JUNO (coral) and HK (gold) are shown. First, comparing with the results for an extended window of $T_w = 24\,\text{h}$, although both the background and the signal counts increase, if the signal rate exceeds the background rate, the alert time can be advanced. 
Indeed, in JUNO, which has a low background rate, using $T_w = 24\,\mathrm{h}$ actually results in an earlier alert timing. On the other hand, for HK, where the background rate is higher, our results indicate that $T_w = 12\,\mathrm{h}$ is more ideal for issuing an alert.

Next, we discuss the dependence on $R_{\text{false}}$. Comparing with the results for $R_{\text{false}} = 10/\text{yr}$, which corresponds to allowing a ten-fold increase in the false alarm probability, we find that the alert times are intuitively advanced for all models.

\bibliography{reference}

\end{document}